\documentclass[fleqn,usenatbib]{mnras}

\usepackage{newtxtext,newtxmath}  

\usepackage[T1]{fontenc}
\usepackage{ae,aecompl}

\usepackage{graphicx}	
\usepackage{amsmath}	

\defcitealias{GH2005}{GH2005}
\defcitealias{GH2008}{GH2008}
\defcitealias{Grassitelli+2018}{G+2018}
\defcitealias{Sander+2020}{S+2020}
\defcitealias{VdK2005}{VdK2005}

\title[Massive He star winds and WR-type mass loss]{On the nature of massive helium star winds and Wolf-Rayet-type mass loss}

\author[A. A. C. Sander \& J. S. Vink]{
Andreas A. C. Sander$^{1}$\thanks{E-mail: Andreas.Sander@armagh.ac.uk}
and 
J. S. Vink$^{1}$
\\
$^{1}$Armagh Observatory and Planetarium, College Hill, Armagh BT61 9DG, Northern Ireland, UK%
}

\date{Accepted 2020 September 01. Received 2020 September 01; in original form 2020 July 14}

\pubyear{2020}

\begin{document}
\label{firstpage}
\pagerange{\pageref{firstpage}--\pageref{lastpage}}
\maketitle

\begin{abstract}
The mass-loss rates of massive helium stars are one of the major uncertainties in modern astrophysics. Regardless of whether they were stripped by a binary companion or managed to peel off their outer layers by themselves, the influence and final fate of helium stars -- in particular the resulting black hole mass -- highly depends on their wind mass loss as stripped-envelope objects. While empirical mass-loss constraints for massive helium stars have improved over the last decades, the resulting recipes are limited to metallicities with the observational ability to sufficiently resolve individual stars. Yet, theoretical efforts have been hampered by the complexity of Wolf-Rayet (WR) winds arising from the more massive helium stars.
In an unprecedented effort, we calculate next-generation stellar atmosphere models resembling massive helium main sequence stars with Fe-bump driven winds up to $500\,M_\odot$ over a wide metallicity range between $2.0$ and $0.02\,Z_\odot$. We uncover a complex $\Gamma_\text{e}$-dependency of WR-type winds and their metallicity-dependent breakdown. The latter can be related to the onset of multiple scattering, requiring higher $L/M$-ratios at lower metallicity. Based on our findings, we derive the first ever theoretically-motivated mass-loss recipe for massive helium stars. We also provide estimates for LyC and \ion{He}{ii}\,ionizing fluxes, finding stripped helium stars to contribute considerably at low metallicity. In sharp contrast to OB-star winds, the mass loss for helium stars scales with the terminal velocity. While limited to the helium main sequence, our study marks a major step towards a better theoretical understanding of helium star evolution.
\end{abstract}




\section{Introduction}
  \label{sec:intro}

Massive stars are important drivers in the Universe. They are relevant for star-formation studies both nearby and far-away \citep[e.g.][]{Vink2020,Leitherer2020,Stanway2020}.
They are likely an important source for cosmic hydrogen (H) and helium (He) re-ionization, but their amount of ionizing flux strongly depends on the stellar wind mass-loss rate $\dot{M}$, and therefore on the host galaxy's metallicity ($Z$). While mass-loss rate predictions for H-burning stars have been produced for decades, for He-burning stars, the field is still in its infancy, with the first set of hydrodynamically consistent predictions only appearing earlier this year \citep{Sander+2020}.

In addition to the dense winds of classical massive Wolf-Rayet (WR) stars, another type of He-burning star has recently returned to the limelight. These binary-stripped He stars \citep{Paczynski1967,Podsiadlowski+1992} have even been discussed as the prime source of cosmic re-ionization \citep[e.g.][]{Stanway+2016,Goetberg+2020}. Yet, the ionizing flux budget of such binary-stripped stars is highly uncertain. Depending on their parameters, stripped stars could either have weak or absent winds, allowing a large fraction of ionizing flux to escape, or drive considerable mass outflows, which would enable a detection via emission lines appearing in their companion's spectrum, but diminish their ionizing flux budget. Given the lack of observational material in this regime, the determination of mass-loss rates and ionizing fluxes from sophisticated theoretical calculations is of major importance.

Apart from their role in setting the correct amount of ionizing radiation as a function of $Z$, stellar wind mass-loss rates are also a key ingredient for enriching the interstellar medium (ISM) with nuclear-processed material (contributing to the yields), as well as for determining massive star evolution, including the stars' final fates. \citet{Vink2017} and \citet{Gilkis+2019} showed that mass-loss rates of stripped stars are crucial to determine the fraction of stripped supernovae (SNe) of types IIb and Ibc \citep[e.g.][]{Yoon+2012,Groh+2013,Eldridge+2013}. Moreover, \citet{EV2006} and \citet{Belczynski+2010} showed that the WR $\dot{M}(Z)$-dependency of \citet{VdK2005} is the determining factor in establishing the $Z$-dependency of `heavy' Black Holes (BHs) of order 40 $M_{\odot}$ as detected by gravitational waves (GWs) with LIGO/VIRGO. Together with the star formation history, the strong dependence of $\dot{M}(Z)$, in particular for He stars, determines when and where massive BHs can be formed \citep[e.g.][]{DT2003,Hainich+2018,Woosley+2020} and which evolutionary pathways towards the merger of double-compact objects are possible at a certain $Z$ \citep[e.g.][]{Belczynski+2020,Langer+2020,Klencki+2020}.

Our initial computations \citep[][]{Sander+2020} revealed the need to develop a deeper physical understanding of the transition between optically thin and thick He stars, as this is of key importance for GW astronomy, SN progenitors, and He ionization in the Universe. In the H-rich part of the Hertzsprung–Russell diagram (HRD), this transition was studied in \citet{Vink+2011}, but the corresponding He-rich part of the HRD is only recently being investigated.

Massive He stars have been intensively studied in the framework of stellar structure \citep[e.g.][]{L1989,Graefener+2012,Grassitelli+2016} and evolution models \citep[e.g.][]{Langer+1994,Vanbeveren+1998,Georgy+2012,CL2013,ME2016,Woosley2019}. Beside the general importance of $\dot{M}$ for a star's fate, it has become clear that the appearance of massive He stars is considerably affected by WR-type mass loss \citep[e.g.][]{HL1996,Grassitelli+2018,Ro2019} lowering the observed effective temperatures. The launch of WR-type winds due to `hot iron bump' opacities \citep{NL2002,GH2005,Sander+2020} happens way beneath the photosphere which is formed far out in the wind \citep[e.g.][]{HL1996,Hamann+2006}, inhibiting a direct spectroscopic determination of the hydrostatic radii of WR stars.

\citet[][hereafter S+2020]{Sander+2020} have recently shown that the mass loss of classical WR stars not only shows a $Z$-dependency, but also a steep dependence on the Eddington-$\Gamma_\text{e}$, as found 
earlier for H-burning very massive stars (VMS), and discovered that below a critical $\Gamma_\text{e}$-value the mass-loss rates drop significantly, which may explain the disappearance of classical WR stars below a transition luminosity \citep{Shenar+2020}. This implies that extrapolations for lower-mass He stars from empirical recipes obtained from classical WR stars \citep[e.g.][]{NL2000} are highly inaccurate, confirming the earlier pilot study by \citet{Vink2017}. 

\begin{figure}
  \includegraphics[width=\columnwidth]{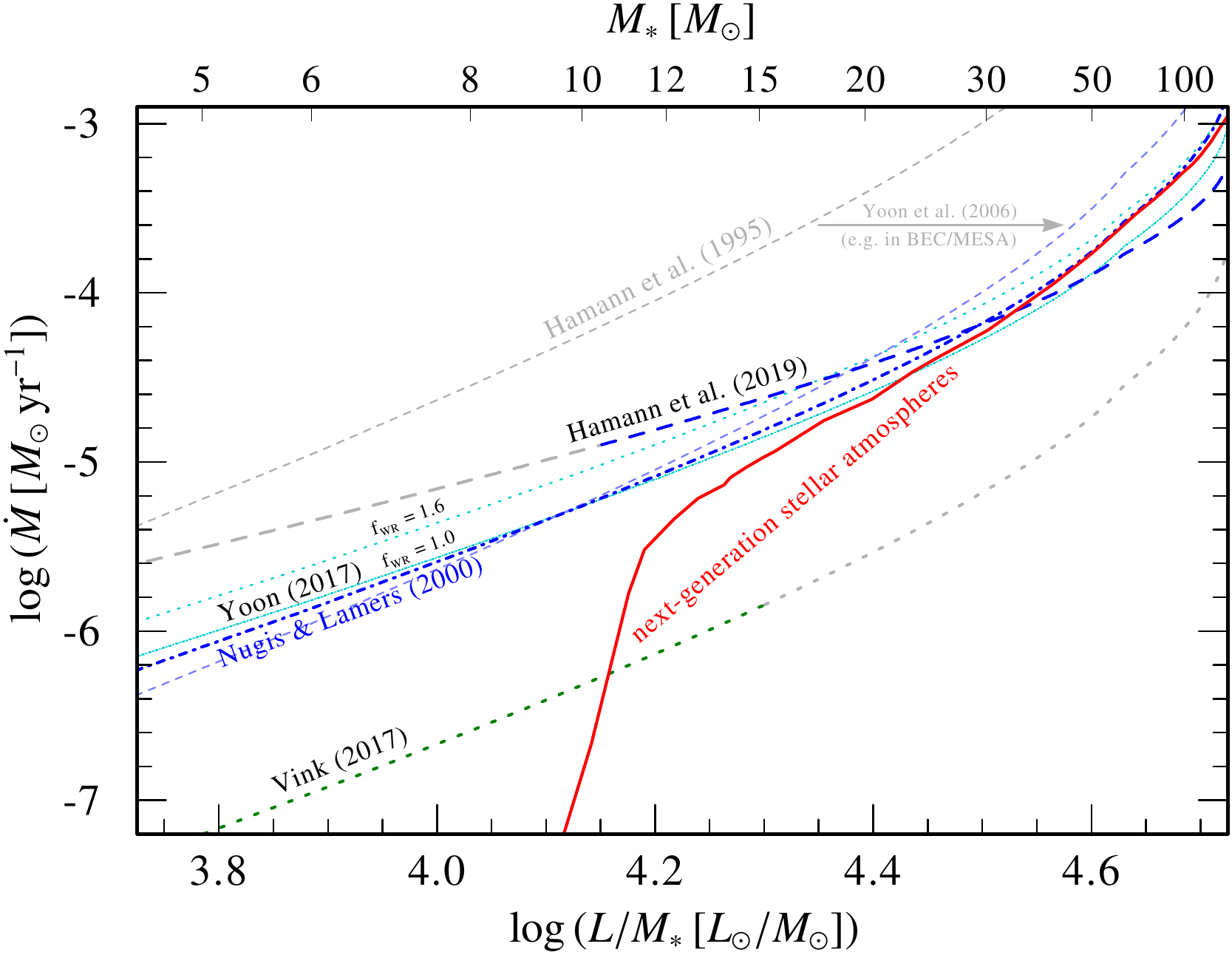}
  \caption{A selection of current mass-loss recipes for massive He stars at $Z_\odot$ applied in stellar evolution models, compared to a curve resulting from our newly calculated set of dynamically consistent atmosphere models.}
  \label{fig:cmp-mdot-recipes}
\end{figure}

Fig.\,\ref{fig:cmp-mdot-recipes} depicts a set of commonly used mass-loss recipes at $Z_\odot$ that straddle both the optically thick, classical WR regime, and the optically thin ``stripped star'' regime. For comparison, we have added the curve resulting from our calculations in this work. The steep decline of $\dot{M}$, which we already noticed in \citetalias{Sander+2020}, showcases one of the main motivations for our present study: Empirical recipes, e.g.\ from \citet{NL2000} or more recently \citet{Hamann+2019}, are based on observed spectra of WR stars, analysed with conventional model atmospheres, where the mass-loss rate is a free parameter. By performing a linear regression over the observed sample or a selected sub-sample, such as hydrogen-free WN stars, a formula for $\dot{M}$ as a function of $L$ or $L/M$ is found. The such determined relations are a representation of the obtained $\dot{M}$ in the studied sample, but their mathematical form does not reflect any deeper insight of the underlying wind physics. Extrapolating such a formula beyond the regime where they were obtained can yield mass-loss rates which can differ significantly from the actual ones, in the worst case by orders of magnitude as demonstrated in Fig.\,\ref{fig:cmp-mdot-recipes}.

In principle, theoretical modelling would allow us to study He star mass loss on a much broader scope, including the study of isolated parameter dependencies or parameter ranges without direct observational constraints. Unfortunately, the inherent complexity of the winds requires a considerable numerical effort \citep[e.g.][]{GH2005}. Consequently, the number of studies so far has been rather limited. \citet{VdK2005} investigated WR mass loss via two Monte Carlo model sequences, each reflecting a prototypical late-type WN and WC star at different $Z$. In a similar temperature regime, but for the very luminous WNh stars, which still have a significant amount of hydrogen, \citet{GH2008} calculated the first sequence of co-moving frame (CMF) models with local dynamical consistency. While their mass-loss recipe accounts for various parameters, it is limited to winds driven by the so-called `cool iron bump', preventing its application to much hotter objects. (Any simple extrapolation to the He star regime would yield mass-loss rates significantly lower than obtained in this work.)
 
While not intended to resemble classical WR stars, the calculations from \citet{Vink2017} were the first effort to derive a theoretical $\dot{M}(L,Z)$-formula for He stars beyond the subdwarf regime, albeit targeting the regime of stripped stars below the mass regime where typically WR stars are observed \citep[see, e.g.,][]{Shenar+2020}. In this work, we do not aim to reproduce low- and intermediate mass He stars as we restrict our calculations to a regime where the winds are launched by the `hot Fe bump'. For cooler stars, the He ZAMS bends to significantly lower temperatures, which is also reflected in the substantial temperature regime difference between \citet{Vink2017} and this work ($50\,$kK vs.\ $141\,$kK). We thus focus on the regime of WR-type mass loss and its onset for more massive He stars. Stripped He stars of lower mass will require their own, tailored investigation.

With the exception of the mass-loss recipe by \citet[][hereafter GH2008]{GH2008}, essentially all current descriptions of $\dot{M}$ for He and classical WR stars are power laws in $L$ (or $L/M$) and $Z$.
 Already \citet{Vink+2001} noted that an overall power-law behaviour in $Z$ is unlikely, e.g.\ due to the saturation of iron lines at higher $Z$. In a prototype study employing a new generation of hydrodynamically consistent atmosphere models with CMF radiative transfer, \citetalias{Sander+2020} demonstrated the complex behaviour of WR mass loss, both along the $L/M$- and the $Z$-dimension.
While power-law-type recipes can be sufficient for some applications, their use can have severe consequences for others, especially when relying on the asymptotic behaviour \citep[see, e.g., the recent comparisons for black hole populations in][]{Woosley+2020}. In this work, we therefore go a step further and calculate large sets of dynamically consistent models with $L$-$M$-combinations based on stellar structure considerations to understand the mass loss of hydrogen-free, massive He stars more accurately. To get a correct grip of the asymptotic behaviour for high $L$ or $L/M$ respectively, we extended our calculations to very high masses way beyond the observed regime of He stars. We will gain fundamental insights into the nature and breakdown of the WR-type mass loss and derive an unprecedented description of $\dot{M}$ for massive He zero age main sequence (He ZAMS) stars. In Sect.\,\ref{sec:powr}, we briefly introduce the underlying model concepts, before discussing all the results in-depth in Sects.\,\ref{sec:resmdot} (mass loss behaviour and recipes), \ref{sec:ztrends} (metallicity trends), and \ref{sec:ionflux} (ionizing fluxes). After our summary and conclusions in Sect.\,\ref{sec:conclusions}, brief appendices address side discussions on the relation of wind efficiency and wind optical depth (Sect.\,\ref{asec:etatau}), an additional test of our $\dot{M}(L)$-recipe (Sect.\,\ref{asec:etamax}), the departure from LTE at the launching point of the wind (Sect.\,\ref{asec:ltedep}), and comparisons to VMS (Sect.\,\ref{asec:bestencmp}) and \citetalias{GH2008} WNh results (Sect.\,\ref{asec:gh2008}).

\section{Stellar atmosphere models}
  \label{sec:powr}
	
In this work, we apply the PoWR code \citep[e.g.][]{GKH2002,HG2003,Sander+2015} to calculate so-called ``next-generation'' stellar atmosphere models, where we couple the results from the non-LTE atmosphere calculations with a consistent solution of the hydrodynamic equation of motion. The radiative transfer is performed in the CMF, thereby implicitly accounting for various effects such as line-overlapping and multiple scattering which are not covered in simpler, but faster calculation methods. A CMF radiative transfer is essential to properly treat the complex WR atmospheres, where both line-overlapping and multiple scattering play a major role. The concept of the next-generation models used in this work has been introduced in \citet{Sander+2017} and \citetalias{Sander+2020}, the latter discussing the application to WR stars.

Our prototypical study of hydrodynamically consistent WR atmosphere models \citepalias{Sander+2020} was focussed on the atmospheric behaviour, where we obtained the $\dot{M}(L/M)$-trend by using the mass-loss rate as a model input and iterated the stellar mass in the atmosphere calculations until a hydrodynamically consistent solution was obtained. This approach is numerically very fast and robust, but can lead to $L$-$M_\ast$-combinations which are not expected from stellar structure and evolution models. In this work, we therefore take a different approach and fix the mass and luminosity of the stars while iterating for $\dot{M}$ to get a hydrodynamically consistent solution. While this approach is slower and can lead to a bit more numerical scatter in the results, it allows us to study $\dot{M}$ for a prescribed set of values for $L$ and $M_\ast$. Given that various parameters influence the results, such as the individual abundances, the choice of $T_\ast$, or the parameters for clumping and micro-turbulence, the complexity of the calculations and the need for manual supervision require to reduce the parameter space. We therefore fix most parameters to one set, including the abundances, which reflect He ZAMS stars and we thus only scale with $Z$. (Different CNO abundances are discussed in Sect.\,\ref{sec:ztrends}.) We choose $L$ and $M_\ast$ such that they are described by the relations for hydrogen-free stars given in \citet{Graefener+2011}. The main input parameters for our models are compiled in Table\,\ref{tab:inputparams}.

\begin{table}
  \caption{Input parameters for our hydrodynamically consistent He star atmosphere models. }
  \label{tab:inputparams}
  \centering
 \begin{tabular}{lc}
      \hline
       Parameter    &  Value(s) \\
      \hline  
      $T_\ast$\,[kK] &  $141$  \\
      $\log\,(L~[\mathrm{L}_\odot])$ &  $4.85\dots7.47$ \\
      $M_\ast\,[\mathrm{M}_\odot]$ &  $7.3\dots500$ \\
			$\varv_\text{mic}$\,[km\,s$^{-1}$] & $30$ \\
      \smallskip
      $D_\infty$     &    $50$  \\

      \multicolumn{2}{l}{\textit{abundances in mass fractions:}}\\
			$X_\text{He}$  &   $1 - 0.014 \cdot Z/Z_\odot$  \\
			$X_\text{C}$   &   $8.7\cdot10^{-5} \cdot Z/Z_\odot$  \\
			$X_\text{N}$   &   $9.1\cdot10^{-3} \cdot Z/Z_\odot$  \\
			$X_\text{O}$   &   $5.5\cdot10^{-5} \cdot Z/Z_\odot$  \\
			$X_\text{Ne}$  &   $1.3\cdot10^{-3} \cdot Z/Z_\odot$  \\
			$X_\text{Na}$  &   $2.7\cdot10^{-6} \cdot Z/Z_\odot$  \\
			$X_\text{Mg}$  &   $6.9\cdot10^{-4} \cdot Z/Z_\odot$  \\
			$X_\text{Al}$  &   $5.3\cdot10^{-5} \cdot Z/Z_\odot$  \\
			$X_\text{Si}$  &   $8.0\cdot10^{-4} \cdot Z/Z_\odot$  \\
			$X_\text{P}$   &   $5.8\cdot10^{-6} \cdot Z/Z_\odot$  \\
			$X_\text{S}$   &   $3.1\cdot10^{-4} \cdot Z/Z_\odot$  \\
			$X_\text{Cl}$  &   $8.2\cdot10^{-6} \cdot Z/Z_\odot$  \\
			$X_\text{Ar}$  &   $7.3\cdot10^{-5} \cdot Z/Z_\odot$  \\
			$X_\text{K}$   &   $3.1\cdot10^{-6} \cdot Z/Z_\odot$  \\
			$X_\text{Ca}$  &   $6.1\cdot10^{-5} \cdot Z/Z_\odot$  \\
      \medskip
			$X_\text{Fe}$  &   $1.6\cdot10^{-3} \cdot Z/Z_\odot$  \\
    \hline
  \end{tabular}
\end{table}

We calculate multiple series of models with an effective temperature of $T_\ast = 141\,$kK at a Rosseland continuum optical depth of $\tau_\text{Ross,cont} = 20$. In most cases, this coincides with the inner boundary of the atmosphere models. However, a boundary of $\tau_\text{Ross,cont} = 20$ turns out to be insufficient for calculating hydrodynamic (HD) models for very high masses ($> 70\,M_\odot$) as their winds are already launched at comparable optical depths. In these cases, we calculate models going further inward with boundary values up to $\tau_\text{Ross,cont} = 100$, but adjusting their input parameters such that we still obtain $T_\ast = 141\,$kK at $\tau_\text{Ross,cont} = 20$.
The effective temperature at $\tau = 2/3$ is an output parameter of our calculations and can be significantly cooler for models with substantial mass loss.
A fixed $T_\ast$ implies that our atmosphere calculations can have different radii and temperatures than predicted by stellar structure models \citep[e.g.][hereafter G+2018]{Grassitelli+2018}. Nonetheless, our choice of $T_\ast$ yields (electron) temperatures at the sonic point comparable to recent structure models by \citetalias{Grassitelli+2018} calculated for He stars between $10$ and $20\,M_\odot$. We also calculate models below $10\,M_\odot$ to study the transition to optically thin winds and the breakdown of WR-type mass loss. However, below a certain $Z$-dependent $\dot{M}$ (or, equivalently, below a certain $L/M$), our choice of $T_\ast$ might not accurately reflect the conditions in the wind-driving region of a He star \citepalias[cf.\ their iron-minimum $\dot{M}$ in][]{Grassitelli+2018}. Therefore, results for stripped He stars below $10\,M_\odot$ and absolute values of $\dot{M}$ beyond the breakdown of WR-type mass loss shall be considered with caution.

The clumping treatment is similar to \citetalias{Sander+2020}, i.e. we apply a depth-dependent microclumping with $D_\infty = 50$ and an exponential onset described by a characteristic velocity $\varv_\text{cl}$ \citep[`Hillier law', cf.\,][]{HM1999} which we fix at $100\,\mathrm{km}\,\mathrm{s}^{-1}$ for all models. The turbulent velocity entering the HD equation is kept at a constant value of $30\,\mathrm{km}\,\mathrm{s}^{-1}$, identical to \citetalias{Sander+2020}. While both clumping and turbulence might vary between different stars and metallicities, the study of their particular influence on the HD solution would quickly fill a paper of its own. Instead, a fixed set of parameters for $D_\infty$ and $\varv_\text{mic}$ allows us to better identify the individual influence of the more fundamental stellar parameters varied in this work.

\section{The mass-loss of massive He stars}
  \label{sec:resmdot}

\subsection{The fundamental role of multiple scattering}
  \label{sec:multiscat}

\begin{figure*}
  \includegraphics[width=\textwidth]{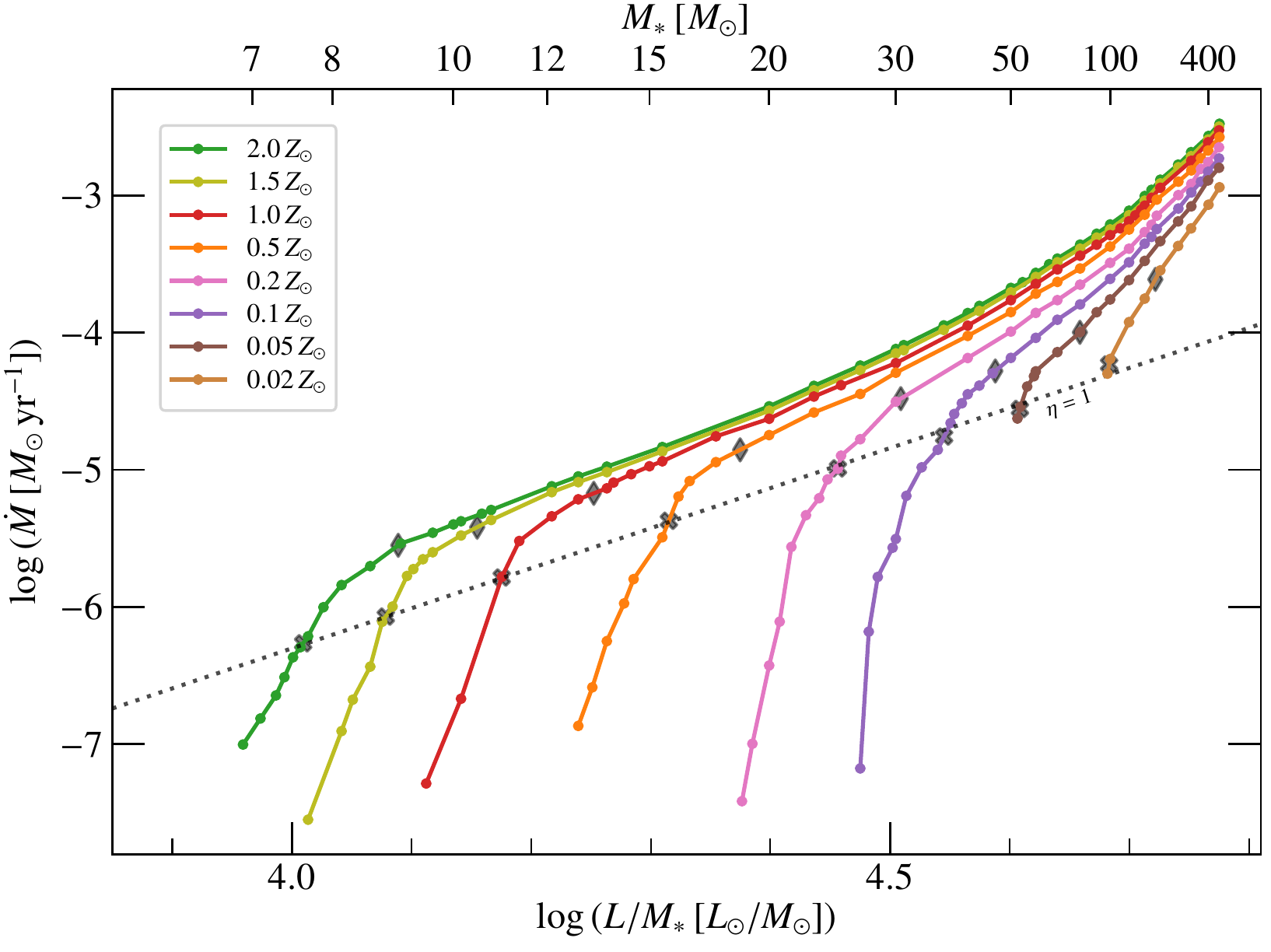}
  \caption{Mass-loss rate $\dot{M}$ as a function of $L/M$ for sets of hydrodynamically consistent model atmospheres with different metallicities $Z$. The onset of multiple scattering $\eta = 1$ in each sequence is marked with a grey cross. Grey diamonds denote the locations of $\eta = 2.5$.}
  \label{fig:mdotldm}
\end{figure*}

\begin{figure}
  \includegraphics[width=\columnwidth]{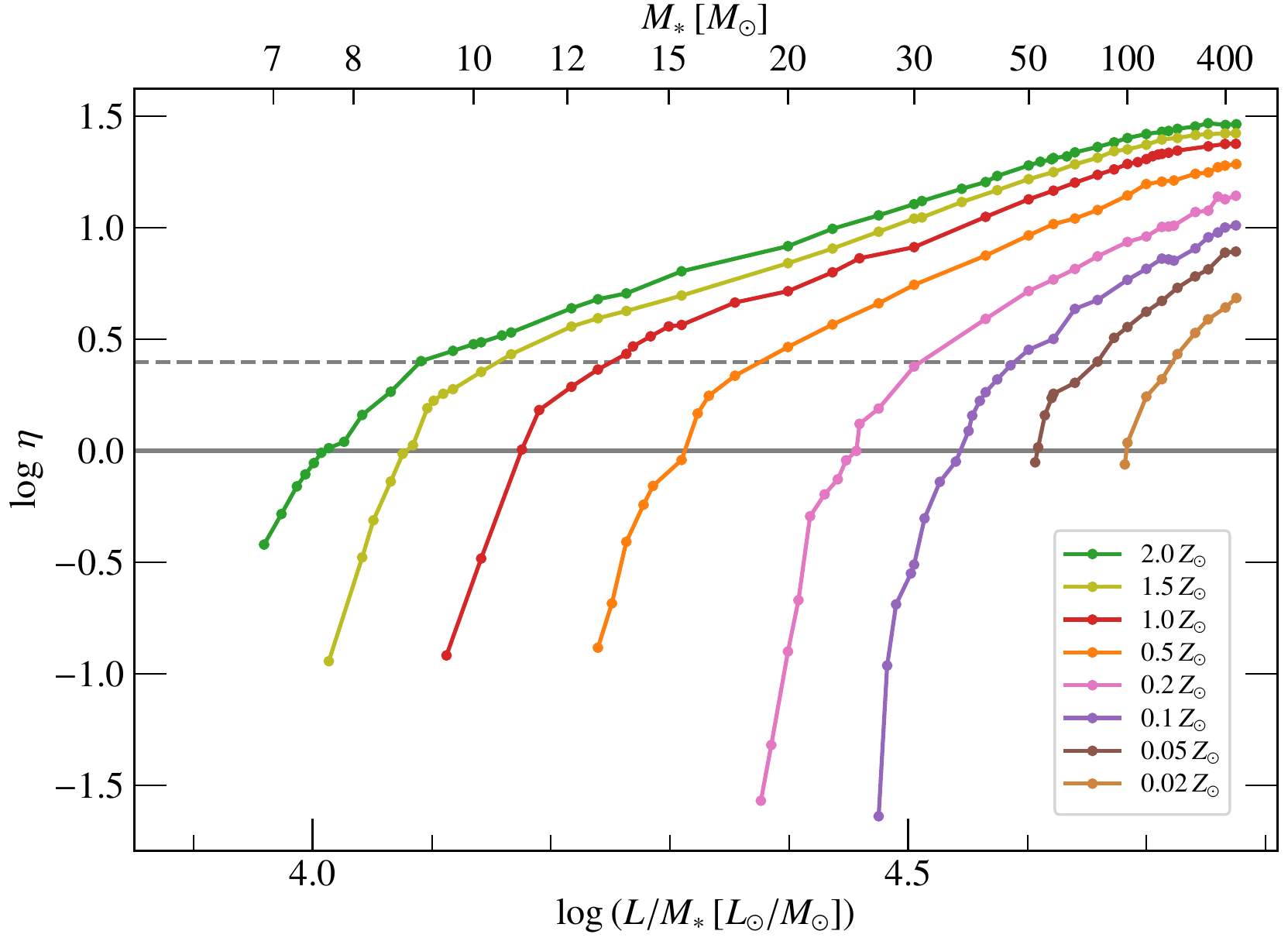}
  \caption{Wind efficiency parameter $\eta$ as a function of $L/M$ for sets of HD model atmospheres with different metallicities $Z$.}
  \label{fig:etaldm}
\end{figure}

An overview of $\dot{M}$ as a function of $L/M$ is given in Fig.\,\ref{fig:mdotldm}. Since we restrict our analysis in this work to models with a fixed $L$-$M$-relation, we can also give the corresponding mass $M$ for each luminosity or $L/M$-ratio, which is shown in the top x-axis. All curves show a similar pattern with a steep increase, followed by a shallower part which eventually becomes steeper again for very high masses. As expected, $\dot{M}$ increases with $Z$, but the differences between the metallicities get considerably smaller for higher masses. The initial rise -- or the decline in $\dot{M}$ towards lower $L/M$ -- corresponds to the transition from an optically thin to an optically thick wind regime.
To get a more quantitative handle on this transition, we study the efficiency of momentum transfer from radiation to gas
\begin{equation}
  \label{eq:eta}
  \eta = \frac{\dot{M}\varv_\infty}{L/c},
\end{equation}
also called `wind-efficiency parameter', which is depicted in Fig.\,\ref{fig:etaldm}. The onset of multiple scattering ($\eta = 1$) is marked by a solid horizontal line. The corresponding location in the curves of $\dot{M}(L/M)$ are indicated with crosses in Fig.\,\ref{fig:mdotldm}. As indicated by the dashed line in Fig.\,\ref{fig:mdotldm}, the onset of multiple scattering follows the linear relation
\begin{equation}
  \label{eq:mdldm-etaunity}
  \left.\log\dot{M}\right|_{\eta = 1} = 2.92 (\pm 0.06) \cdot \left.\log{L/M}\right|_{\eta = 1} - 17.98 (\pm 0.26)\text{.}
\end{equation}
For the $L/M$-values of $\eta = 1$ in the metallicity space, we can further find a linear correlation of $L/M$ with $\log Z$, i.e.:
\begin{align}
  \label{eq:ldm-etaunity-z}
  \left.\frac{L}{M}\right|_{\eta = 1} &= - p \cdot \log\frac{Z}{Z_\odot} + q\text{.} \\
  \nonumber  &\text{with} \log p = 4.286 (\pm 0.005) \text{~and~} \log q = 4.187 (\pm 0.006)\text{.}
\end{align}
More convenient numbers are obtained when replacing the $L/M$-ratios with the Eddington-$\Gamma_\text{e}$, i.e.\
\begin{equation}
  \label{eq:gedddef}
\Gamma_\text{e} = \frac{\sigma_\text{e}}{4 c m_\mathrm{H}G} q_\text{ion} \frac{L}{M} = 10^{-4.51} q_\text{ion} \frac{L/L_\odot}{M/M_\odot}\text{,}
\end{equation}
defining also the ratio of the acceleration due to free electron scattering relative to gravitational acceleration. In WR atmospheres, the number of free electrons per atomic mass unit $q_\text{ion}$ can change in the wind. For the inner part, our models give values of $q_\text{ion} \approx 0.5$.
A direct fit of the $\Gamma_\text{e}$ values calculated from the model atmospheres for $\eta = 1$ yields
\begin{align}
  \label{eq:gedd-etaunity-z}
  \left.\Gamma_\mathrm{e}\right|_{\eta = 1} = &-0.300\,(\pm 0.003) \cdot \log\frac{Z}{Z_\odot} + 0.236\,(\pm 0.003)\text{.}
\end{align}
When converting the coefficients for the $Z$-trends to $L/M$-ratios, the obtained values are close to the results from Eq.\,(\ref{eq:ldm-etaunity-z}). 
The values of $\Gamma_\mathrm{e}$ obtained via Eq.\,(\ref{eq:gedd-etaunity-z}) predict the onset of multiple scattering at different metallicities. This so-called `transition value' \citep{VG2012} has so far only been studied between the regimes of H-burning Of and WNh stars. The values we obtain from our sets of He star models, e.g.\ $\left.\Gamma_\mathrm{e}\right|_{\eta = 1,\odot} \approx 0.24$ and $\left.\Gamma_\mathrm{e}\right|_{\eta = 1,\text{LMC}} \approx 0.33$, are significantly lower than what was previously obtained via theoretical and semi-empirical approaches for the Of/WNh-transition \citep[e.g.][]{Vink+2011,VG2012,Bestenlehner+2014,Bestenlehner2020}. These differences are likely due to the different parameter regimes (e.g.\ $T_\ast$, abundances), but whether the general scaling of $\left.\Gamma_\mathrm{e}\right|_{\eta = 1}(Z)$ is similar cannot be answered without calculating hydrodynamically consistent atmospheres with $\eta = 1$ in other parameter regimes.
Nonetheless, the values of $\left.\Gamma_\mathrm{e}\right|_{\eta = 1}$ will become important for more of our results later on.

\begin{figure}
  \includegraphics[width=\columnwidth]{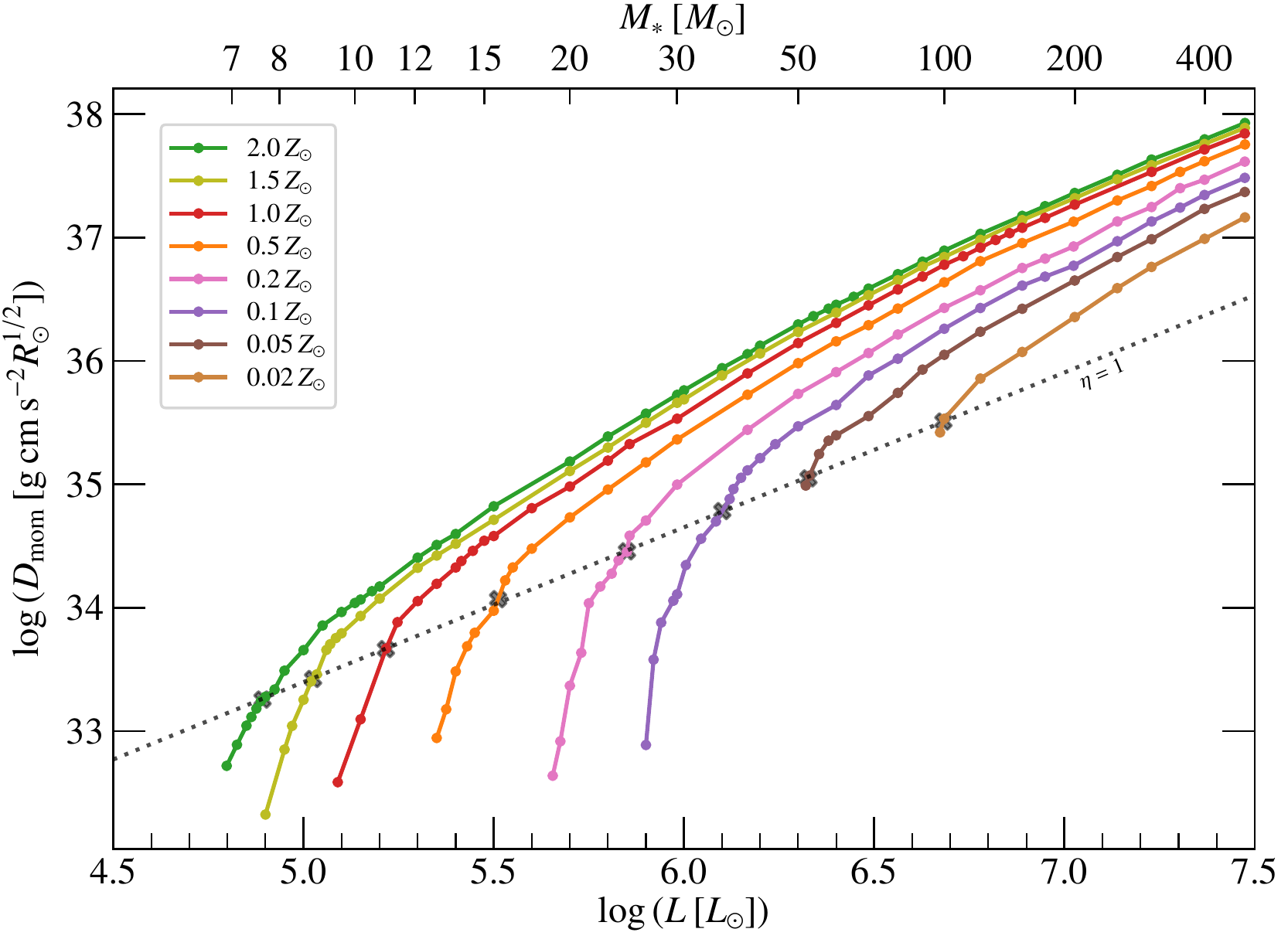}
  \caption{Modified wind momentum $D_\mathrm{mom}$ as a function of $L$ for sets of HD model atmospheres with different metallicities $Z$.}
  \label{fig:dmoml}
\end{figure}

Based on empirical findings \citep{Kudritzki+1995,Kudritzki+1999} and theoretical considerations \citep{Puls+1996}, one can define the so-called `modified wind momentum'   
\begin{equation}
  \label{eq:dmom}
  D_\text{mom} = \dot{M} \varv_\infty \sqrt{R_\ast/R_\odot}
\end{equation}
as a quantity expected to strongly correlate with the stellar luminosity $L$ (``wind-momentum luminosity relation'': WLR). For OBA-stars, where the terminal velocity can be accurately determined, the WLR in the form of $\log D_\text{mom} \propto \log L$ holds quite well, regardless of metallicity \citep[e.g.][]{Kudritzki+1999,Mokiem+2007,Ramachandran+2017}. For WR stars, there is no theoretical prediction but empirical tests of the WLR \citep{Hainich+2015} yielded a significant scatter. In Fig.\,\ref{fig:dmoml}, we plot $D_\text{mom}(L)$ for our model sets at different metallicities. Linear curves (or parts of them) would imply the validity of the WLR. Instead, we obtain a significant bending for all metallicities, underlining the fundamental difference in the nature of WR-type winds compared to those of O stars and BA supergiants. The single scattering limit ($\eta = 1$) is denoted by crosses on each curve. It is likely that a form of the WLR is valid for thin winds below the single scattering limit, but our data points in this regime are more uncertain and too sparse to draw any solid conclusions on the $D_\text{mom}(L)$-slopes.

The positions of the single scattering limit on each of the curves in Fig.\,\ref{fig:dmoml} are not random, as $D_\text{mom}$ and $\eta$ can be related. For values of constant $\eta$, we get $\dot{M}\varv_\infty \propto L$. The constant $T_\ast$ in our model sets further implies $R_\ast \propto L^{1/2}$. We thus expect $\left.D_\text{mom}\right|_{\eta = \mathrm{const.}} \propto L^{1.25}$ and use this as a sanity check for our models. Indeed, a linear fit to our set of interpolated points for $\eta = 1$ yields good agreement ($1.254 \pm 0.009$). The corresponding relation is denoted as a dotted line in Fig.\,\ref{fig:dmoml}. Since this slope is inherent to all lines of constant $\eta$, one can simply shift this line to compare models for different metallicities, but the same $\eta$. The bending in the $D_\text{mom}(L)$-curves further hints that their asymptote might be a line of constant $\eta$, implying there would be a maximum wind efficiency for each metallicity. We will discuss this further below and -- from a different perspective -- in appendix Sect.\,\ref{asec:etamax}.

Despite the interesting results emerging from our models for the values corresponding to $\eta = 1$, it is also evident from Fig.\,\ref{fig:mdotldm}, that there is no clear `transition value' with respect to the slope of the curves in the $L/M$-direction, neither for $\eta$, nor for the flux-weighted optical depth of the wind, which we discuss in appendix Sect.\,\ref{asec:etatau} including its connection to $\eta$. The $\eta$-curves tend to flatten at higher values of $L/M$ for a given metallicity $Z$ with the `kink' at values up to $\eta \approx 2.5$, indicated by a dashed line in Fig.\,\ref{fig:etaldm}. For values of $\eta > 1$, there is no similar trend to Eq.\,(\ref{eq:ldm-etaunity-z}), which is illustrated by the diamonds marking $\eta = 2.5$ in Fig.\,\ref{fig:mdotldm}. 
 
From calculating HD stellar structure models, \citetalias{Grassitelli+2018} found a minimum $\dot{M}$ for winds driven by the hot Fe bump ($\dot{M}_\text{Fe}$). Our atmospheric results qualitatively align with these structural results in the thick-wind regime. However, for the optically thin regime, our assumption of a fixed $T_\ast$ implies that we always obtain winds driven by the hot Fe bump, while \citetalias{Grassitelli+2018} obtain inflated solutions for lower $M$ with winds launched further out at temperatures lower than those associated with this bump. Considering that their minimum $\dot{M}_\text{Fe}$ is also $Z$-dependent, the fact that the hot Fe bump is inducing envelope inflation rather than launching a stellar wind is expected to be more prominent at higher $Z$. Our absolute results for $\dot{M}$ and $\varv_\infty$ below the breakdown of WR-type mass loss for $Z \geq Z_\odot$ therefore have to be treated with caution. Inspecting Fig.\,\ref{fig:mdotldm} yields that this $Z$-limit aligns with our lower $M$-applicability limit of approximately $10\,M_\odot$.

\subsection{Terminal velocity trends}
  \label{sec:vinf}

\begin{figure*}
  \includegraphics[width=0.32\textwidth]{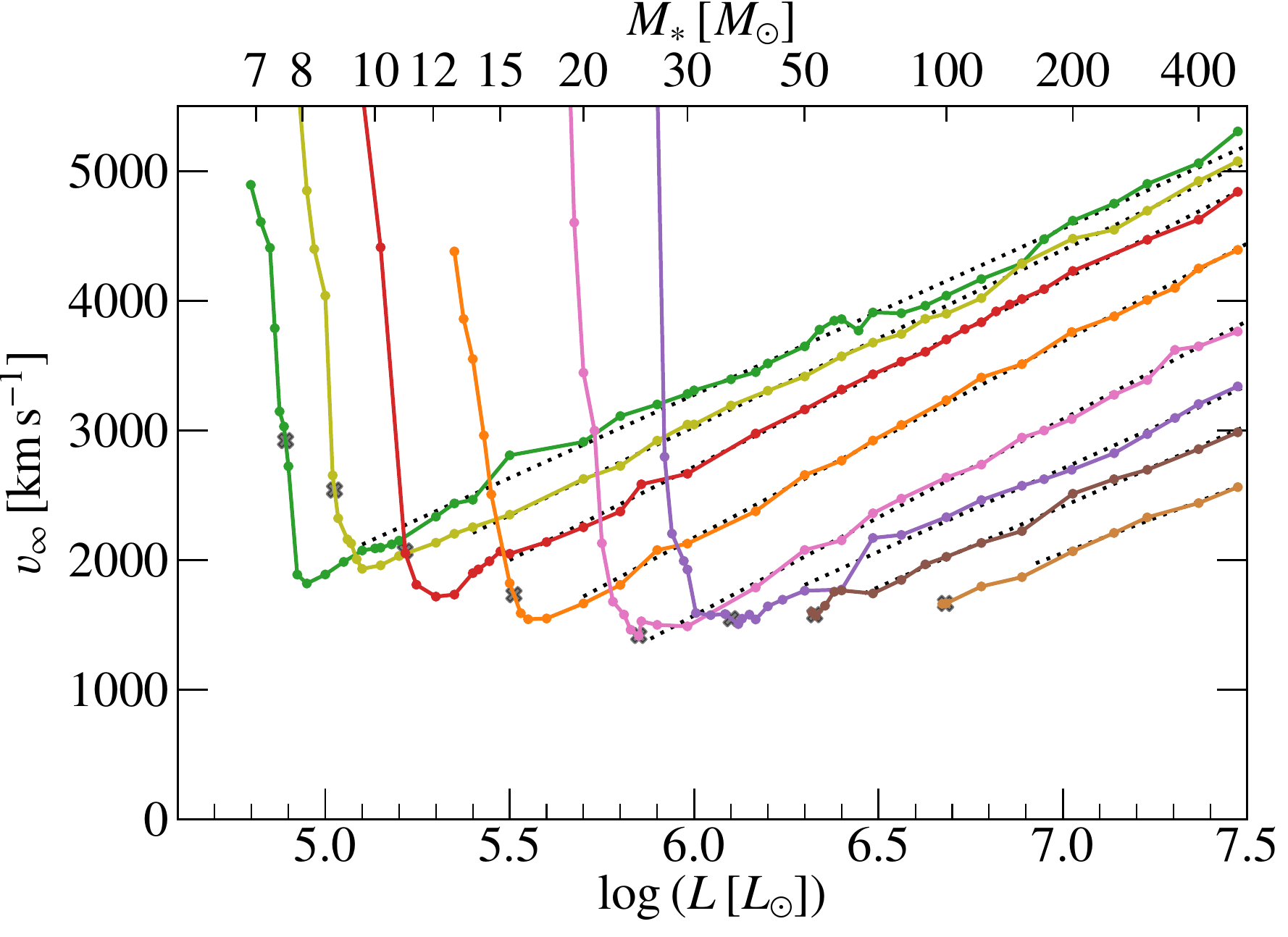} \hfill
  \includegraphics[width=0.32\textwidth]{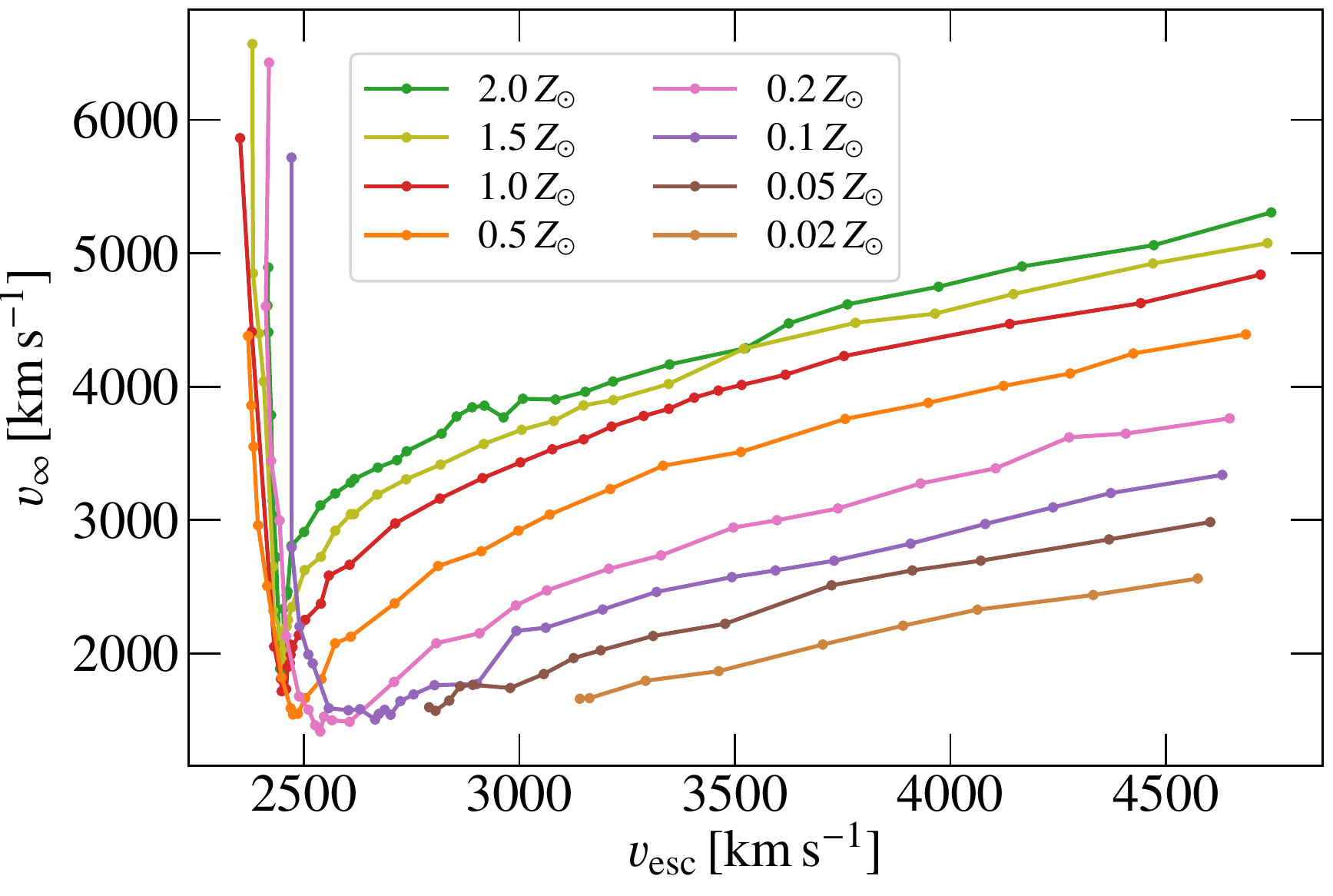} \hfill
  \includegraphics[width=0.32\textwidth]{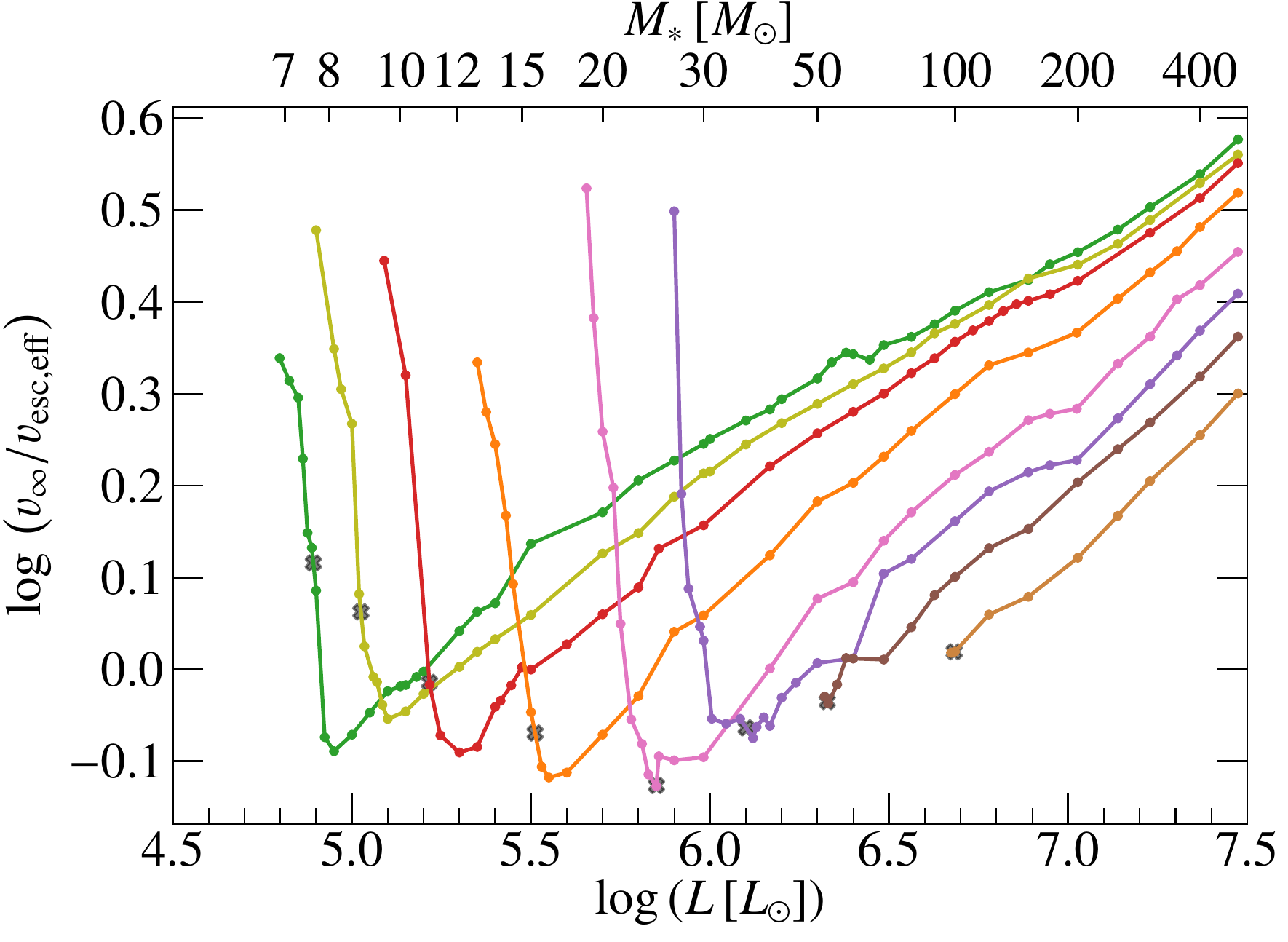}
  \caption{Left panel: Terminal velocity $\varv_\infty$ as a function of $L$. Middle panel: $\varv_\infty$ as a function of $\varv_\text{esc}$. Right panel: Ratio of $\varv_\infty$ and the effective escape velocity as a function of $L$.}
  \label{fig:vinftrends}
\end{figure*}

The terminal velocities of our model sequences are depicted as a function of $L$ in the left panel of Fig.\,\ref{fig:vinftrends}. In contrast to $\dot{M}$, $\varv_\infty$ is subject to a higher uncertainty, both physically and numerically. While observational constraints of $\varv_\infty$ are more straight-forward than for $\dot{M}$, the resulting HD model values rely on 
various factors, including the precise abundances, the clumping stratification, numerical choices for the outer boundary treatment, or the temperature correction method. In atmosphere models without HD consistency, these issues are usually not important as their impact on the emergent spectrum is often marginal, especially as the wind stratification is predefined by choosing a fixed $\varv_\infty$ together with a fixed velocity stratification -- typically a $\beta$-law. In our models, $\varv_\infty$ is obtained by integrating the hydrodynamic equation of motion from the critical point outwards, thereby also following any imprint of e.g.\ changes in the ionization stratification. Even for the regime of optically thick winds, the data curves for each $Z$ in Fig.\,\ref{fig:vinftrends} show a considerable scatter. Nonetheless, the amount of data allows us to conclude that in the limit of optically thick winds, $\varv_\infty$ scales with $\log L$, i.e. 
\begin{equation}
  \varv_\infty = m(Z) \cdot \log L/L_\odot + k(Z)\text{.}
\end{equation}
There is considerable scatter along the $Z$-dimension for both coefficients, making it hard to draw any firm conclusions. If not simply due to scatter, $m(Z)$ seems to increase with $Z$ up to an inferred maximum around $\log Z/Z_\odot \approx -0.5$ before decreasing again. But due to the uncertainties, we refrain from drawing any more quantitative conclusions than $\varv_\infty \propto \log L$ for winds with sufficient density.

When transitioning to less dense winds, the left panel of Fig.\,\ref{fig:vinftrends} reveals a dramatic change in the behaviour of $\varv_\infty$. With the winds becoming more optically thin, the terminal wind velocity approaches a minimum before eventually increasing sharply. This increase in $\varv_\infty$ corresponds to the steep drop in $\dot{M}$ and $\eta$ discussed above. Regardless of metallicity, $\varv_\infty$ never falls below $\varv_{\infty,\text{min}} \approx 1500\,\mathrm{km\,s}^{-1}$. While the absolute slopes of $\dot{M}$ and $\varv_\infty$ in the thin-wind regime and the value of $\varv_{\infty,\text{min}}$ are connected to our choice of $T_\ast$ and thus uncertain, we can conclude that for H-free stars with winds driven by the hot Fe bump there should not be any object with $\varv_\infty \ll \varv_{\infty,\text{min}}$. Moreover, our result implies that stars with $\varv_\infty \approx \varv_{\infty,\text{min}}$ are in a transition regime and thus might be different in some aspects from other classical WR stars, e.g.\ in that they do not adhere to relations derived from the assumption of pure LTE at the critical point.

To test our prediction of $\varv_{\infty,\text{min}}$, we inspect the analysed sample of Galactic \citep{Hamann+2006,Hamann+2019} and LMC WN \citep{Hainich+2014} stars. Indeed, all of the analysed hydrogen-free WNs have $\varv_\infty > 1500\,\mathrm{km\,s}^{-1}$, apart from two early-type WNs in the LMC having a slightly lower value of $\varv_\infty \approx 1300\,\mathrm{km\,s}^{-1}$. A more serious exception, however, are the WC9 stars as several of them have terminal velocities down to $\varv_\infty \approx 1000\,\mathrm{km\,s}^{-1}$ \citep{Sander+2012,Sander+2019}. Whether this is just an issue of fine-tuning our models for these type of stars (e.g.\ in terms of abundances and $T_\ast$) to bring $\varv_{\infty,\text{min}}$ down to $1000\,\mathrm{km\,s}^{-1}$ or a signature of a different kind of wind regime in WC9s remains unclear and will have to be investigated in more tailored studies. Nonetheless, the postulate of $\varv_{\infty,\text{min}}$ as such, which actually has a mild $Z$-dependence and should increase with higher metallicity, remains an important outcome and explains why we do not find any early-type WR stars with low terminal velocities ($< 1000\,\mathrm{km\,s}^{-1}$), regardless of the host galaxy.

We further investigate the relation of $\varv_\infty$ with $\varv_\text{esc} := \sqrt{2 G M_\ast / R_\text{crit}}$. In the middle panel of Fig.\,\ref{fig:vinftrends}, $\varv_\infty$ is shown as a function of $\varv_\text{esc}$, revealing a complex behaviour. After entering the dense wind regime, $\varv_\infty$ increases monotonically with $\varv_\text{esc}$. While the asymptotic behaviour could be linear, the overall slope in this regime is neither linear, nor a power-law nor a simple logarithm. There is also a clear shift along the $Z$-dimension. In the thin-wind regime, the curves from different $Z$ are overlapping and there is little distinction with metallicity. This curvature in the slopes of $\varv_\infty(\varv_\text{esc})$ is likely also the reason why Eq.\,(\ref{eq:goetztauscale}) is not a fully sufficient description to relate $\eta$ and $\tau_F(R_\text{crit})$. \citet{Graefener+2017} assumed $\varv_\infty \approx \sqrt{1 - \Gamma_\text{w}} \varv_\infty$ in their calculations with $\Gamma_\text{w}$ being a representative value for $\Gamma$ in the wind. While one can mathematically always find such a representative value, the complexity of the slope here shows that the selection of such a value would actually have to be a function of a varying quantity itself, such as $L$, $L/M$, or $\tau_F(R_\text{crit})$, thus spoiling the intended simplification. 

When discussing the issue of wind driving, the term $\varv_\text{esc}$ is also often used to denote the \emph{effective} escape velocity
\begin{equation}
  \label{eq:vesceff}
	\varv_\text{esc,eff} := \sqrt{\frac{2 G M_\ast}{R_\text{crit}}\left[1 - \Gamma_\text{e}(R_\text{crit})\right]} = \varv_\text{esc}\sqrt{1 - \Gamma_\text{e}(R_\text{crit})} \text{.}
\end{equation}
When plotting the ratio of $\varv_\infty$ to $\varv_\text{esc,eff}$ in the right panel of Fig.\,\ref{fig:vinftrends}, in this case as a function of $L$, it becomes clear that -- unlike in CAK \citep{Abbott1982lineacc} -- there is no regime where we have a constant $\varv_\infty/\varv_\text{esc,eff}$ over a considerable parameter range. Instead, we find a constant increase of $\varv_\infty/\varv_\text{esc,eff}$ with $L$ after the dense wind regime has been reached. Despite some numerical scatter in the results for $\varv_\infty$, it is clear that there is a curvature in the slopes of $\varv_\infty/\varv_\text{esc,eff}(L)$, so we cannot describe the behaviour by a simple power law. This is in contrast to the conclusions which were reached by \citet{NL2000} when evaluating empirical results combined with theoretical considerations. \citet{NL2000} obtained a slightly negative slope of $-0.13$ for $\log\,\varv_\infty/\varv_\text{esc,eff}$ versus $\log L/L_\odot$ for WN stars, while our results point to positive slopes around $\approx 0.3$ when ignoring the curvature, which roughly coincides with their regression result for WC stars. Empirical sets of WN stars always show a mixture of evolutionary stages and abundances, making it hard to isolate an underlying behaviour for a quantity like $\varv_\infty$. Moreover, only a smaller fraction of the parameter domain we see in Fig.\,\ref{fig:vinftrends} is mapped in the observations. Consequently, the data entering empirical relations between quantities for WR stars are commonly subjected to an inherent scatter. To remedy this situation and to get an insight of which $L$- and $M$-regimes are realized in nature, a combination of theoretical modelling efforts and tailored spectral analyses will be indispensable.

\subsection{The similarity of Wolf-Rayet winds}
  \label{sec:mdottrans}

As recently discussed in \citet{Shenar+2020}, one can describe the onset of WR-type spectral appearance with the help of a so-called `transformed radius'
  \begin{equation}
  \label{eq:rt}
 R_{\mathrm{t}} = R_{\ast} \left[ \frac{\varv_{\infty}}{2500\,\mathrm{km/s}} 
 \left/ \frac{\dot{M} \sqrt{D} }{10^{-4} M_{\odot}/\mathrm{yr}} \right. 
 \right]^{\frac{2}{3}}
  \end{equation} 
This quantity was invented by \citet{SHW1989} to reflect the finding that models with different mass-loss rates $\dot{M}$ will yield an almost identical spectrum, when also shifting their stellar radii $R_\ast$ by a certain amount. A similar observation was made for $\varv_\infty$, yielding the semi-empirical relation (\ref{eq:rt}), which was later adjusted to also account for (optically thin) clumping \citep{HK1998}. As discussed in \citet{HG2004}, there is also a parameter degeneracy between $R_{\mathrm{t}}$ and the stellar temperature $T_\ast$ for very dense winds with (almost) identical spectra along contours of $R_\mathrm{t} \propto T_\ast^{-2}$. For a set with a given luminosity $L$, terminal velocity $\varv_\infty$, and clumping factor $D$, this corresponds to a constant $\dot{M}$. 

While $R_\mathrm{t}$ has the dimension of a radius, its actual value does not reflect any physically significant radius. It is therefore more convenient in our context to express the invariance of WR-type spectra with the so-called `transformed mass-loss rate'	
		\begin{equation}
		  \label{eq:mdott}
		  \dot{M}_\text{t} = \dot{M} \sqrt{D} \cdot \left( \frac{1000\,\text{km/s}}{\varv_\infty} \right) \left( \frac{10^6 L_\odot}{L} \right)^{3/4}\text{,}
		\end{equation}
 		introduced by \citet{GraefenerVink2013}. The value of $\dot{M}_\text{t}$ can be understood as the mass-loss rate $\dot{M}$ the star would have, if it had a smooth wind (i.e., $D = 1$), a 
		terminal wind velocity of $1000\,$km/s and a luminosity of $10^6\,L_\odot$.
 Inserting all definitions, one can show 
\begin{equation}
  \dot{M}_\text{t} = \frac{2}{5} \left( \frac{R_\odot T_\odot^2}{R_\mathrm{t} T_\ast^2} \right)^{3/2} 10^{-4} M_{\odot}\,\mathrm{yr}^{-1}
\end{equation}		
with $T_\odot$ denoting $T_\ast$ for the sun. Thus, for a constant value of $T_\ast$, the condition $R_\mathrm{t} = \mathrm{const.}$ directly implies also $\dot{M}_\mathrm{t} = \mathrm{const.}$, i.e. the concepts of $\dot{M}_\mathrm{t}$ and $R_\mathrm{t}$ are equivalent here.

\begin{figure}
  \includegraphics[width=\columnwidth]{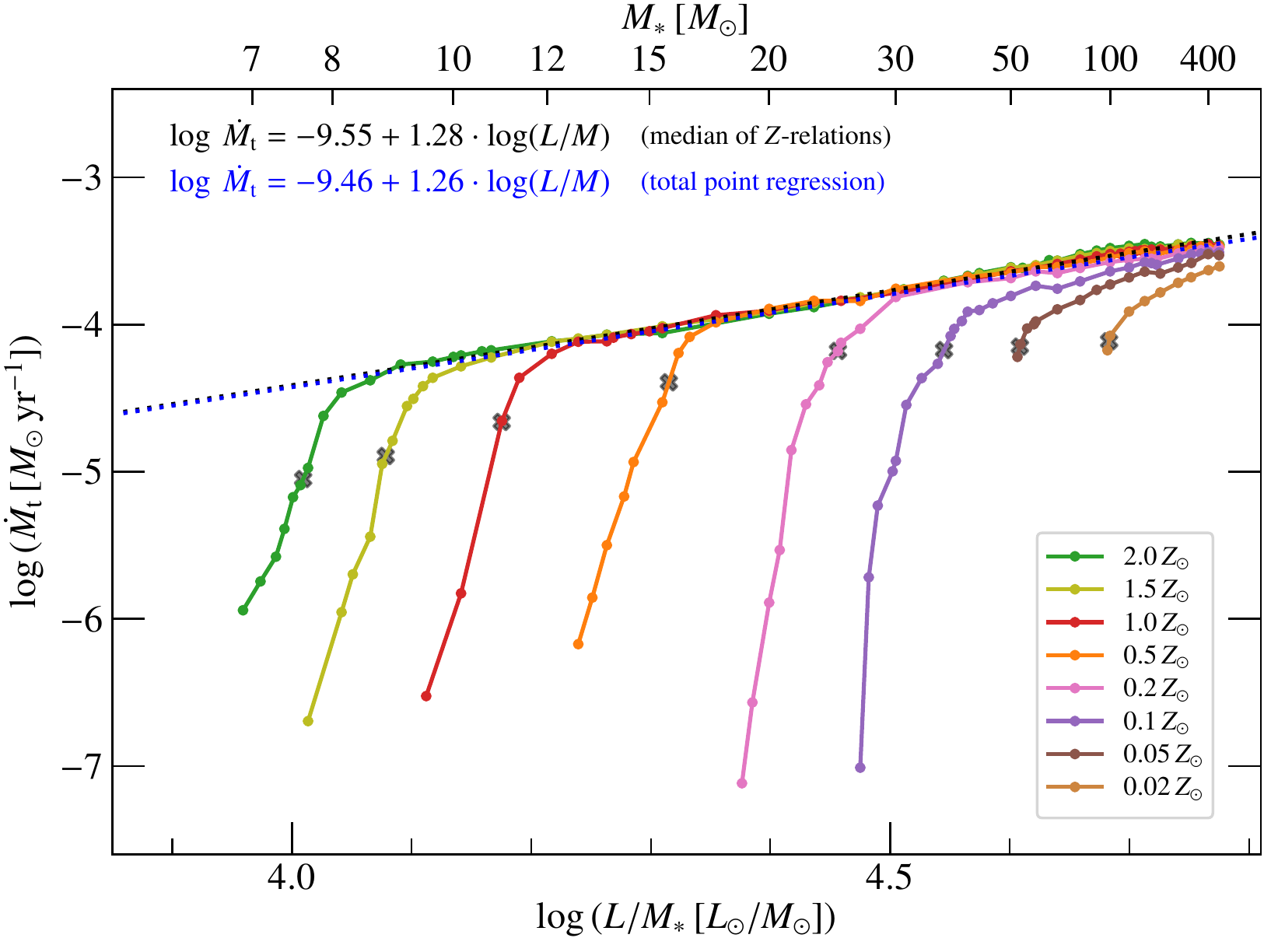}
  \caption{Transformed mass-loss rate $\dot{M}_\text{t}$ as a function of $L/M$ for all calculated HD model sequences.}
  \label{fig:mdottldm}
\end{figure}

\begin{figure}
  \includegraphics[width=\columnwidth]{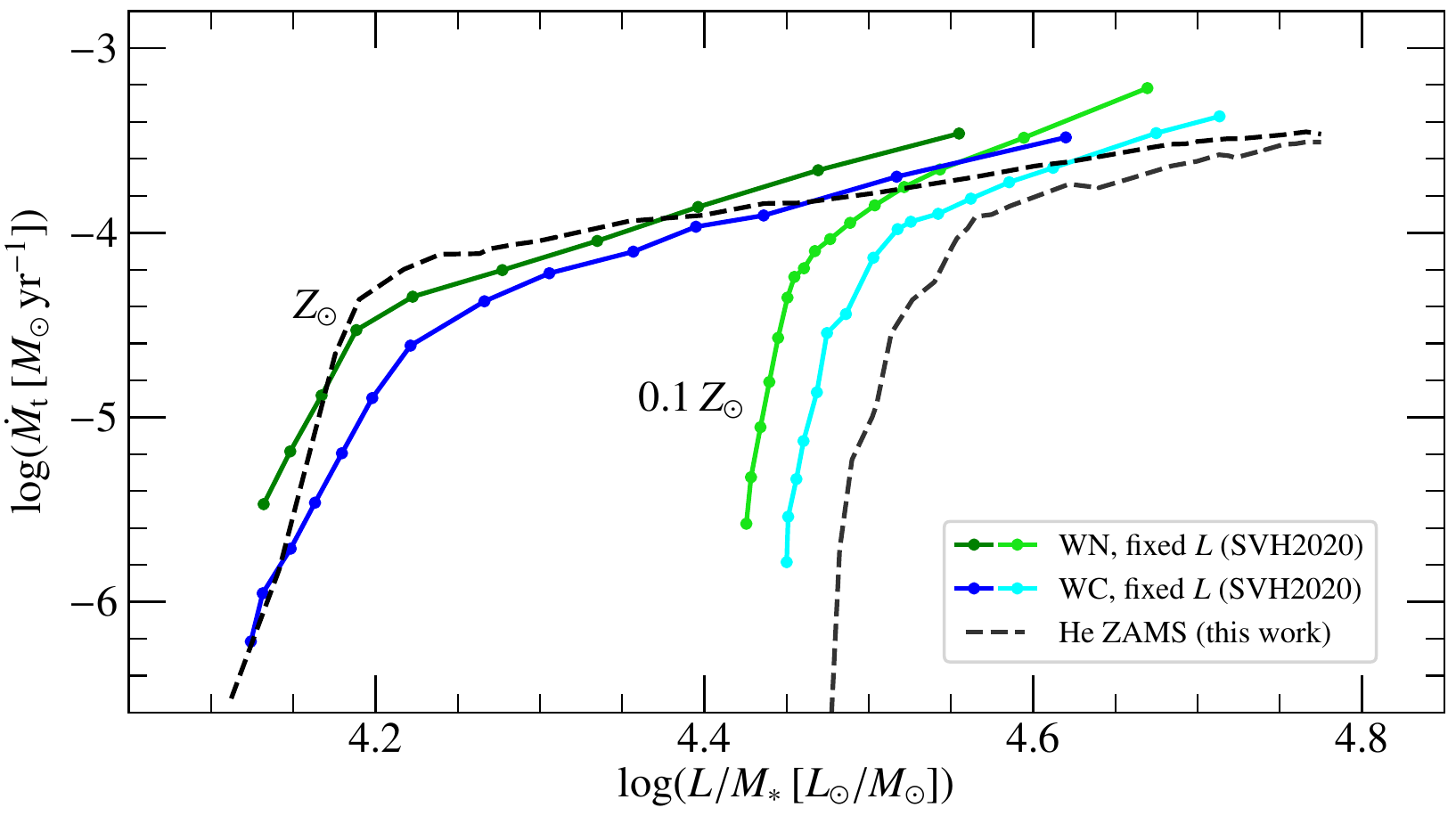}
  \caption{Comparison with the results from \citetalias{Sander+2020} (green: their WN, blue: their WC, black: this work) hints that the linear behaviour of $\log \dot{M}_\mathrm{t}$ versus $\log \left(L/M\right)$ is universal, but the slope depends on the particular values of the applied $L$ and $M$.}
  \label{fig:svh2020cmp}
\end{figure}

Plotting the transformed mass-loss rate $\dot{M}_\text{t}$ against the $L/M$-ratio shows the intriguing result depicted in Fig.\,\ref{fig:mdottldm}: While the onset of WR-type mass loss still strongly depends on the metallicity $Z$, all the models essentially align along the same linear curve in the $\log \dot{M}_\text{t}$-$\log (L/M)$-plane. This does not happen for $\dot{M}(L/M)$, meaning that the actual mass-loss for WR stars changes with $Z$, but the scaling with $\varv_\infty^{-1}$ in $\dot{M}_\text{t}$ nullifies these differences. $\dot{M}/\varv_\infty$ is a measure for the wind density, which seems to be -- approximately -- conserved once the regime of WR-type mass loss is fully reached. As a direct consequence, we get the result that for WR stars at different $Z$, but with the same $L$ and $M$, $\varv_\infty$ scales linear with $\dot{M}$. We will discuss and test this prediction later in Sect.\,\ref{sec:ztrends}.

The factor of $L^{-3/4}$ in the definition of $\dot{M}_\mathrm{t}$ is responsible for the linear appearance of the $\dot{M}_\text{t}(L/M)$-curve. The curve seems to flatten for the highest masses, but neglecting this as well as the significantly deviating data for $Z < 0.1\,Z_\odot$, yields the relation
\begin{equation}
  \label{eq:mdotldmfit}
  \log \dot{M}_\text{t} = 1.26 (\pm 0.04) \cdot \log(L/M) - 9.46 (\pm 0.16)
\end{equation}
for the `pure' WR regime. Instead of creating an overall dataset to determine Eq.\,(\ref{eq:mdotldmfit}), we could perform individual fits per metallicity and then derive their median. Both results are shown in Fig.\,\ref{fig:mdottldm} and yield very similar coefficients. 
In Fig.\,\ref{fig:svh2020cmp}, we compare our findings for $Z_\odot$ and $0.1\,Z_\odot$ to the results from the WN and WC models by \citetalias{Sander+2020} in Fig.\,\ref{fig:svh2020cmp}. It is immediately evident that the linearity of $\dot{M}_\text{t}(L/M)$ in the regime of WR-type mass loss is not a coincidence as the curves show even less scatter than our new results, which is most likely due to the higher numerical stability when iterating for $M$ instead of $\dot{M}$. Thus, while the particular slope and the location of the breakdown depend on the chemical composition and the particular $L$-$M$-combinations, the concepts arising from our work are of fundamental nature and can likely be transferred to other wind regimes with similar conditions.

Since $\dot{M}_\text{t} = \mathrm{const.}$ corresponds to $R_\text{t} = \mathrm{const.}$ as $T_\ast$ is fixed in our study, this implies the spectra for WR stars of the same $L$ and $M$ at different $Z$ look very similar to each other \citep[$R_\text{t} = \mathrm{const.}$, see][]{SHW1989}. In particular, this means that the normalized (emission) spectra of WR stars with the same $L$ and $M$ in different galaxies are kind of `scaled' versions of themselves and therefore will get the same WR subtype classification. However, this does not automatically imply the opposite, namely that stars of the same subtype must have similar stellar parameters. While empirical studies hint that this could be true for WC stars \citep[e.g.][]{Sander+2012}, the situation is much more diverse for WN stars \citep[e.g.][]{Hamann+2006,Hainich+2014,Shenar+2019}, even when considering those which are hydrogen-free. Unfortunately, empirical studies -- at least with classical atmosphere models -- are  subject to a degeneracy of the solution in terms of $R_\ast$ for dense winds, interestingly also along lines with $\dot{M} \approx \text{const}$. Analyses with hydrodynamically consistent models are a major step to break this degeneracy, but would likely require a considerable amount of tailored calculations for each object and thus are beyond the scope of the present work.

\subsection{Towards a meaningful mass-loss recipe for He stars}
  \label{sec:mdotrecipe}
	
\begin{figure}
  \includegraphics[width=\columnwidth]{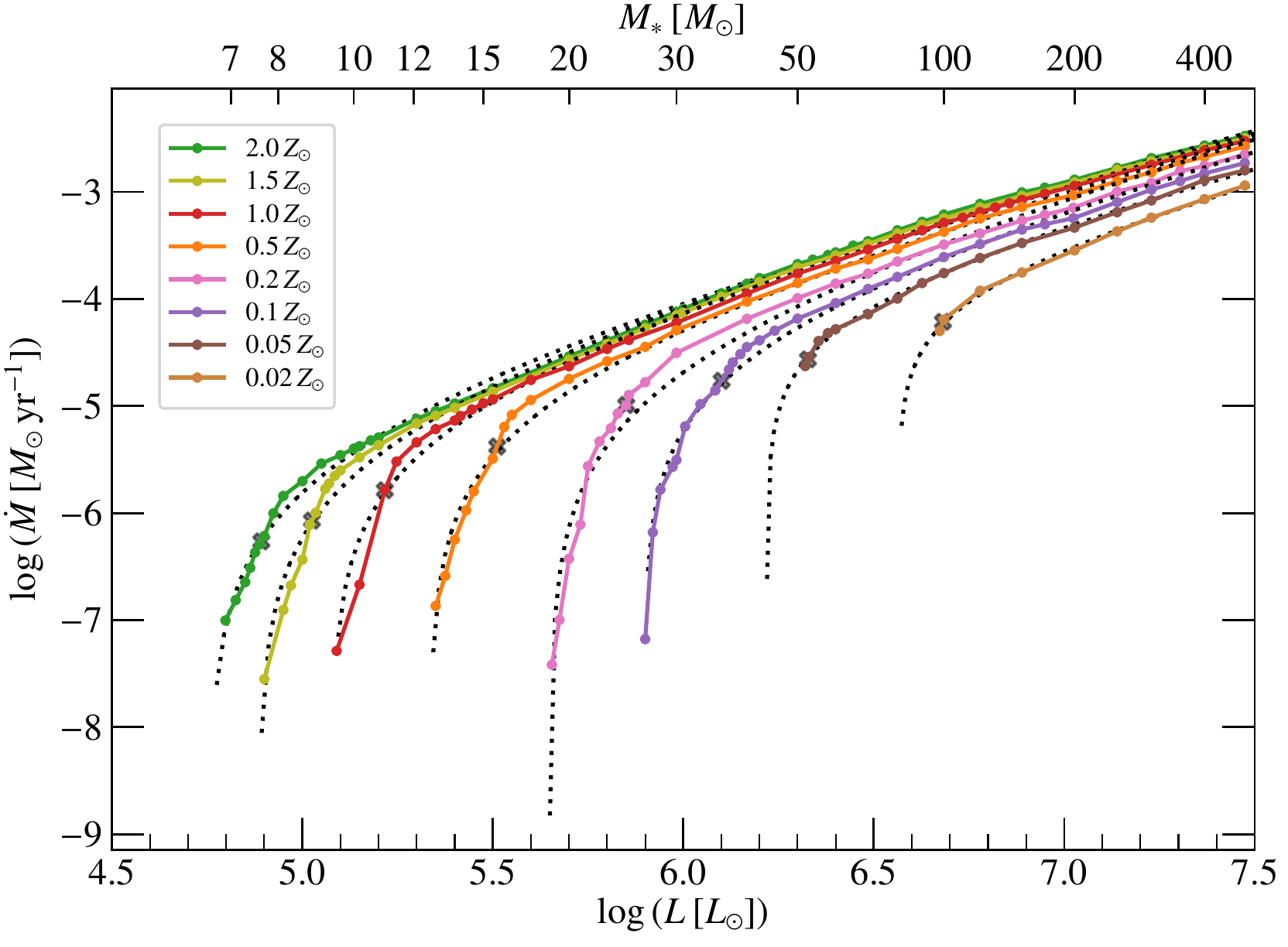}
  \caption{Mass-loss rate $\dot{M}$ as a function of $L$ for model sequences of different $Z$ with fits from Eq.\,(\ref{eq:mdotlrecipe}) overplotted in grey dashed lines.}
  \label{fig:mdotl}
\end{figure}

Utilizing our insights gained from the set of dynamically consistent atmosphere models, we aim to derive a meaningful $\dot{M}$-recipe for He ZAMS stars with WR-type mass loss.  As argued in \citetalias{Sander+2020}, a recipe of type $\dot{M}(L/M)$ or $\dot{M}(\Gamma_\text{e})$ is preferred from HD considerations \citep[see also][]{NL2002,GH2008,Graefener+2011} and thus recommended, but we also derive an $\dot{M}(L)$-recipe for comparison purposes.

\subsubsection{WR-type mass loss as a function of luminosity}
  \label{sec:mdotl}

To get an idea for a recipe for $\dot{M}(L)$, we can combine our previous findings for $\dot{M}_\mathrm{t}$ and $\varv_\infty$ with the insight that a major part of the $\log\,\dot{M}$-curves are becoming linear when plotted over $\log\,(\log L/L_\odot)$. We further know that the additional factor of $L^{-3/4}$ in $\dot{M}_\mathrm{t}$ helped to get a mostly linear slope in Fig.\,\ref{fig:mdottldm}. Assuming that the different `breakdown' locations of $\dot{M}$ can be described by a shift in the outer logarithm, this leads to a description of the form 
\begin{equation}
  \label{eq:mdotdlog}
  \log \dot{M} = \alpha \cdot \log \left( \log L - \log L_{0} \right) + \frac{3}{4} \log \frac{L}{10 L_{0}} + \log \dot{M}_{10} \text{,}
\end{equation}
or equivalently
\begin{equation}
  \label{eq:mdotlrecipe}
  \dot{M} = \dot{M}_{10} \left( \log \frac{L}{L_0} \right)^{\alpha} \left(\frac{L}{10 L_{0}}\right)^{3/4}\text{.}
\end{equation}
$L_0$ denotes the asymptotic limit for which there is theoretically zero mass loss. The exponent $\alpha$ describes the curvature of the breakdown, and $\dot{M}_{10}$ is the mass-loss rate for $L = 10 L_{0}$. Inspecting the plot of our data together with the derived fits in Fig.\,\ref{fig:mdotl}, we get a good representation of the overall behaviour including the breakdown for all metallicities. Unfortunately, the prediction of $\dot{M}$ in the transition region is often a bit too low. Moreover, the mass-loss rate in the regime directly above the onset is a bit too high for super-solar $Z$. One could branch-out the concept from Eq.\,(\ref{eq:mdotdlog}) further by introducing a fourth parameter instead of fixing the factor $3/4$ in the second term, but this only leads to minor improvements in the slopes and a slightly worse asymptotic behaviour for the highest masses.
The relatively low number of free parameters is a strength of our $\dot{M}(L)$-recipe which also ensures a mathematically smooth description down to $L_{0}$. In practice, $\dot{M}$ will likely not approach zero at $L_0$, but evolve around very low numbers as we yield in some additional test calculations. The solutions thus reflect our finding of a breakdown of the mechanism leading to WR-type mass loss when approaching $L_0$. Around there, a different regime of very thin winds takes over with boundaries that have yet to be constrained.

Our recipe in the form of Eq.\,(\ref{eq:mdotdlog}) further provides an explanation for the fact that empirical \citep[e.g.][]{Hainich+2015,Hamann+2019} and theoretical \citep[e.g.][]{Vink+2000,Vink2017} studies yield power laws for $\dot{M}(L)$ with a lot of scatter in their exponent. Performing a Taylor expansion of the outer decadic logarithmic term in Eq.\,(\ref{eq:mdotdlog}) until the linear order reads
\begin{equation}
  \log \left(x - a\right) \approx \log\left(x_0 - a\right) + \frac{1}{\left(x_0 - a\right) \ln\left(10\right)} \left(x - x_0\right)
\end{equation}
for an expansion around $x_0$. A typical empirical study performed for WR stars would explore a region with $L \approx 10\,L_{0}$. Performing an expansion of Eq.\,(\ref{eq:mdotdlog}) with $x_0 = \log(10\,L_{0}) = 1 + \log L_{0}$ immediately yields
\begin{align}
  \log \dot{M} & \approx \frac{\alpha}{\ln 10} \left[\log L - (1 + \log L_{0}) \right] + \frac{3}{4} \log \frac{L}{10 L_{0}} + \log \dot{M}_{10} \\
	             & \equiv \tilde{\alpha} \log L + \tilde{\beta}
\end{align}
with new constants $\tilde{\alpha}$ and $\tilde{\beta}$ that essentially only depend on the choice of the expansion point $x_0$. Thus, in particular the parameter $\tilde{\alpha}$, describing the slope of the power law, crucially depends on the chosen luminosity range. At lower metallicity, $L_{0}$ becomes larger and studies will likely probe regions with $L < 10\,L_{0}$, thus yielding a steeper power law than at higher metallicity.

\begin{figure*}
  \includegraphics[width=0.32\textwidth]{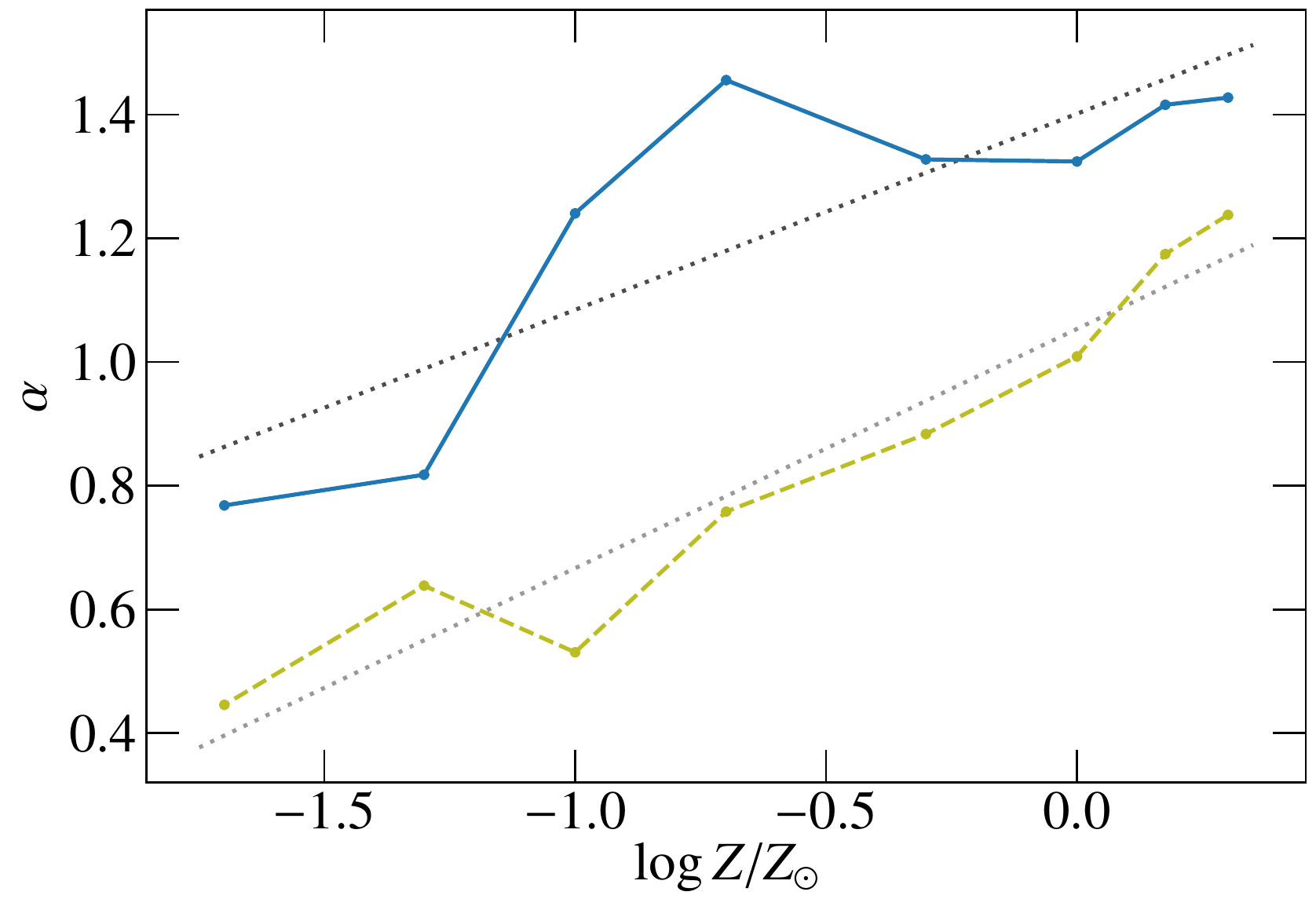} \hfill
  \includegraphics[width=0.32\textwidth]{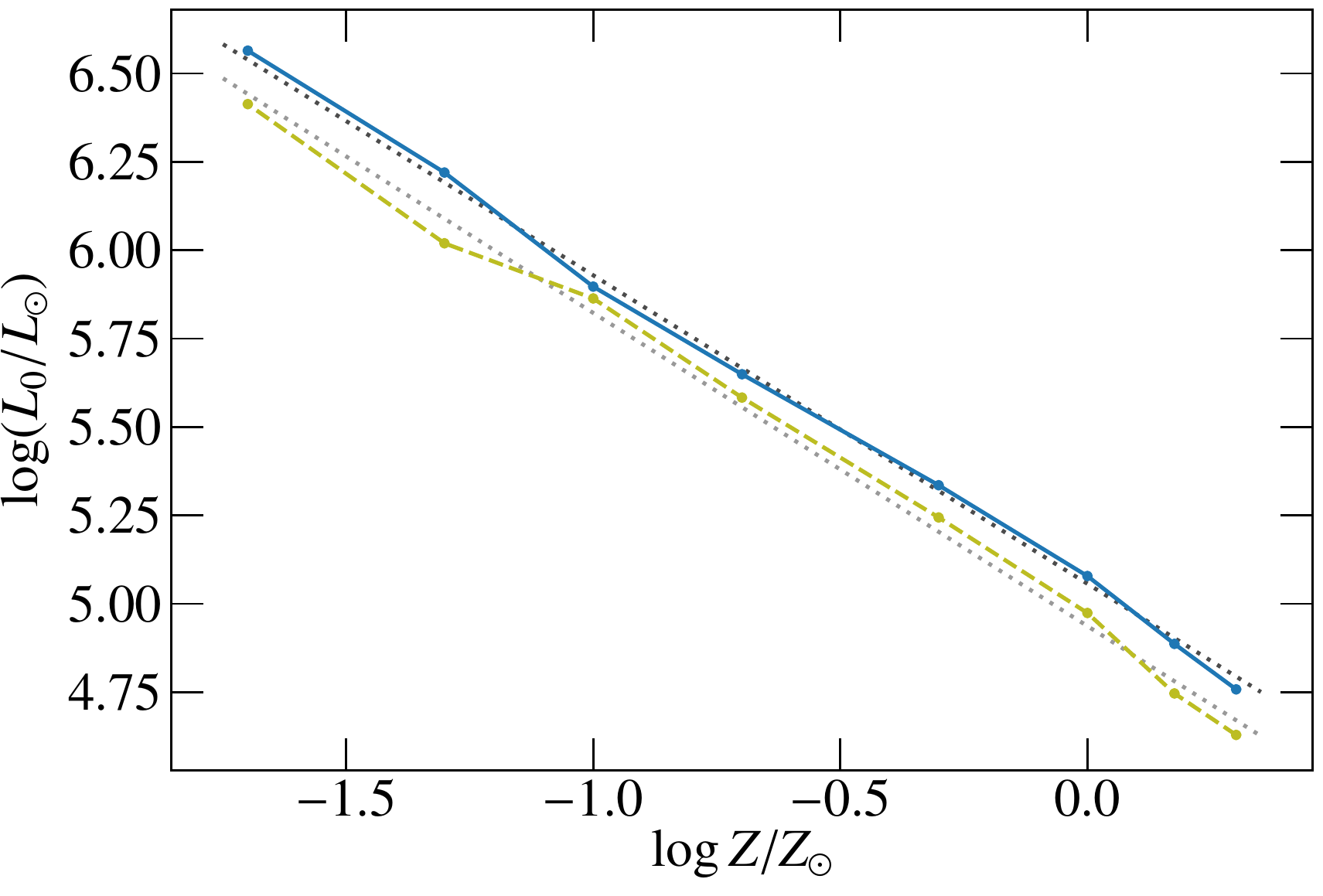} \hfill
  \includegraphics[width=0.32\textwidth]{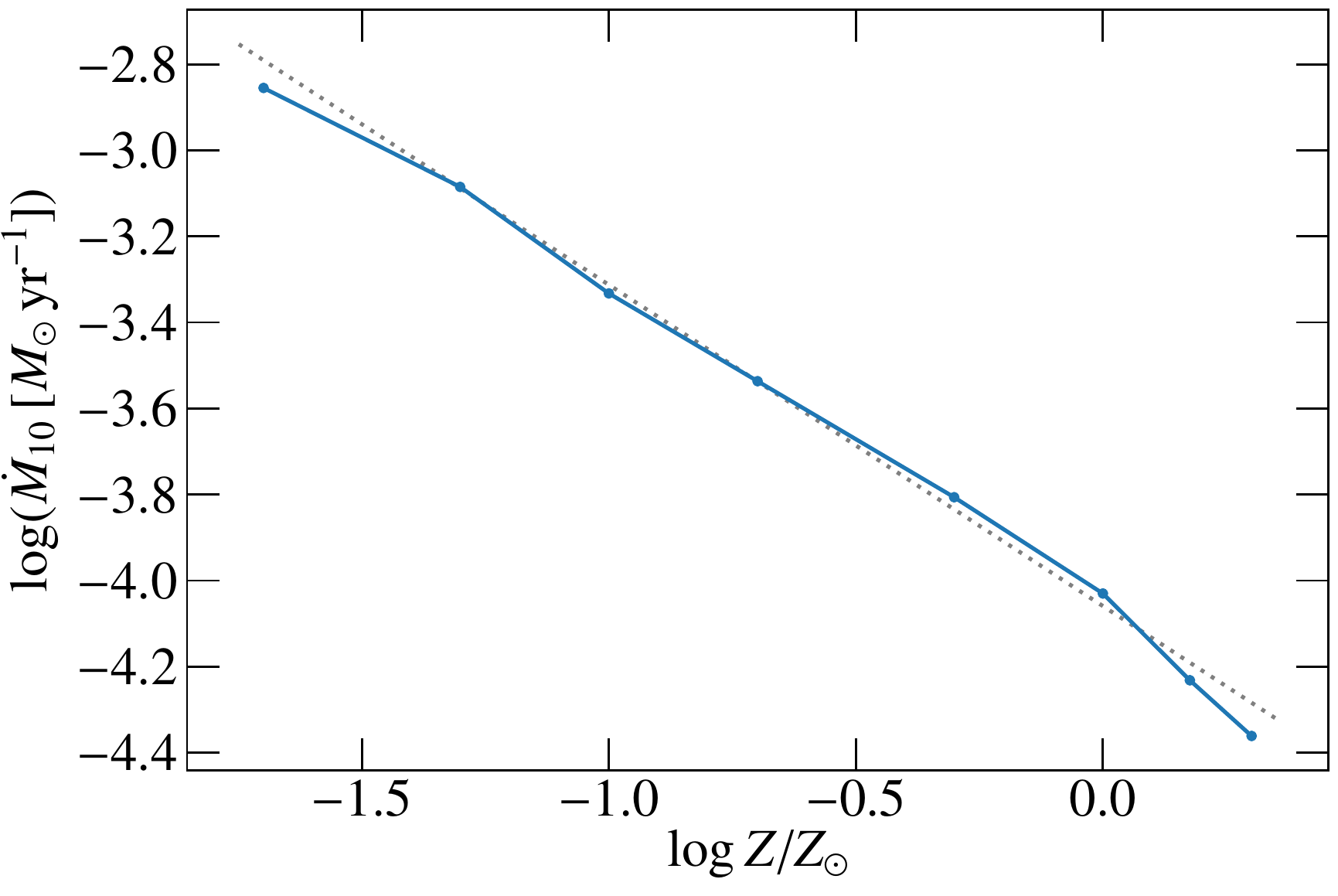}
  \caption{Metallicity-dependency of the fit coefficients $\alpha$ (left panel), $L_0$ (middle panel), and $\dot{M}_{10}$ (right panel) inherent to the $\dot{M}(L)$ mass-loss recipe denoted in Eq.\,(\ref{eq:mdotlrecipe}). The blue curve denotes the actual values derived from fitting the $\dot{M}(L)$ datasets with grey dashed lines indicating the best fit. As the parameters $\alpha$ and $L_0$ can also be obtained via fitting $\eta(L)$ (see appendix Sect.\,\ref{asec:etamax}), these results are indicated by light green dashed curves (including a best fit).}
  \label{fig:mdlrfit}
\end{figure*}

To investigate the $Z$-dependence of our three fit parameters in  Eq.\,(\ref{eq:mdotdlog}), the panels in Fig.\,\ref{fig:mdlrfit} depict trends for $\alpha$, $L_0$, and $\dot{M}_{10}$. The first two panels show an additional line as these parameters can also be determined in an alternative way by fitting $\eta(L)$ (cf.\ appendix Sect.\,\ref{asec:etamax}). The parameter $\alpha$ (left panel in Fig.\,\ref{fig:mdlrfit}) tends to increase with $Z$, but shows a lot of scatter and the results from the two methods differ by more than a factor of two for some metallicities. In contrast, both $L_0$ (middle panel in Fig.\,\ref{fig:mdlrfit}) and $\dot{M}_{10}$ (right panel in Fig.\,\ref{fig:mdlrfit}) show a much smoother behaviour and decrease with increasing $Z$. Independent of the fit method, $L_0$ seems to follow a power-law with a $\approx -0.87$ slope. With $\dot{M}_{10}(Z)$ also being sufficiently described by a power law, we get the following relations:
\begin{align}
  \alpha =&~0.32 (\pm 0.08) \cdot \log\frac{Z}{Z_\odot} + 1.40 (\pm 0.07) \\
	\log L_0/L_\odot =& -0.87 (\pm 0.02) \cdot \log\frac{Z}{Z_\odot} + 5.06 (\pm  0.02) \\
	\log \left(\frac{\dot{M}_{10}}{M_\odot\,\mathrm{yr}^{-1}}\right) =& -0.75 (\pm 0.02) \cdot \log\frac{Z}{Z_\odot} -4.06 (\pm  0.02)
\end{align}

While we could in principle try to add more terms to get a more fine-tuned $\dot{M}(L)$-recipe, this would spoil the purpose of a basic description with rather meaningful parameters. Instead, we will focus on a $\dot{M}(\Gamma_\text{e})$-type recipe, which better reflects the nature of WR-type mass loss.

\subsubsection{WR-type mass loss as a function of $\Gamma_\mathrm{e} \propto L/M$}
  \label{sec:mdotgamma}

\begin{figure}
  \includegraphics[width=\columnwidth]{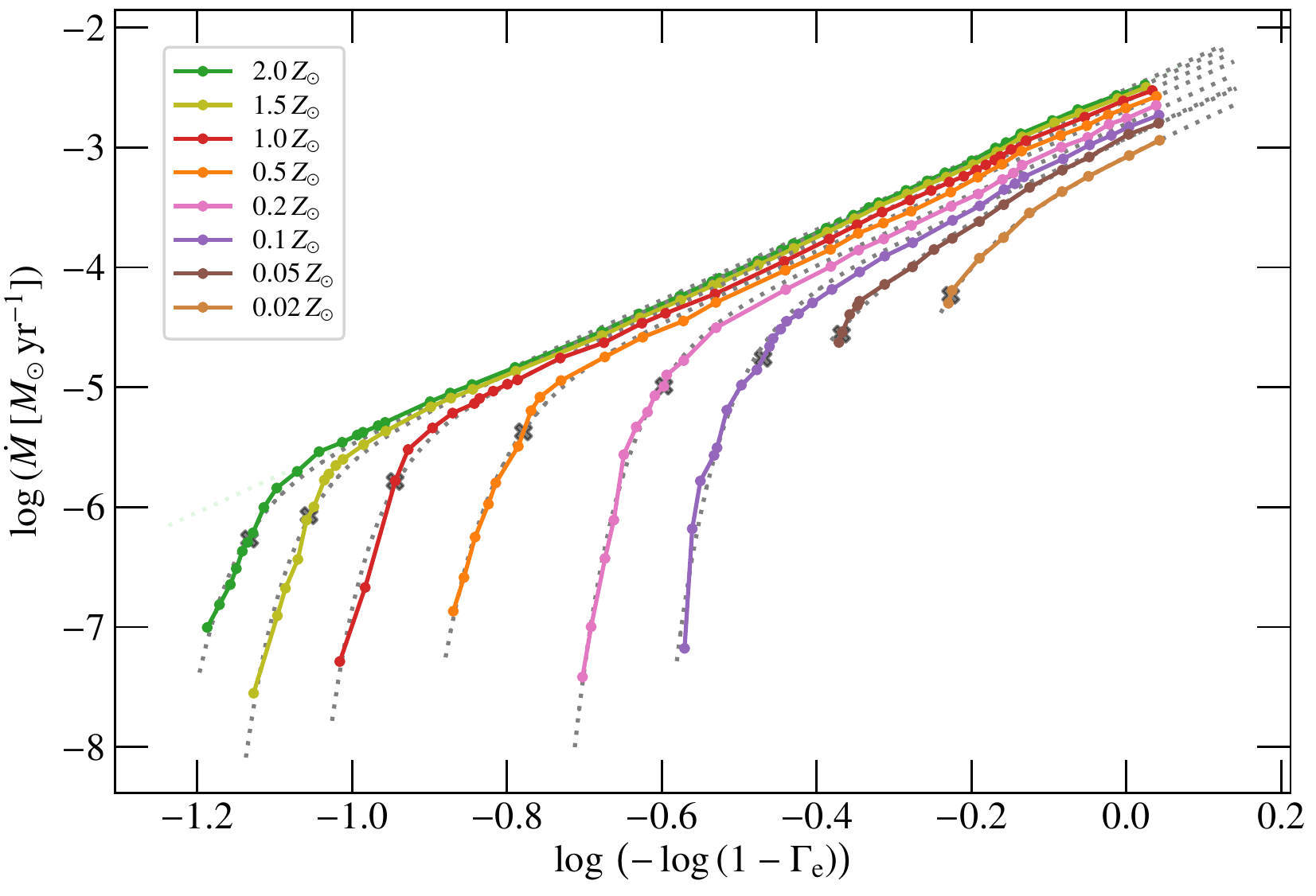}
  \caption{$\dot{M}(-\log(1-\Gamma_\text{e}))$ for all model sequences with fits according to Eq.\,(\ref{eq:mdemrecipe}) denoted as grey dashed lines. In the fits, we assume that the linear part of the curves have the same $Z$-independent slope.}
  \label{fig:mdot-lmemgedd}
\end{figure}

 For an $\dot{M}(\Gamma_\text{e})$-recipe, the recent formula for VMS by \citet{Bestenlehner2020} would be a natural candidate to consider. However, as we outline in appendix section \ref{asec:bestencmp}, such a formula cannot capture the breakdown of WR-type mass loss. Nonetheless, the considerations in appendix section \ref{asec:bestencmp} reveal the need to consider $\dot{M}$ as a function of $\log\left(1-\Gamma_\text{e}\right)$ to obtain a sufficient description without any arbitrary transition inside the pure WR-wind regime. When plotting this in a sufficient double-logarithmic form as shown in Fig.\,\ref{fig:mdot-lmemgedd}, we obtain in the asymptotic limit of high $\Gamma_\mathrm{e}$ the relation
\begin{align}
  \label{eq:mdasymp}
  \log \dot{M} &= a \cdot \log\left[-\log\left(1-\Gamma_\text{e}\right)\right] + d\\
	\nonumber    &= a \cdot \log\left[\log\frac{1}{1-\Gamma_\text{e}}\right] + d\text{.}
\end{align}
The pure WR-wind regime is fully described by Eq.\,(\ref{eq:mdasymp}). Fit results of the asymptotic behaviour yield that of the two parameters $a$ and $d$, only the latter notably depends on the metallicity $Z$.  Moreover, a description of $\dot{M}$ according to Eq.\,(\ref{eq:mdasymp}) also approaches infinity for $\Gamma_\text{e} \rightarrow 1$, similar to a recipe of the form $\dot{M} \propto (1-\Gamma_\mathrm{e})^{-a}$. 

In reality, the $\dot{M}$-trend described by Eq.\,(\ref{eq:mdasymp}) is altered by the Z-dependent breakdown of WR-type mass loss. Given our $\dot{M}(L)$-recipe, a natural extension of Eq.\,(\ref{eq:mdasymp}) to account for this breakdown would be a multi-parameter formula with parameters inside the outer logarithm, i.e.
\begin{equation}
   \label{eq:breakdownrecipeconcept}
  \log \dot{M} =  a \cdot \log\left[-b\cdot\log\left(1-\Gamma_\text{e}\right) + c \right] + d\text{.}
\end{equation}
Such a formula allows for another hard limit at $\Gamma_\text{e} > 0$, thereby also covering the breakdown of WR-type mass loss. Unfortunately, the intrinsic bending of the curve around the important transition regime provides only an insufficient description with considerable deviations in $\dot{M}$. Instead, we can represent the transition of the wind regimes much more accurately with an additional term describing an exponential decline. In total, this leads to a mass-loss recipe of the form
\begin{align}
  \label{eq:mdemrecipeconcept}
  \log \dot{M} &=  a \cdot X - c \cdot 10^{-b\cdot X} + d \\
  \nonumber	     &\mathrm{with~} X := \log\left[-\log\left(1-\Gamma_\text{e}\right)\right]\text{.}
\end{align}
The first and third term in Eq.\,(\ref{eq:mdemrecipeconcept}) are taken from Eq.\,(\ref{eq:mdasymp}) and describe the linear slope for higher $X$ observed in Fig.\,\ref{fig:mdot-lmemgedd} with a $Z$-dependent offset $d$. The second term adds an exponential decline towards lower $X$ and requires both a scaling and a shift in $X$. The latter has been re-written as a factor $c$ in front of the exponential term in Eq.\,(\ref{eq:mdemrecipeconcept}). Inserting the definition of $X$, the recipe reads
\begin{equation}
  \label{eq:mdemrecipe}
  \log \dot{M} =  a \cdot \log\left[-\log\left(1-\Gamma_\text{e}\right)\right] - c \cdot \left[-\log\left(1 - \Gamma_e\right)\right]^{-b} + d\text{.}
\end{equation}

To get a more meaningful number which reflects the exponential decline of WR-type mass loss, we can combine $c$ with the stretching factor $b$ to introduce the `breakdown-indicator'
\begin{equation}
  \label{eq:bdindicator}
  \Gamma_{\text{e},\text{b}} = 1 - 10^{-10^{\frac{1}{b} \log \frac{c}{c_\text{b}}}}
\end{equation}
with a constant $c_\text{b}$ to be defined. A reasonable measure is to set $c_\text{b} = \log(2) \approx 0.3$, so that $\Gamma_{\text{e},\text{b}}$ reflects the $\Gamma_\text{e}$-value for which $\dot{M}$ deviates from the pure WR regime by a factor of two. For $\Gamma_{\text{e}} < \Gamma_{\text{e},\text{b}}$, this deviation then grows exponentially. A fit for the values derived for $\Gamma_{\text{e},\text{b}}$ yields the relation
\begin{equation}
  \Gamma_{\text{e},\text{b}} = -0.319 (\pm 0.009) \cdot \log(Z/Z_\odot) + 0.244 (\pm 0.008) \text{.}
\end{equation}
This relation almost coincides with Eq.\,(\ref{eq:gedd-etaunity-z}), allowing us the important conclusion 
\begin{equation}
  \Gamma_{\text{e},\text{b}} = \left.\Gamma_\mathrm{e}\right|_{\eta = 1}\text{.}
\end{equation}
Consequently, we can quantitatively tie the onset of multiple scattering with the onset or breakdown of WR-type mass loss and introduce an alternative to Eq.\,(\ref{eq:mdemrecipeconcept}) with a simpler exponential term:
\begin{equation}
   \label{eq:breakdownrecipe}
  \log \dot{M} =  a \cdot \log\left[-\log\left(1-\Gamma_\text{e}\right)\right] - \log(2) \cdot \left(\frac{\Gamma_{\text{e},\text{b}}}{\Gamma_\text{e}}\right)^c + d\text{.}
\end{equation}

\begin{figure}
  \includegraphics[width=\columnwidth]{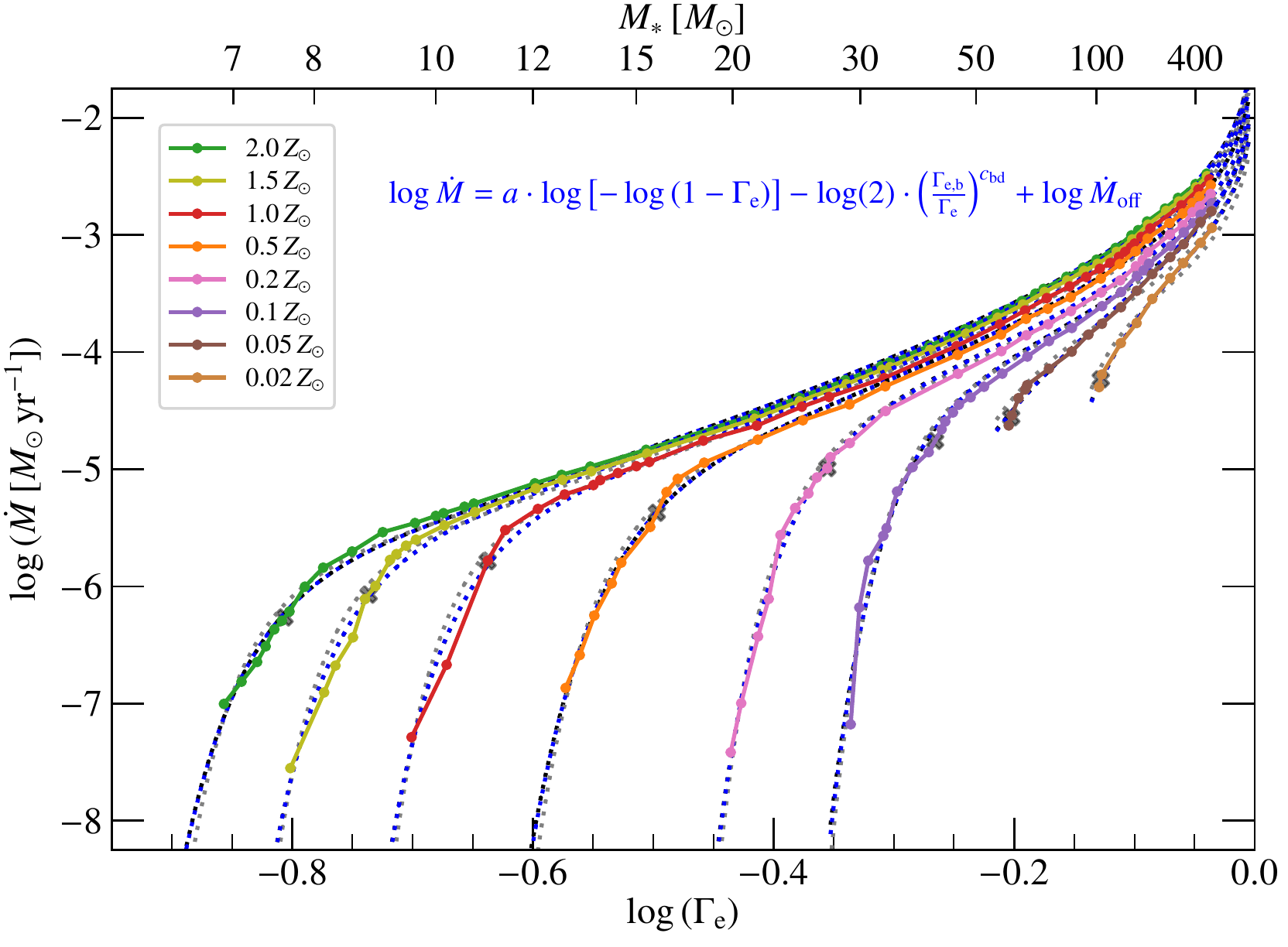}
  \caption{$\dot{M}(\Gamma_\text{e})$ for all model sequences with fits from different recipes. The black dashed lines denote fits according to Eq.\,(\ref{eq:mdemrecipe}) while the blue-dashed lines indicate a fit following Eq.\,(\ref{eq:breakdownrecipe}). For comparison, also a fit with fixed values for $\Gamma_{\text{e},\text{b}}$ in Eq.\,(\ref{eq:breakdownrecipe}) is shown (gray dashed lines).}
  \label{fig:mdot-gedd-recipe}
\end{figure}

\begin{figure*}
  \includegraphics[width=0.32\textwidth]{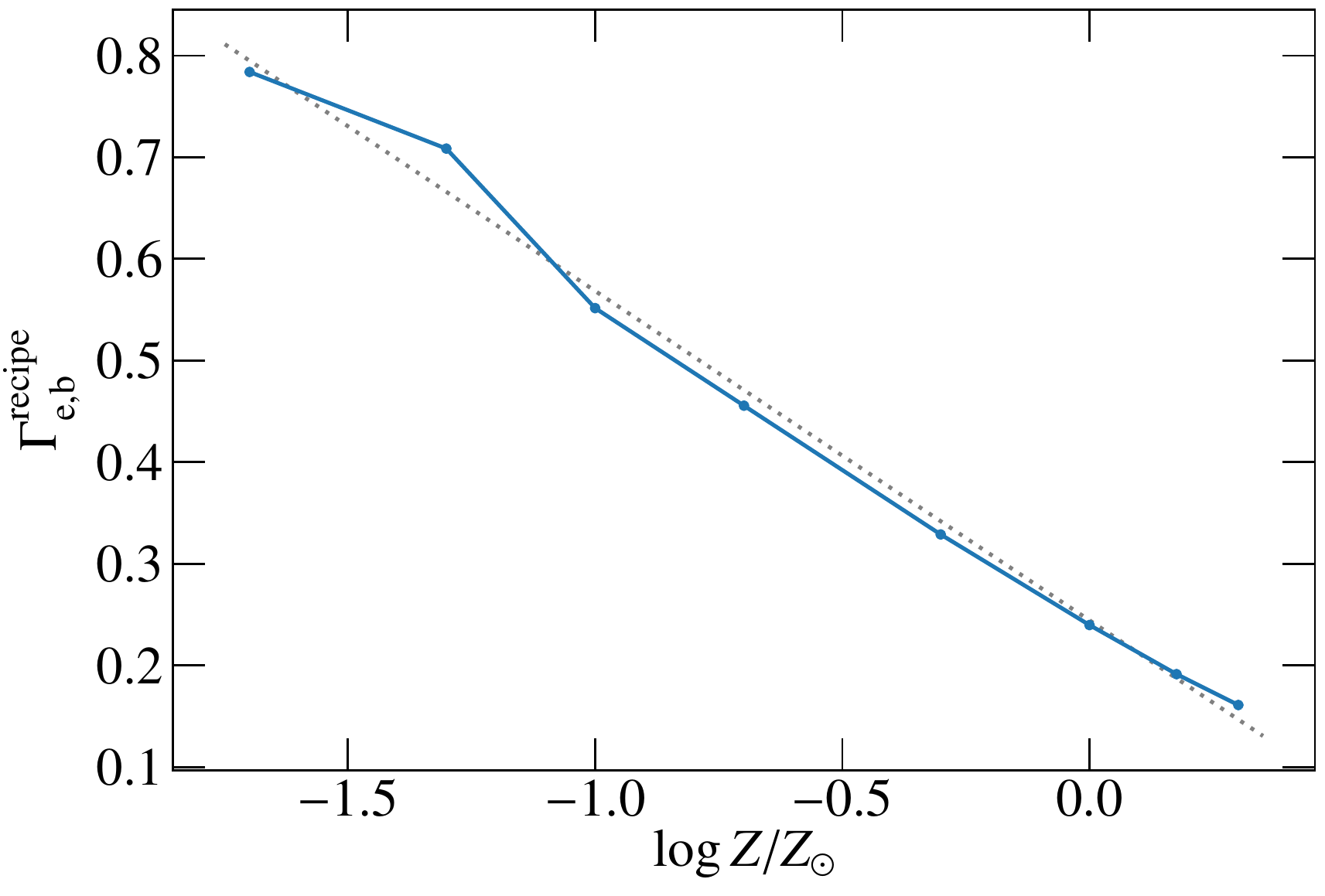} \hfill
  \includegraphics[width=0.32\textwidth]{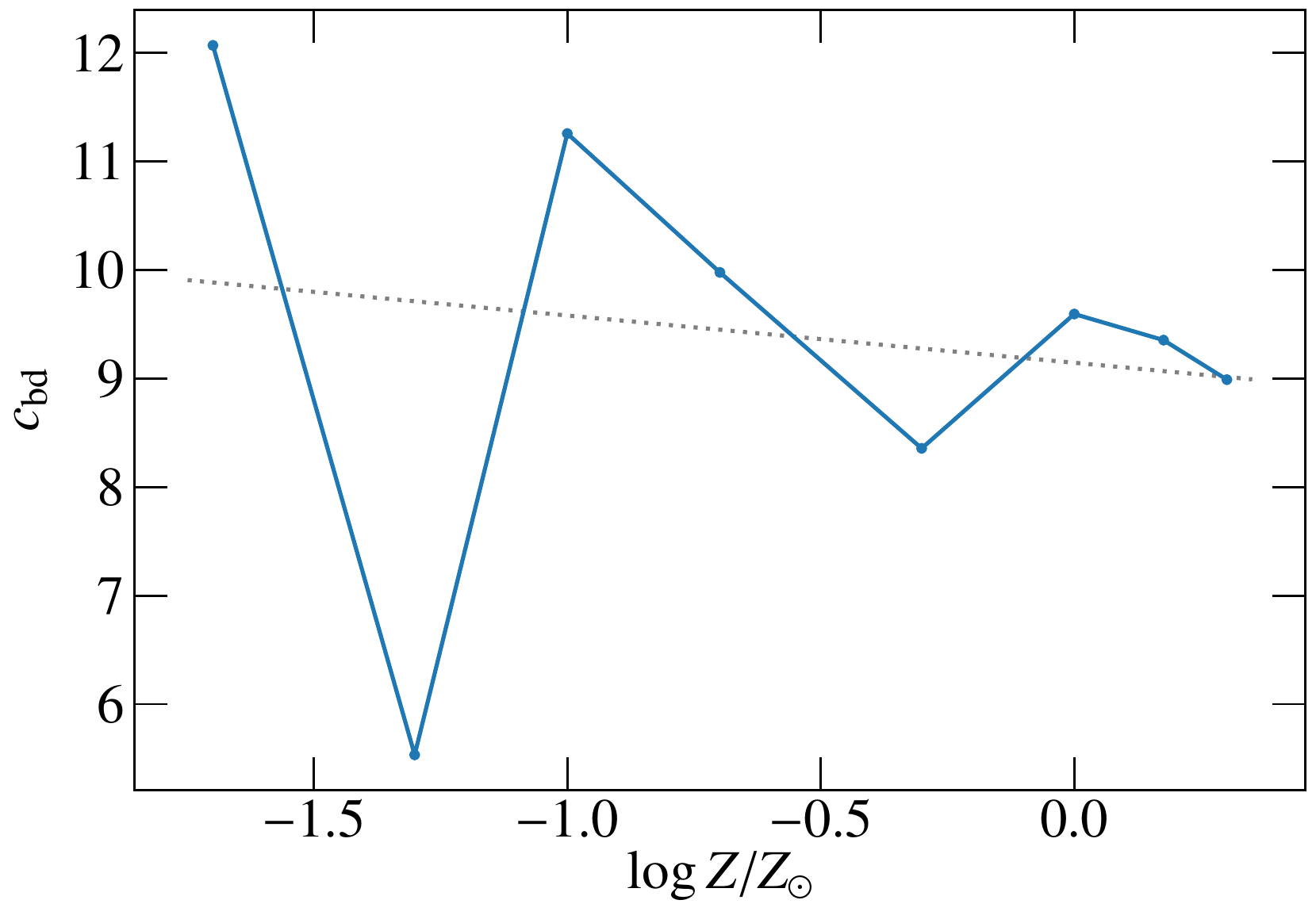} \hfill
  \includegraphics[width=0.32\textwidth]{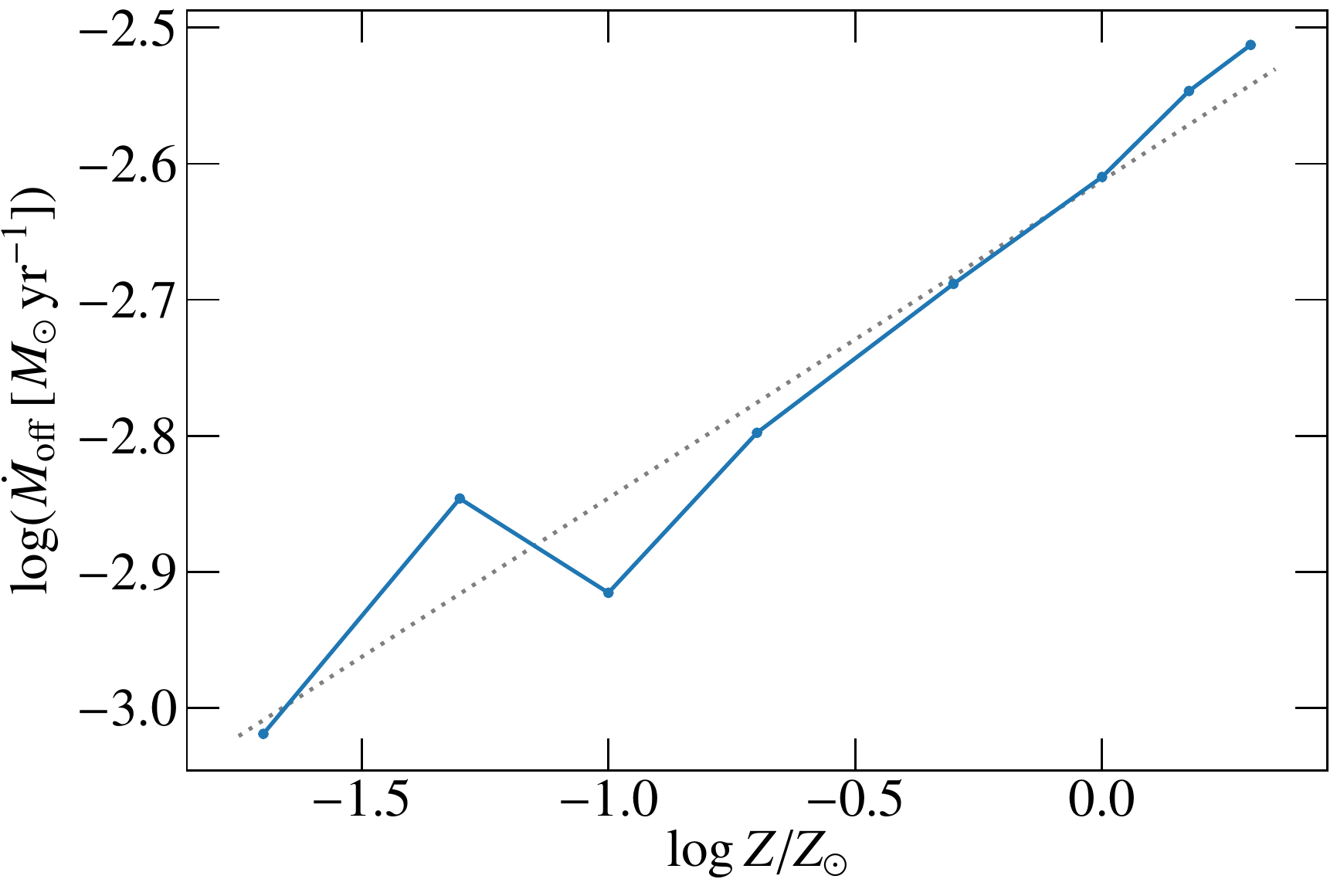}
  \caption{Metallicity-dependency of the fit coefficients $\Gamma_{\text{e},\text{b}}$ (left panel), $c_\text{bd}$ (middle panel), and $\dot{M}_\text{off}$ (right panel) inherent to the $\dot{M}(\Gamma_\text{e})$ mass-loss recipe denoted in Eq.\,(\ref{eq:breakdownrecipe}). The blue curve denotes the actual values derived from fitting the $\dot{M}(\Gamma_\text{e})$ datasets with grey dashed lines indicating the best fit.}
  \label{fig:mdbdfit}
\end{figure*}

We test our recipes and their underlying implications by a two-step process, where we first fit the asymptotic part to get
\begin{equation}
  \label{eq:aparam}
  a = 2.932 (\pm 0.016)\text{.}
\end{equation}
We then keep the same $a$ as a fixed parameter when fitting the whole datasets according to Eqs.\,(\ref{eq:mdemrecipe}) and (\ref{eq:breakdownrecipe}). Both fits yield excellent results, as depicted in Fig.\,\ref{fig:mdot-gedd-recipe}, and correctly capture the complete asymptotic behaviour, including the breakdown of WR-type mass loss towards low $\Gamma_\text{e}$. 
A regression of the $\Gamma_{\text{e},\text{b}}$-values derived from the recipe fits resulting from Eq.\,(\ref{eq:breakdownrecipe}) yields a $Z$-dependence of 
\begin{equation}
  \label{eq:geddbdfit}
  \Gamma_{\text{e},\text{b}} = -0.324 (\pm 0.011) \cdot \log(Z/Z_\odot) + 0.244 (\pm 0.010) \text{,}
\end{equation}
which is depicted in the left panel of Fig.\,\ref{fig:mdbdfit}. The coefficients in Eq.\,(\ref{eq:geddbdfit}) are close to their counterparts in Eqs.\,(\ref{eq:gedd-etaunity-z}) and (\ref{eq:bdindicator}). Fixing the $\Gamma_{\text{e},\text{b}}$-values a-priory to the empirical values of $\left.\Gamma_\mathrm{e}\right|_{\eta = 1}$ does not improve the fit quality and bears the risk of putting too much emphasis on values which are not directly calculated, but stem from the interpolation between two models close to $\eta = 1$. The parameter $d \equiv \log \dot{M}_\text{off}$, which denotes $\dot{M}(\Gamma_\text{e} \approx 0.9)$, also shows a clear $Z$-trend (cf.\ right panel of Fig.\,\ref{fig:mdbdfit}) which can be described by
\begin{equation}
  \log \dot{M}_\text{off} = 0.23 (\pm 0.04) \cdot \log(Z/Z_\odot) - 2.61 (\pm 0.03) \text{.}
\end{equation}
The trend for the exponent $c$, which captures the `acceleration' of the breakdown, is less conclusive. The middle panel of Fig.\,\ref{fig:mdbdfit} reveals a considerable scatter for the exponent $c$ (denoted $c_\text{bd}$) even on a logarithmic scale, but also a certain tendency towards a roughly constant value. This is confirmed in our linear fit 
\begin{equation}
  \label{eq:cbdparam}
  c_\text{bd} = -0.44 (\pm 1.09) \cdot \log(Z/Z_\odot) + 9.15 (\pm 0.96) 
\end{equation}
which yields a rather robust constant with a highly uncertain, possibly even vanishing $Z$-trend. Given that the models in the regime of $\eta \ll 1$ are the most uncertain in our study, the ambiguity in $c_\text{bd}$ is not a surprise. Therefore, the absolute values of $\dot{M}$ more than an order of magnitude below $\dot{M}(\Gamma_{\text{e},\text{b}})$ should be handled with caution and regarded as uncertain. Nonetheless, it is important to keep in mind that the breakdown of $\dot{M}$ by orders of magnitude as such is real due to the transition of the wind regime \citepalias[cf.][]{Sander+2020}.

In short, our $\dot{M}(\Gamma_\text{e})$-recipe does an excellent job in describing the mass loss of our calculated model sets as it sufficiently reproduces the obtained curves and correctly accounts for the asymptotic behaviour, despite the uncertainty of the absolute values for $\eta \ll 1$. Our recipe can even explain our earlier finding of $\left.\log \dot{M}\right|_{\eta = 1} \propto \left.\Gamma_\text{e}\right|_{\eta = 1} = \Gamma_{\text{e},\text{b}}$ (cf.\ Eq.\,\ref{eq:mdldm-etaunity}). We write $\dot{M}(\Gamma_{\text{e},\text{b}})$ and use $\ln(1-\Gamma_\text{e}) \approx -\Gamma_\text{e}$ since $\Gamma_{\text{e},\text{b}} \ll 1$ to obtain
\begin{equation}
  \left.\log\dot{M}\right|_{\eta = 1} \approx a \log\left(\frac{\Gamma_{\text{e},\text{b}}}{\ln (10)}\right) - \log(2) + d.
\end{equation}

\begin{figure}
  \includegraphics[width=\columnwidth]{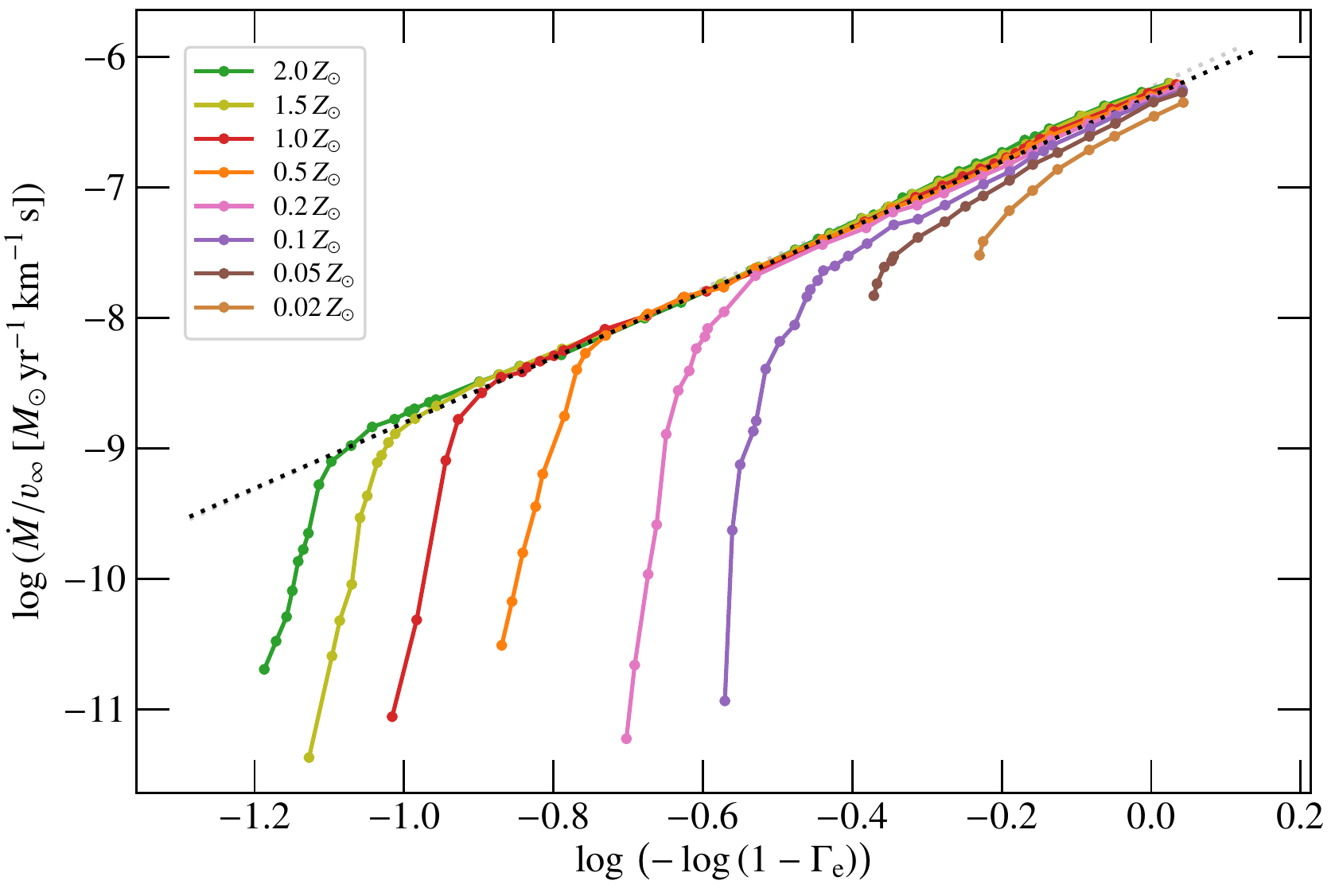}
  \caption{The ratio of $\dot{M}$ and $\varv_\infty$ as a function of $-\log(1-\Gamma_\text{e})$, depicting an essentially $Z$-independent slope for pure WR-type winds.}
  \label{fig:mdotdvinf-emgedd}
\end{figure}

\subsubsection{Relating mass loss and terminal velocity in WR-type winds}
  \label{sec:mdotvinf}
	
As mentioned above, a recipe in the form of $\dot{M}(\Gamma_\text{e})$ has the disadvantage that the stellar mass (or the $L/M$-ratio) must be known. We can avoid this with a completely different kind of recipe, which, unfortunately, is only valid in the pure WR-wind regime. In Fig.\,\ref{fig:mdotdvinf-emgedd}, we plot $\log(\dot{M}/\varv_\infty)$ as a function of $\log\left[-\log\left(1-\Gamma_\text{e}\right)\right]$. The collapsing of all curves onto one line in the pure WR-wind regime is expected from our findings for $\dot{M}_\mathrm{t}$. Moreover, the linear slope in this regime allows for an easy, metallicity-independent fit of the form
\begin{equation}
  \log\frac{\dot{M}}{\varv_\infty} = k \cdot \log\left[-\log\left(1-\Gamma_\text{e}\right)\right] + l\text{.}
\end{equation}
Far away from the breakdown ($\Gamma_\text{e} \gg \Gamma_\text{e,b}$), Eq.\,(\ref{eq:breakdownrecipe}) reduces to
\begin{equation}
  \log\dot{M} = a \cdot \log\left[-\log\left(1-\Gamma_\text{e}\right)\right] + d(Z)\text{.}
\end{equation}
Combining these two equations allows us to eliminate the complex $\Gamma_\text{e}$-dependency, yielding the simple relation
\begin{equation}
  \label{eq:vinfmdotscaling}
  \log \varv_\infty = \left(1 - \frac{k}{a}\right) \log \dot{M} - l + \frac{k}{a} d(Z)\text{.}
\end{equation}
Hence, in the limit of pure WR-type mass loss, $\log \varv_\infty$ scales linear with $\log \dot{M}$. Ideally, the scaling factor is even identical at different $Z$ as only the shift in Eq.\,(\ref{eq:vinfmdotscaling}) has a $Z$-dependent term. This would provide us with a powerful observational mass-loss diagnostic only depending on the observable $\varv_\infty$, once the coefficients from Eq.\,(\ref{eq:vinfmdotscaling}) are known, either from theoretical models or from observational gauging. Unfortunately, this is an asymptotic relation, since we have not used the full Eq.\,(\ref{eq:breakdownrecipe}), but only the high-$\Gamma_\text{e}$ limit.

\begin{figure}
  \includegraphics[width=\columnwidth]{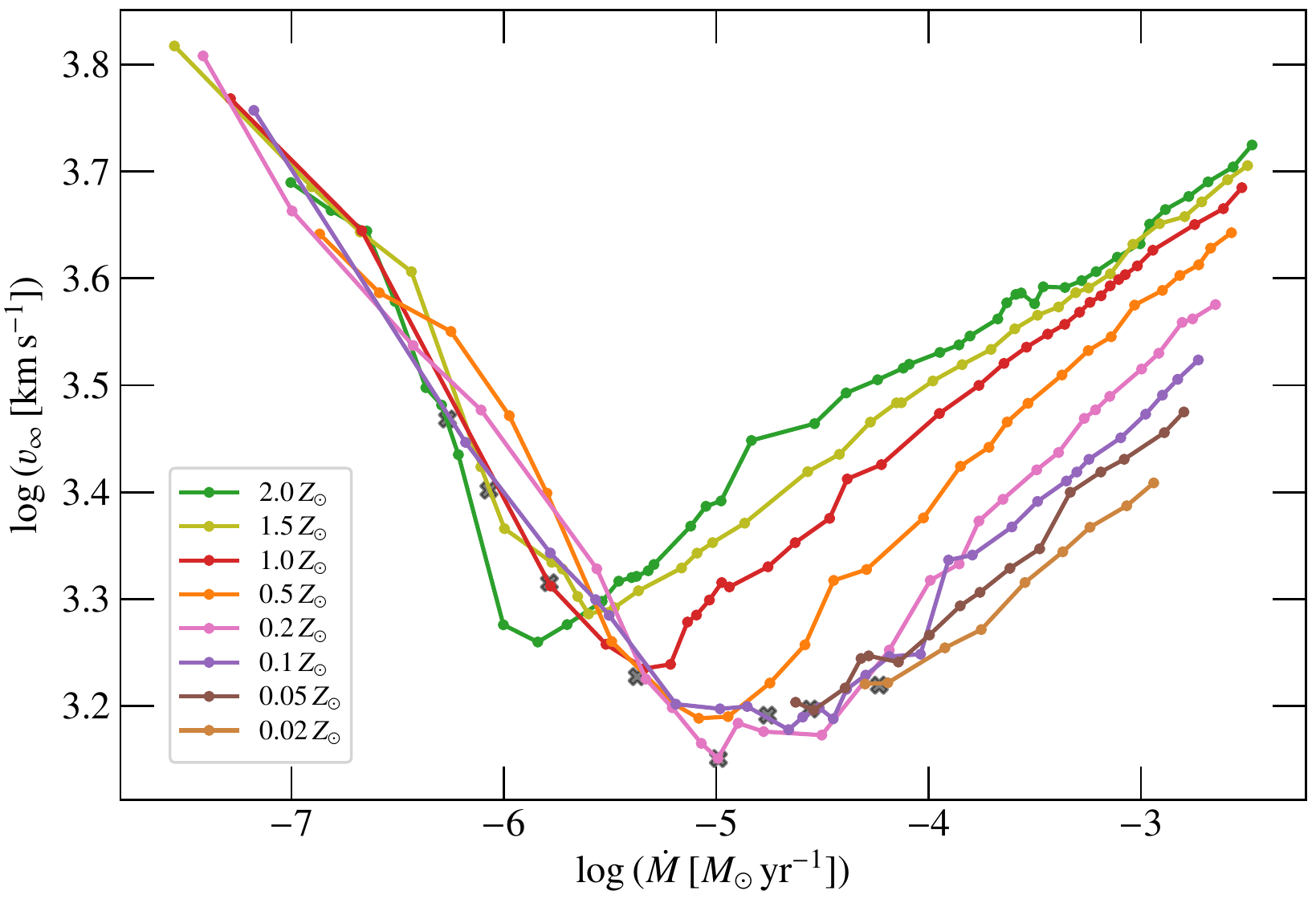}
  \caption{Terminal velocity $\varv_\infty$ for our HD model sequences as a function of their mass-loss rate $\dot{M}$, illustrating the fundamental difference between optically thin and optically thick, WR-type winds.}
  \label{fig:vinf-mdot}
\end{figure}

A general increase of $\dot{M}$ with increasing $\varv_\infty$ was already found by \citet{Graefener+2017} in numerical wind models assuming a $\beta$-law, while their analytic approximation yielded a non-monotonic behaviour. In Fig.\,\ref{fig:vinf-mdot}, we now plot $\varv_\infty(\dot{M})$ of our models to check whether the proportionality suggested by Eq.\,(\ref{eq:vinfmdotscaling}) is noticeable in our results. While there is some clear scatter in our models for $\varv_\infty$, in particular the curves for higher metallicity display a clearly linear scaling of $\log \varv_\infty$ with $\log \dot{M}$. For the lower metallicities, this is also visible once higher mass-loss rates are reached. From Eq.\,(\ref{eq:vinfmdotscaling}) we would expect all slopes to be the same, but  Fig.\,\ref{fig:vinf-mdot} illustrates that this ideal situation is not fulfilled in our model sequences. Still, they could give rise to a recipe in the form of $\log \dot{M} = \tilde{k}(Z) \cdot \log \varv_\infty + \tilde{l}(Z)$ with $Z$-dependent coefficients $\tilde{k}$ and $\tilde{l}$. Of course, the coefficients could be susceptible to various quantities fixed in this study, such as $T_\ast$, the clumping stratification, or the chemical composition. For example, the $(\tilde{k},\tilde{l})$-coefficients for WC stars would differ from the ones for hydrogen-free WN stars.
In any case, the clear switch from a negative to a positive correlation between $\dot{M}$ and $\varv_\infty$ in Fig.\,\ref{fig:vinf-mdot} underlines the fundamental difference between optically thick and optically than stellar winds, or -- in other words -- regimes with and without multiple scattering.

\section{Explicit Metallicity Trends}
  \label{sec:ztrends}

As an alternative to investigating the $Z$-dependency of our recipes, we can take a direct look at our set of models along the $Z$-dimension. Given our choice of abundances, the $Z$-dependencies in both models and recipes essentially reflect the iron abundance and thus could be written as a $Z_\text{Fe}$- (or $X_\text{Fe}$-) dependency. Some recipes, such as those from \citet{NL2000} or \citet{Tramper+2016}, also include $Y$-dependencies, but for a study of hydrogen-free stars with $Y=1-Z$ and a general scaling of all metals, an explicit $Y$-dependency would introduce a degeneracy. Moreover, we see general issues with explicit $Y$-dependencies in mass-loss recipes. An explicit $Y$-term could lead to the misleading interpretation that He is a major contributor to $\dot{M}$, which is not the case. Instead, the role of He is usually an indirect one as it contributes a larger number of free electrons compared to heavier elements.

Before discussing the explicit effects of scaling all elements contributing to $Z$, we briefly investigate the influence of increased and decreased CNO abundances. In contrast to \citetalias{Sander+2020}, we do not assume an empirical CNO mixture in this work, but instead stick to CNO abundances derived from evolutionary models, in line with our concept of investigating the He ZAMS. Different CNO abundances can have an indirect influence on the results of our study due to their effect on the free electron budget and their contribution to the total optical depth. To get a first measure of this impact, we run a test calculation with empirical CNO abundances for a model where we would expect a significant impact, namely a massive $70\,M_\odot$ He star model at $Z_\odot$. The derived $\dot{M}$ for a model with typical empirical WN abundances (i.e.\ $X_\text{N} = 0.015$) is only $0.02\,$dex higher than our corresponding He star model. In the outer wind, the higher $X_\text{N}$ leads to an increase of about $2.5\%$ in $\varv_\infty$. 

\begin{figure}
  \includegraphics[width=\columnwidth]{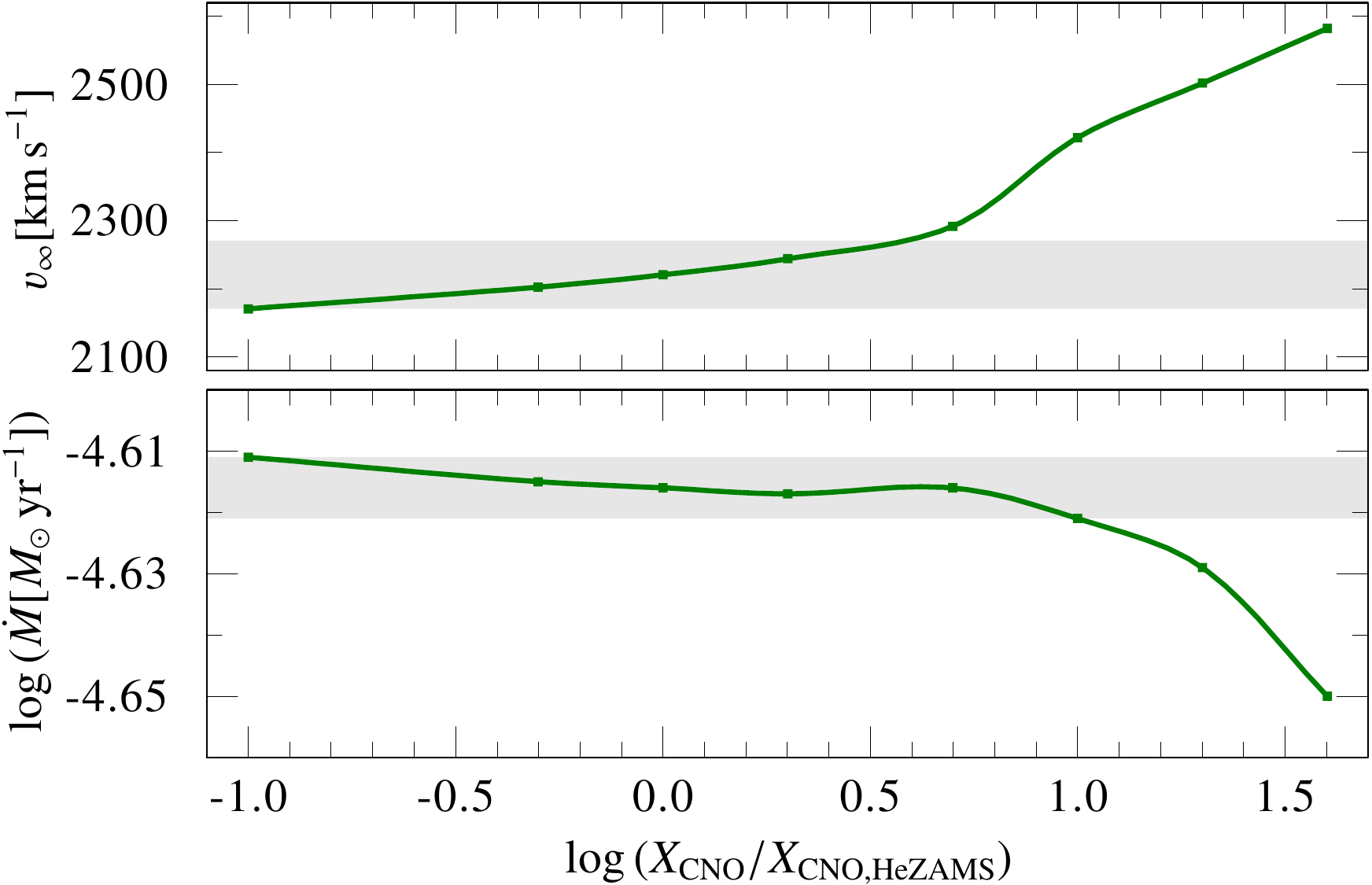}
  \caption{Terminal velocity (upper panel) and mass-loss rate (lower panel) for He star models with $20\,M_\odot$ and different CNO abundances. All other metals have been kept at their solar values as denoted in Table\,\ref{tab:inputparams}.}
  \label{fig:mdot-cno-abu}
\end{figure}
To quantify the impact in a range that corresponds more to the observed WN population, we run a series of models with $20\,M_\odot$, where we only vary the CNO abundance, scaling those three elements between $0.1$ to $40$ times of our standard value. The result is depicted in Fig.\,\ref{fig:mdot-cno-abu} and confirms our findings of the $70\,M_\odot$ test case. Even when scaling the CNO abundance up and down by a factor of $10$, the mass-loss rate $\dot{M}$ does not change by more than a factor $0.005\,$dex, which is indicated by the shaded area in the lower panel. For larger CNO abundances, the effect gets larger due to the depletion of He and the resulting reduction in the free electron budget \citepalias[cf.\ the abundance effect discussion in][]{Sander+2020}. Nonetheless, the difference corresponds to not more than $0.02\,$dex when reducing the He fraction from $0.99$ to $0.63$. The terminal velocity $\varv_\infty$ is a bit more affected as the larger portions of CNO provide additional opacity in the outer wind, but stays in a region of $\pm 50\,\mathrm{km\,s}^{-1}$ (grey area in the upper panel of Fig.\,\ref{fig:mdot-cno-abu}) when decreasing CNO by a factor of $10$ or increasing it by a factor of $4$. 
Thus, while the spectral imprint could be quite significant, we conclude that the particular choice of the precise CNO abundances should not significantly alter our derived results, as long as they represent a mixture after CNO equilibrium, but before any further He star evolution (e.g.\ to the WC stage). \citetalias{Sander+2020} calculated the impact of different C and O abundances in WC star atmospheres, obtaining a decrease in $\dot{M}$ for higher C or O, with the exact changes in $\dot{M}$ depending on whether C or O is added. Thus, we argue that at least for major individual contributors in WR stars, such as C and O in WC stars or hydrogen in WN stars, recipes with explicit elemental abundance terms will likely be necessary. 

\begin{figure}
  \includegraphics[width=\columnwidth]{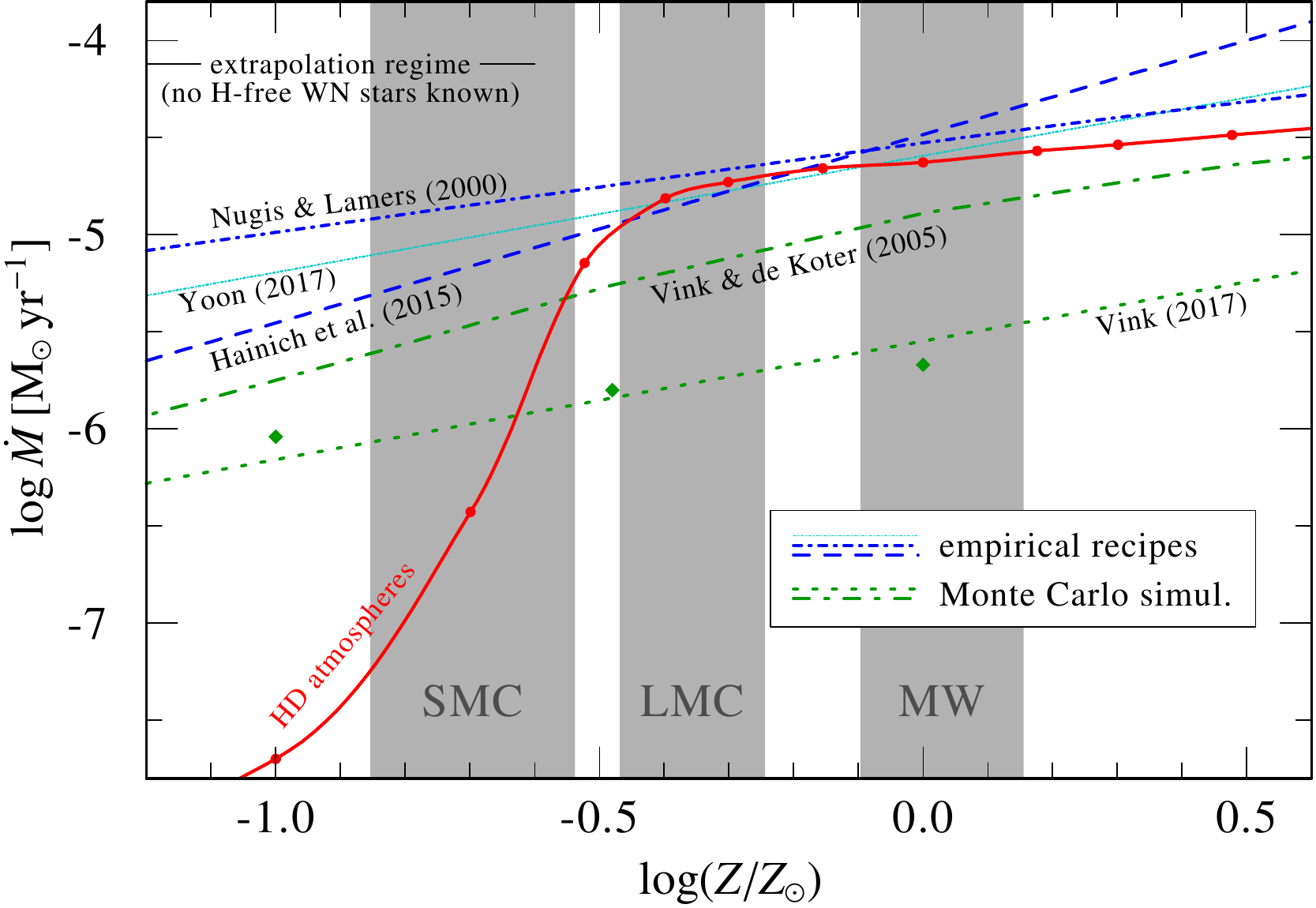}
  \caption{Mass-loss rate $\dot{M}$ as a function of metallicity $Z$ for our hydrogen-free atmosphere models with $20\,M_\odot$, compared to different empirical and theoretical recipes evaluated for $X_\text{H} = 0$. The green diamonds denote the individual models from \citet{Vink2017}.}
  \label{fig:cmp-mdot-z-recipe}
\end{figure}

In Fig.\,\ref{fig:cmp-mdot-z-recipe}, we plot $\dot{M}(Z)$ for a set of HD atmosphere models for $20\,M_\odot$ and compare it to various empirical and theoretical descriptions. While absolute values have to be taken with care due to the underlying uncertainties (e.g.\ choice of $T_\ast$, $M$-$L$-relation, assumed clumping recipe, no fine-tuning of the abundances), the general trends are important as they will have major impacts when applied e.g.\ in evolutionary models or population synthesis. As imminent from Fig.\,\ref{fig:cmp-mdot-z-recipe}, all empirical descriptions, which have been derived for WN stars, have severe problems when extrapolating them, in particular towards lower $Z$. At higher $Z$, the recipe from \citet{NL2000}, albeit having a slightly steeper slope, comes closest to our result. At the same time, it also fails to reproduce the breakdown of $\dot{M}$ below $Z_\text{LMC}$, leading to a massive over-prediction of He star mass loss at low $Z$, i.e.\ also in the early Universe. The steeper relation by \citet{Hainich+2015} correctly captures the behaviour around $Z_\text{LMC}$, where $\dot{M}(Z)$ starts to deviate from the flat slope at higher $Z$. The formula by \citet{Yoon2017} is a compromise of \citet{Tramper+2016} -- essentially a slight extension of \citet{Hainich+2015} -- and \citet{NL2000}. Consequently, it performs well in the observationally constrained regime between the metallicities of the Milky Way and the Magellanic Clouds. However, all the empirical recipes are power-laws in $Z$, which are -- as our HD models demonstrate -- insufficient in capturing the full behaviour of $\dot{M}(Z)$.

Previous theoretical relations have shortcomings as well. Both Monte-Carlo descriptions were calculated for a cooler temperature regime of $50\,$kK. In particular, the recipe from \citet{Vink2017} -- another power law -- was never intended for WR stars, but for stripped lower-mass He stars. Still, it provides an important comparison for the low-$Z$ regime where we do not obtain WR-type winds. The theoretical description from \citet[][hereafter VdK2005]{VdK2005} was based on the parameters of a late-type WN star. While often only their power-law approximation for the $Z$-dependency ($Z^{0.85}$) is used in mass-loss recipes, we are plotting their actual raw dataset in Fig.\,\ref{fig:cmp-mdot-z-recipe}, allowing us to study deviations from a power law. The mass-loss rates in \citetalias{VdK2005} were calculated for models with $20\,M_\odot$, similar to our set, but assumed a slightly lower luminosity of $\log L/L_\odot = 5.62$ instead of our $5.7$. This implies a lower $L/M$-ratio and thus it is no surprise that their mass-loss rates are lower than our results, even at higher $Z$. As evident from Fig.\,\ref{fig:cmp-mdot-z-recipe}, the models in \citetalias{VdK2005} do not show the steep drop in $\dot{M}$ around SMC metallicity. This is likely due to the assumption of fixed $\varv_\infty$ and the limitation to global consistency in \citetalias{VdK2005}.
 As our locally consistent models reveal \citepalias[cf.\ Figs.\,13 and 16 in][]{Sander+2020}, not all line opacities (significantly) affect $\dot{M}$, but mainly those from the Fe group. Opacities (e.g.\ from CNO) only available in the outer wind increase $\varv_\infty$, but do not affect $\dot{M}$. Without this local treatment and the fixed assumption for $\varv_\infty$, the models from \citetalias{VdK2005} use the available opacity for raising $\dot{M}$, thus obtaining too high mass-loss rates. In retrospect, the individual results for the stripped-star models in \citet{Vink2017}, employing a local dynamical approach \citep[cf.][]{MV2008}, already hinted towards a non-power-law behaviour in $\dot{M}(Z)$, but did not indicate any transition as drastic as we obtain in our CMF calculations. Evidence for a complex $\dot{M}(Z)$-behaviour of WR-type mass loss was further obtained by \citetalias{GH2008}, albeit for WNh stars (see also appendix Sect.\,\ref{asec:gh2008}).

The exemplary slope of the dynamically consistent models in Fig.\,\ref{fig:cmp-mdot-z-recipe} provides a first insight on the nature of the $\dot{M}$-transition regime along the $Z$ dimension. Similar to what we find along the $L$-dimension, a moderate slope -- actually much more flat than for $\dot{M}(L)$ -- in the high-$Z$ regime gets gradually steeper when approaching the optically thin limit. While the data in the $L$-dimension hint that $\dot{M}$ might approach a breakdown luminosity $L_{0}$, the behaviour seems to be more complex along the $Z$-axis. Nonetheless, there is a drop in $\dot{M}$ by several orders of magnitude in a rather narrow $Z$-range. Therefore, it is well justified to also label this as a `breakdown' of WR-type mass loss, now below a certain metallicity regime.
 
\begin{figure}
  \includegraphics[width=\columnwidth]{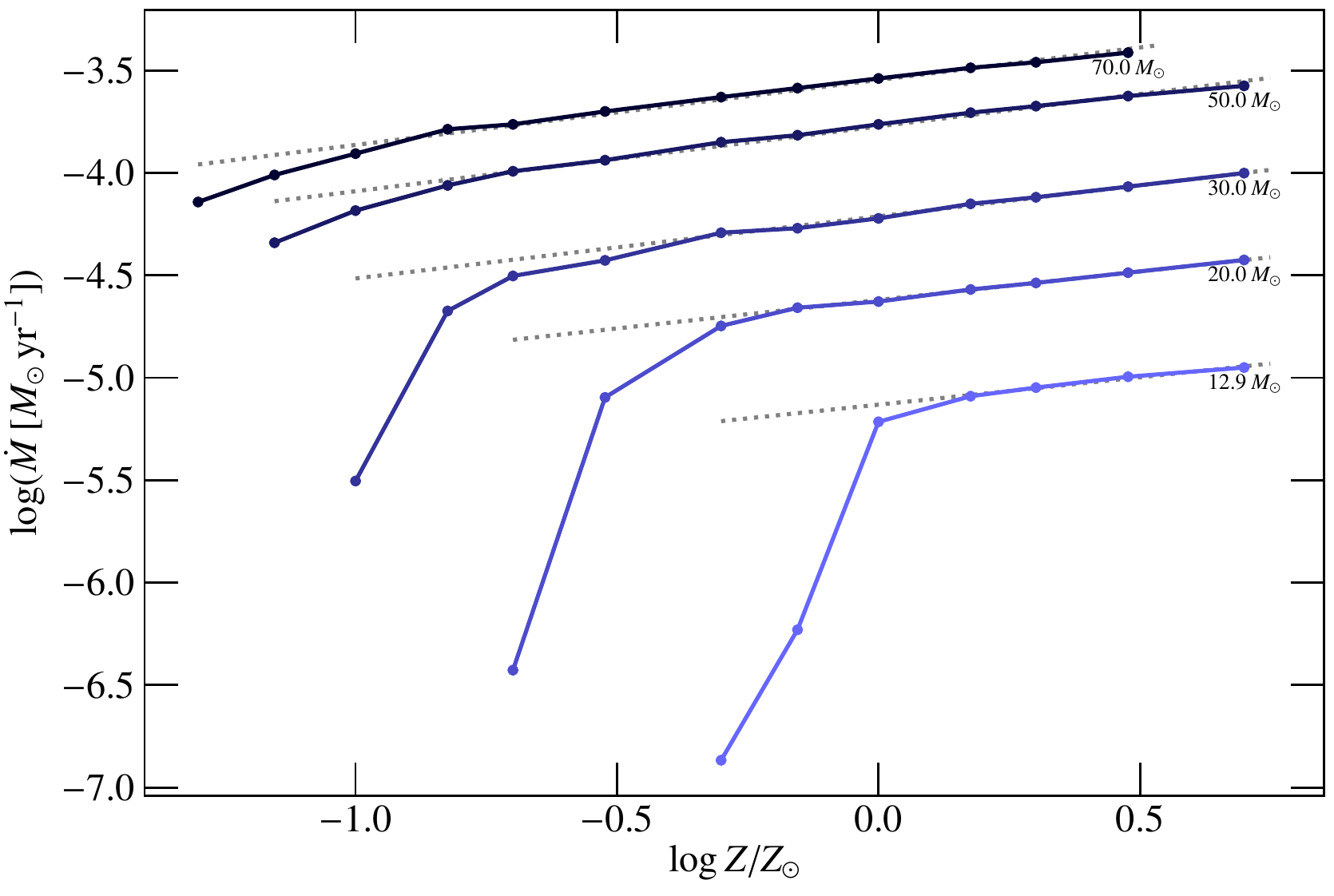}
  \caption{Mass-loss rate $\dot{M}$ as a function of $Z$ for selected He star models. The grey dashed lines denote a fit in the `pure' WR-wind regime.}
  \label{fig:mdotz}
\end{figure}

In Fig.\,\ref{fig:mdotz}, we show $\dot{M}(Z)$ for models with masses of $12.9$, $20$, $30$, $50$, and $70\,M_\odot$, covering an $L/M$-span between approximately $4.24$ and $4.64$. In particular for the lower-$L/M$ curves, we can see the same qualitative behaviour as for the WN- and WC-prototype models in Fig.\,14 in \citetalias{Sander+2020}, namely a dramatic `breakdown' of $\dot{M}$ towards lower $Z$. As also evident from the curves for different $Z$ in Sect.\,\ref{sec:resmdot}, this `breakdown' of WR-type mass loss with lower $Z$ depends on the mass itself and does not happen at the same $Z$ for all curves. This immediately implies that the $Z$-dependency in a mass-loss recipe for WR stars cannot by written in the form of a simple factor, independent of other stellar parameters. In particular, the typical treatment in the form of $\dot{M} \propto Z^{\gamma}$, is clearly insufficient. 

While the overall behaviour of $\dot{M}(Z)$ is complex and the data points are too sparse to properly check the `shape', we can once again identify the `breakdown' being determined by the location of the single scattering limit. Interestingly, in the regime of multiple scattering and high mass-loss, $\dot{M}(Z)$ can be nicely reproduced by a power law with a -- surprisingly -- shallow slope of $\gamma \approx 0.3$\footnote{The actual fits yield values from $0.27(\pm 0.02)$ for $12.9\,M_\odot$ to $0.32(\pm 0.01)$ for $70\,M_\odot$, hinting at a possible slight increase with $L/M$.}. This is roughly on the order of the flattening found by \citet{VdK2005} for their $\dot{M}_\text{WN}$-curves at $Z/Z_\odot > 3$. However, we find this shallow slope in a much broader $Z$-regime, which was so far never obtained in empirical or theoretical WN recipes. For WC stars, \citet{Tramper+2016} obtained a $Z_\text{Fe}$-exponent of $0.25$. While our models do not resemble the CNO abundances of WC stars, the other abundances, in particular for the Fe group, are comparable. \citetalias{Sander+2020} demonstrated that WC stars have a different absolute $\dot{M}$ for a given $L$ and $M_\ast$ compared to WN stars, but their mass-loss rates follow a similar trend. We would therefore expect a similarly shallow $\dot{M}(Z)$-slope for WCs. Given these considerations, the large difference in the $Z_\text{Fe}$-exponents ($1.3$ for WNs, $0.25$ for WCs) obtained by \citet{Tramper+2016} is surprising. Their low slope for the WCs could just be coincidence, but WC stars might in fact be a better tracer of the pure WR-wind regime. WC stars are generally believed to have evolved from WN stars, further peeling away some of their outer material until the products of He burning are visible on the surface. Given that He stars are already quite compact, the stripping from WN to WC is commonly attributed to stellar winds and occurs naturally in stellar evolution models with higher WN mass-loss rates. Thus, one can argue that WC stars likely stem from those WN stars with stronger $\dot{M}$, which are in turn more likely to be in the pure WR-wind regime. Therefore, even the group of observed H-free WN stars might be more heterogeneous in terms of their internal wind regimes than the group of observed WC stars. 

The flat power law with $\gamma \approx 0.3$ also provides an important insight into the occurrence of WR stars in different galaxies. Once the regime of WR-type mass loss is reached, there is not an enormous difference in $\dot{M}$ for different metallicities, essentially just a factor of $2$ for an order of magnitude in metallicity. However, the onset barrier for this regime is highly $Z$-dependent. Thus, at lower $Z$ such as in the Magellanic Clouds, He stars of lower masses will no longer reach this regime, while the more massive ones do and appear almost like their Milky Way counterparts. This phenomenon, which we explain only qualitatively here, has recently been discussed from a more empirical perspective in \citet{Shenar+2020}. To get a proper grip on it, we will have to study the spectral imprint, in particular in the transition regime, where our flat $\gamma$ is no longer valid, but stars might still have emission-line spectra.

\begin{figure}
  \includegraphics[width=\columnwidth]{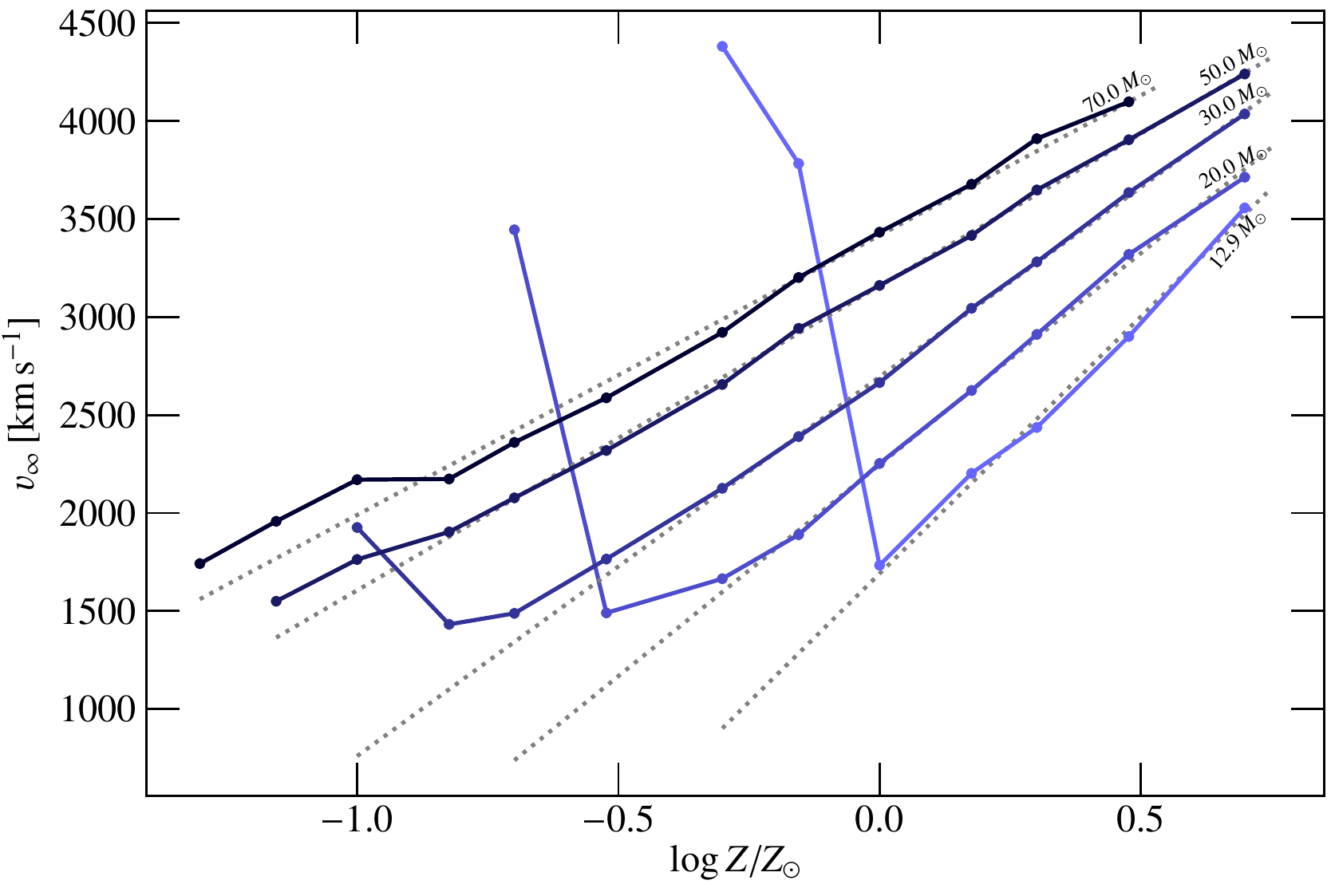}
  \caption{Terminal wind velocity $\varv_\infty$ as a function of $Z$ for selected models.}
  \label{fig:vinfz}
\end{figure}

\begin{figure}
  \includegraphics[width=\columnwidth]{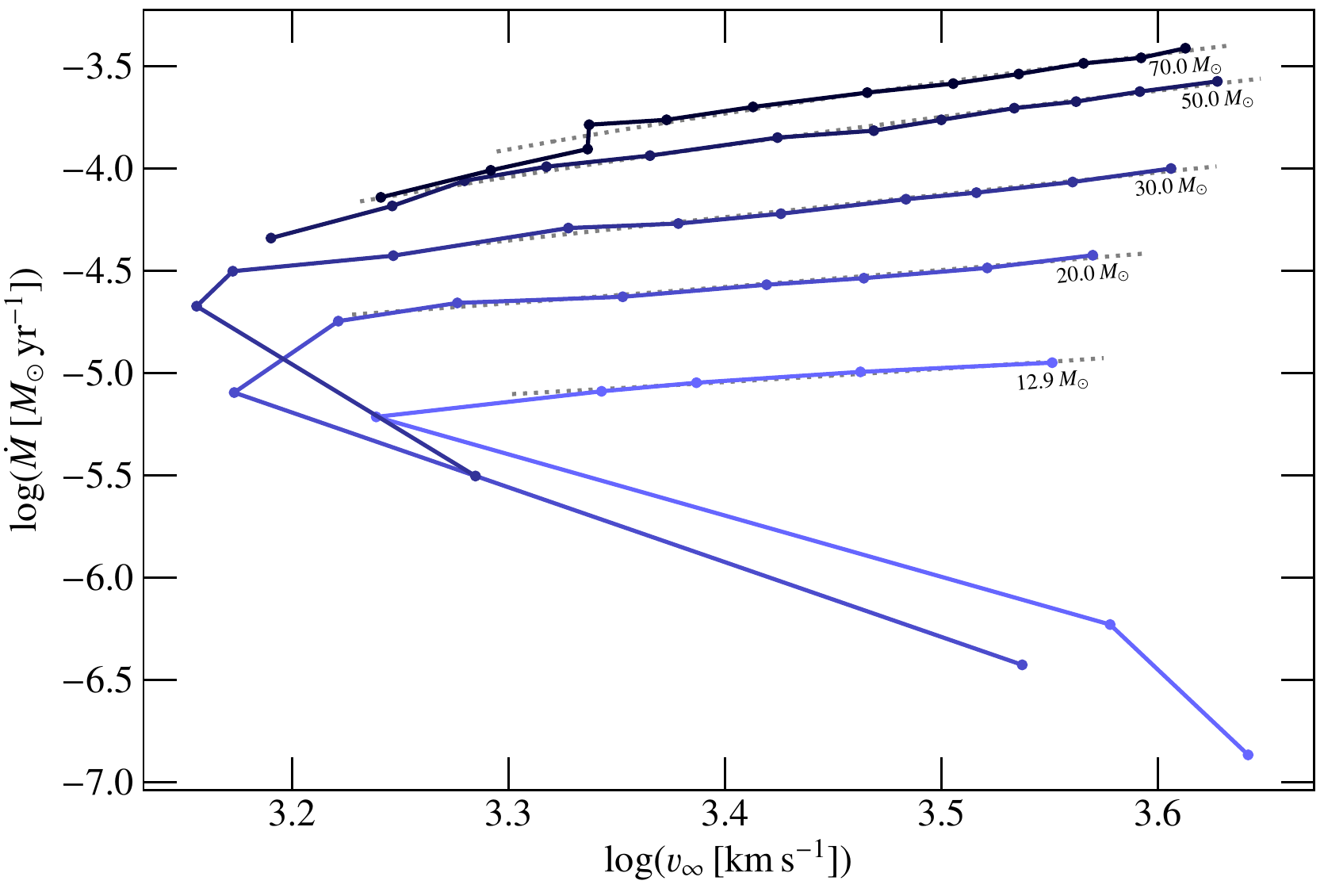}
  \caption{Mass-loss rate $\dot{M}$ versus terminal velocity $\varv_\infty$ for selected He star models. Solid lines connect models with same $M$, but different $Z$, while grey dotted lines denote a linear fit of $\dot{M}(\varv_\infty)$ in the pure WR-wind regime.}
  \label{fig:mdotvinfz}
\end{figure}

The behaviour of $\varv_\infty(Z)$ is depicted in Fig.\,\ref{fig:vinfz}. Comparable to the results for $\varv_\infty(L)$ in Sect.\,\ref{sec:vinf}, we find $\varv_\infty \propto \log Z$ for each set of $L/M$ in the high mass-loss regime. This could be seen as a surprise, as $\dot{M}_\mathrm{t} = \mathrm{const.}$ for $(L/M) = \mathrm{const.}$ at different $Z$ implies $\dot{M} \propto \varv_\infty$ and thus one might assume that $\varv_\infty(Z) \propto \dot{M}(Z) \propto Z^{\gamma}$. However, $\dot{M} \propto \varv_\infty$ is fulfilled, as we illustrate in Fig.\,\ref{fig:mdotvinfz}, but does not imply $\dot{M}(Z) \propto \varv_\infty(Z)$. Once again, Figs.\,\ref{fig:vinfz} and \ref{fig:mdotvinfz} underline the fundamental difference between the optically thick and thin regime. Moreover, Fig.\,\ref{fig:mdotvinfz} underlines that our finding of $\log \dot{M} \propto \log \varv_\infty$, which we also obtained for the mass (or luminosity) domain, seems to be an inherent feature of WR-type mass loss as such.

\begin{figure}
  \includegraphics[width=\columnwidth]{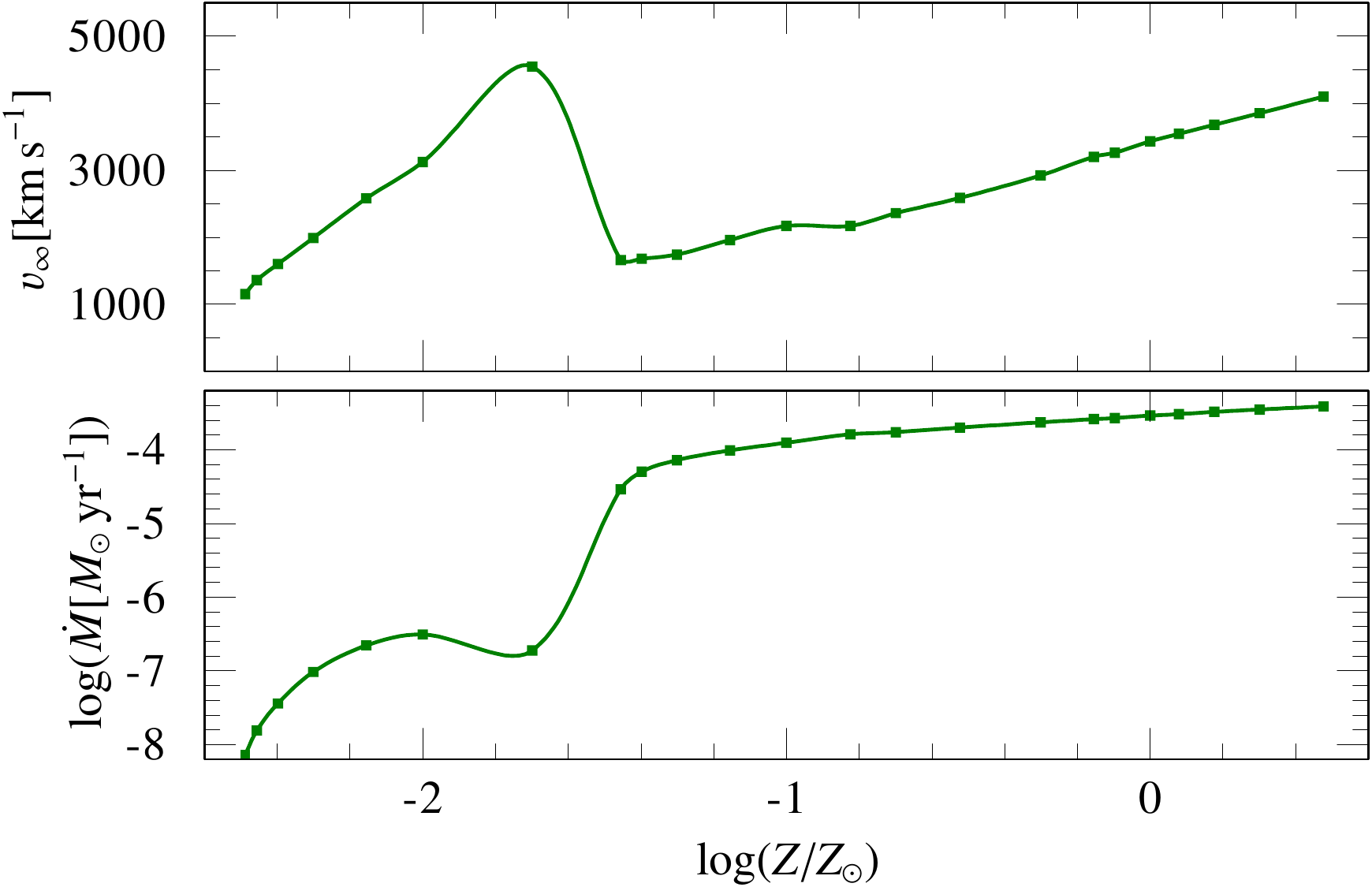}
  \caption{Terminal wind velocity $\varv_\infty$ (upper panel) and mass-loss rate $\dot{M}$ (lower panel) as a function of $Z$ for a set of models with $70\,M_\odot$.}
  \label{fig:m70ztrends}
\end{figure}

The behaviour we see in Figs.\,\ref{fig:mdotz} and \ref{fig:vinfz} cannot simply be extrapolated to very low $Z$ without further considerations. While the increase of $\varv_\infty$ at lower $Z$ in Fig.\,\ref{fig:vinfz} is also due the transition to optically thin winds and the corresponding switch to higher Fe ions as drivers of the outer wind, similar to the $L$-direction (cf.\ Fig.\,\ref{fig:vinftrends}), any reduction in $Z$ (and thus Fe) also means a removal of important wind-driving opacities. Therefore, $\varv_\infty$ is expected to decrease again at even lower $Z$. To verify this assumption, we calculate a series of further low-$Z$ models for $70\,M_\odot$. While $70\,M_\odot$ He stars are probably not the most prototypical example, their stronger winds compared to e.g.\ a $20\,M_\odot$ model makes them numerically more favourable.

The upper panel in Fig.\,\ref{fig:m70ztrends} depicts the $\varv_\infty$-trend for the $70\,M_\odot$ models down to $\log Z/Z_\odot = -2.4$ and indeed confirms our assumption that $\varv_\infty$ must eventually decrease again when transitioning to lower and lower metallicities. These curves will likely look a bit different for other mass ranges, so we refrain from deducing any kind of recipe. Nonetheless, Fig.\,\ref{fig:m70ztrends} illustrates that along the $Z$-dimension, a non-monotonic behaviour can generally be expected. This also applies to $\dot{M}(Z)$, which is shown in the lower panel of Fig.\,\ref{fig:m70ztrends}. The steep drop in $\dot{M}$ does not continue forever, something we did not experience along the $L$-dimension -- at least in our considered parameter range. Moreover, $\dot{M}$ does not decrease monotonically with $Z$, but shows kind of a `rebound' before eventually decreasing again with lower $Z$. We attribute this behaviour to our choice of a fixed clumping recipe with a dependence on a characteristic velocity $\varv_\text{cl}$. This dependence introduces a clumping onset that will (slightly) shift with $Z$. While for a monotonic behaviour of $\varv_\infty(Z)$, this shift of the clumping onset would also be monotonic, it can actually move in both directions if $\varv_\infty$ has minima and maxima, thereby also causing a non-monotonic behaviour in $\dot{M}(Z)$. Of course, if clumping is connected to $Z$-dependent phenomena such as sub-surface convection \citep[e.g.][]{Cantiello+2009}, the assumption of a fixed clumping recipe would no longer be sufficient, in particular at very low metallicities. For a smooth wind, there might be no local maxima or minima in $\dot{M}(Z)$. However, any deeper investigation of clumping and its influence on the derived relations will require its own study.

\section{He stars as sources of ionizing flux}
  \label{sec:ionflux}

\begin{figure}
  \includegraphics[width=\columnwidth]{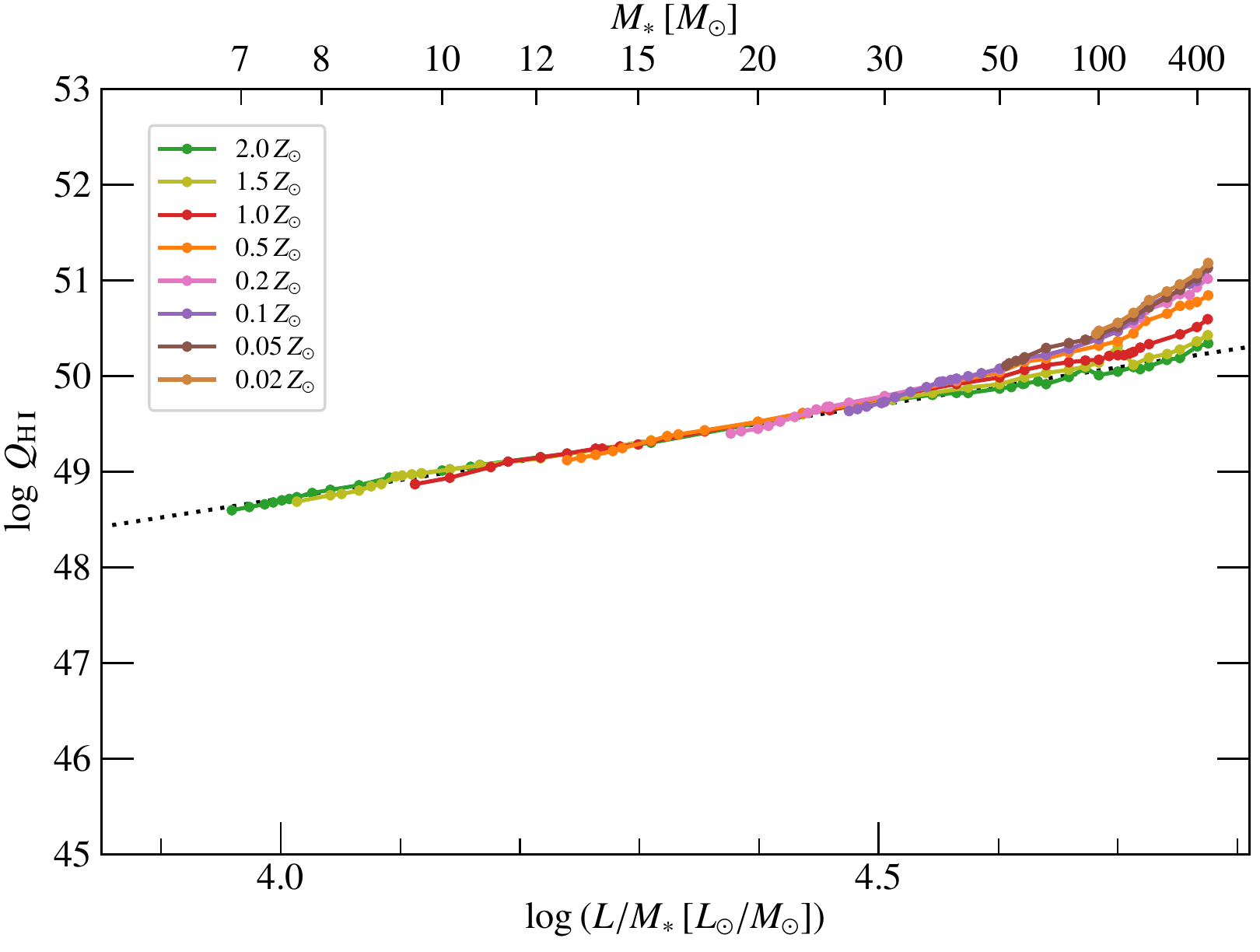}
  \caption{Number of hydrogen ionizing photons per second $Q_{\ion{H}{i}}$ as a function of $L/M$ for different metallicities $Z$. The black dotted line denotes a linear fit for the $2\,Z_\odot$-dataset.}
  \label{fig:ionfluxh1}
\end{figure}

\begin{figure}
  \includegraphics[width=\columnwidth]{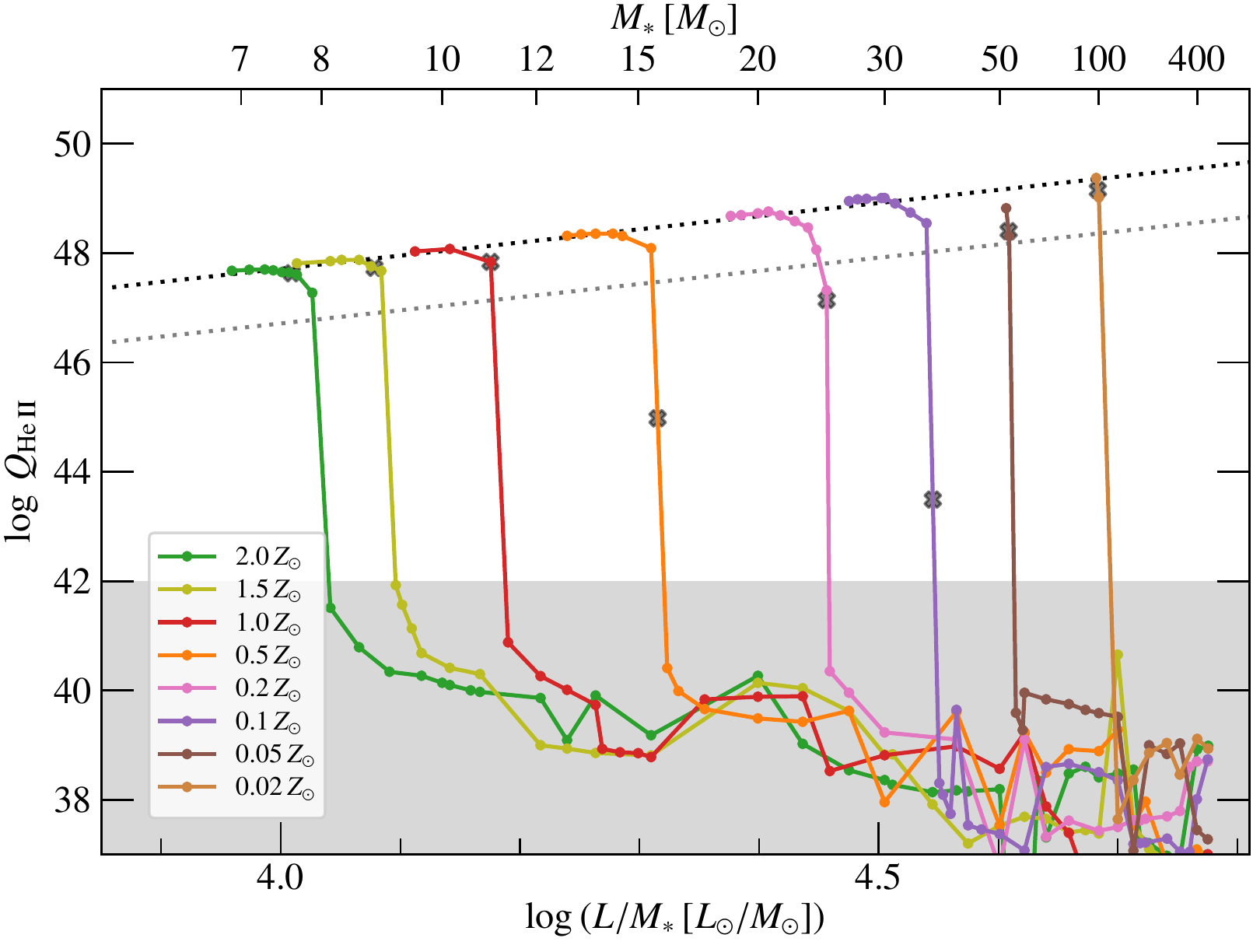}
  \caption{Number of helium ionizing photons per second $Q_{\ion{He}{ii}}$ as a function of $L/M$ for different metallicities $Z$. The black dotted line denotes a linear fit for the thin-wind limit. The grey dotted line shows the same relation shifted by one magnitude, providing an estimate of the minimum ionizing flux below the cut-off limits. The filled crosses denote the crossing of the single scattering limit ($\eta = 1$). Absolute numbers in the low-flux region (grey-shaded area) are more uncertain as they are susceptible to the numerical treatment of boundary conditions in the model atmospheres.}
  \label{fig:ionfluxhe2}
\end{figure}

With their high temperatures, He stars are a major source of ionizing fluxes. As illustrated in Fig.\,\ref{fig:ionfluxh1}, all the models in our study produce a hydrogen-ionizing flux on the order of $10^{49}$ Lyman continuum (LyC) photons per second, regardless of metallicity. Only for very massive He stars with $M > 30\,M_\odot$, the different amount of metals starts to have an effect.
Not accounting for the deviations at very high masses at lower metallicity, this uniform behaviour can be well described by the linear relation
\begin{equation}
  \label{eq:qhi}
  \log Q_{\ion{H}{i}} = 1.96 (\pm 0.02) \cdot \log (L/M\,[L_\odot/M_\odot]) + 40.88 (\pm 0.08)
\end{equation}
which is derived from the data for $Z = 2\,Z_\odot$.

For ionizing \ion{He}{ii}, the situation is quite different. As shown in Fig.\,\ref{fig:ionfluxhe2}, a considerable flux of \ion{He}{ii} ionizing photons can only leave the star if the wind is sufficiently thin, corresponding to lower mass-loss rates. As we investigated for $\dot{M}$, this transition to thin winds is highly dependent on metallicity. When examining the models in the transitions region, we find that the evanescence of \ion{He}{ii} ionizing flux coincides with $\eta$ crossing unity, with a slight exception for the highest $Z$ in our sample, where the ionizing flux does not vanish until $\eta \approx 1.2$. Below the single scattering limit, the ionizing fluxes gradually approach a relation described by 
\begin{equation}
  \label{eq:qheii}
  \log Q_{\ion{He}{ii}}^\text{max} = 2.41 (\pm 0.08) \cdot \log \left(\frac{L}{M} \left[\frac{L_\odot}{M_\odot}\right]\right) + 38.09 (\pm 0.33)\text{,}
\end{equation}
which provides a formula for the maximum \ion{He}{ii} ionizing flux (in photons per second) for a given $L/M$. A grey dotted line in Fig.\,\ref{fig:ionfluxhe2} denotes the same relation shifted downwards by one order of magnitude, providing a lower estimate of $\log Q_{\ion{He}{ii}}$ in the regime where there is already a substantial ionizing flux, but the wind is not transparent enough to have reached $Q_{\ion{He}{ii}}^\text{max}$. The essential breakdown of $\log Q_{\ion{He}{ii}}$ by orders of magnitude can be described by
\begin{equation}
  \label{eq:qheiibreakdown}
	\log \left.\left(\frac{L}{M} \left[\frac{L_\odot}{M_\odot}\right]\right)\right|_\text{cutoff} = -0.31 (\pm 0.04) \cdot \frac{Z}{Z_\odot} + 4.58 (\pm 0.04)
\end{equation}
with $\log Q_{\ion{He}{ii}} < 42$ for $L/M > \left.L/M\right|_\text{cutoff}$. The results for both $Q_{\ion{H}{i}}$ and $Q_{\ion{He}{ii}}$ are of course subject to the underlying assumption of a fixed $T_\ast$ in our models. As discussed above, stellar structure models for the He ZAMS \citepalias{Grassitelli+2018} expect considerably lower values of $T_\ast$ for masses lower than approximately $10\,M_\odot$. In these cases, the true ionizing fluxes will be lower than what is obtained by Eqs.\,(\ref{eq:qhi}) and (\ref{eq:qheii}). Given that also $\dot{M}$ might be higher at lower $T_\ast$, any further quantification for $M < 10\,M_\odot$ will require a separate calculation of HD models tailored to lower-mass stripped stars.

\section{Summary and conclusions}
  \label{sec:conclusions}
	
In this work, we present a set of next-generation stellar atmosphere models for massive He stars with luminosities, masses, and abundances representing the high-mass part of the He ZAMS. In a pioneering study, we cover eight different metallicities between $2.0\,Z_\odot$ and $0.02\,Z_\odot$ with He star masses up to $500\,M_\odot$. Due to the local hydrodynamical consistency in our models, we can obtain mass-loss rates and wind stratifications from a given set of stellar parameters, allowing us to derive an $\dot{M}$-recipe for massive He stars with $M \geq 10 M_\odot$ from first principles. We obtain two distinct regimes representing optically thick and thin winds with a complex transition regime around the onset of multiple scattering. While there is no `kink' in the derived $\dot{M}$-recipe, an exponential breakdown of WR-type mass loss sets in when approaching the transition regime. The transition goes along with considerable shifts in the lead wind-driving ions towards higher ionization stages and a minimum in the derived $\varv_\infty$. In the regime of optically thick winds, $\varv_\infty$ increases with $\dot{M}$, while the opposite is obtained in the thin-wind regime.

All of our models investigate a regime where the winds are launched at the so-called `hot Fe bump', meaning that radiation pressure caused by Fe M-shell opacities are the decisive contribution on top of electron scattering. In the outer part, various line opacities play a role, though Fe remains the leading driver at all considered metallicities. In the transition regime, the He continuum contribution becomes important throughout the wind with relative importance increasing outwards. This additional opacity is probably vital to explain the smooth regime transitions obtained in our study.

In the regime of `pure' WR-type mass loss, the wind density measure $\dot{M}/\varv_\infty$ is conserved, independent of metallicity $Z$. In particular, WR-type winds follow a $Z$-independent linear relation between $\log \dot{M}_\mathrm{t}$ and $\log L/M$ with a potential turnover at several hundred solar masses. Furthermore, a linear relation without a turnover is obtained for $\log(\dot{M}/\varv_\infty)$ as a function of $\log[-\log(1-\Gamma_\text{e})]$. Numerical uncertainty in $\varv_\infty$ and our clumping recipe with a $\varv_\text{cl}$-term add some scatter to these `ideal' relations in our data. There is a `break away' from these relation towards lower masses with characteristic values depending on the metallicity $Z$. The `thin wind'-regime might also adhere to a ($Z$-dependent) power-law in $\dot{M}_\mathrm{t}(L/M)$, but with a much steeper slope than in the WR regime. For very high masses, we can infer that the wind efficiency $\eta$ eventually reaches a maximum, which increases with metallicity $Z$. The absolute numbers in the regime of optically thin winds have to be taken with care. Presently, no He stars with thin winds and $M > 10\,M_\odot$ are known. While our models are necessary to study the breakdown of WR-type mass loss, their results will only be applicable if the objects have zero hydrogen and their winds are driven by the hot Fe bump.

We obtain two $\dot{M}$-recipes for massive He stars, one describing $\dot{M}$ as a function of $L$, and another one describing $\dot{M}$ as a function of $\Gamma_\text{e} \propto L/M$. For most purposes, we recommend to use the $\dot{M}(\Gamma_\text{e})$-recipe -- Eq.\,(\ref{eq:breakdownrecipe}) with coefficients denoted in Eqs.\,(\ref{eq:aparam}) to (\ref{eq:cbdparam}) -- as it not just provides a better representation of the model data, but also reflects the nature of WR-type mass loss as an $L/M$-dependent quantity more accurately. The new $\dot{M}(\Gamma_\text{e})$-recipe covers the full complexity unveiled by our model sequences. It consists of a `linear' and a `breakdown' term with the latter being parametrized by the ($Z$-dependent) onset of multiple scattering. 
Both $\dot{M}$-recipes are available online\footnote{The script is available at \texttt{https://armagh.space/asander}} via a \textsc{Python} script, which issues a warning if an output value is in the more uncertain breakdown regime.

A direct investigation of $\dot{M}(Z)$ reveals that for a fixed set of $L$ and $M$, the mass loss in the pure WR-wind regime follows a power-law $\dot{M} \propto Z^\gamma$ with a shallow slope of $\gamma \approx 0.3$. This is much lower than commonly assumed and qualitatively explains the similarity of WR stars in galaxies with different $Z$. Closer to the breakdown regime, which is $L/M$-dependent, the $\dot{M}(Z)$-slope becomes more and more steep, further explaining why at lower metallicities WR stars are only seen at higher luminosities. 

When inspecting the terminal velocities, we discover a lower limit of $\varv_\infty \approx 1500\,\mathrm{km\,s}^{-1}$. This limit approximately coincides with the transition from optically thick to thin winds. While lower-mass He stars might have completely different winds \citepalias[cf.][]{Grassitelli+2018} not covered in this work, the high $\varv_\infty$-values in massive H-free WR stars could be taken as an indirect proof of their winds being launched by the hot Fe bump.
Given the uncertainty of the clumping factor ($D_{\infty}$), our lower $\varv_\infty$-limit is in qualitative agreement with observations of He stars with optically thick winds, but raises questions about the driving onset of late-type WC, in particular WC9, winds.

Our terminal velocities show a non-monotonic trend in $Z$, with an increase of $\varv_\infty$ when transitioning to thin winds, but then again a decrease at even lower $Z$. While our set of models is not sufficient to quantify this behaviour, we note that this non-monotonic behaviour is not accounted for in any current recipe, and might be important in galactic contexts, e.g.\ for the dynamical evolution of galaxies.

While all He stars are major sources of LyC photons (i.e.\ hydrogen ionizing flux), mass loss and \ion{He}{ii} ionizing flux are complementary quantities, i.e.\ both cannot be large at the same time. At low $Z$, only He stars extremely close to $\Gamma_\text{e} = 1$ will have a considerable mass loss, while those with lower $\Gamma_\text{e}$ instead show transparent winds and intense UV radiation \citep[TWUIN, cf.][]{Szecsi+2015}, contributing significant amounts of \ion{He}{ii} ionizing flux. The onset of WR-type mass loss and the corresponding breakdown in \ion{He}{ii} ionizing flux increases with decreasing metallicity and approximately coincides with the onset of multiple scattering in the wind. In a low-$Z$ dwarf such as I\ Zw\ 18 with $Z \approx 0.02\,Z_\odot$, this onset would only happen at a He ZAMS mass of $\approx 100\,M_\odot$.

Together with our findings for $\varv_\infty$, we can conclude that extragalactic narrow \ion{He}{ii} emission (e.g.\ at 1640\,\AA) observed in low-$Z$ galaxies most likely cannot be related to classical WR stars (assuming their winds are driven by the hot Fe bump). Instead, such narrow \ion{He}{ii} emission is either an indicator of H-burning very massive stars \citep[see][]{GV2015} or has a nebular origin, which could very well stem from stripped He stars with transparent winds.

All results in this work are based on models of H-free stars at the onset of central He burning with $T_\ast = 141\,$kK, i.e.\ with winds driven by the hot Fe bump. In contrast, the observed WR population is a heterogeneous mixture of objects in different evolution stages. WN stars come in two flavours, with and without hydrogen. While those without cannot be in the stage of core-H burning, those with hydrogen can in principle be in various stages. WC and WO stars are further evolved and thus differ in chemical composition and $L/M$-ratio. The best observational counterparts to our models presented in this work are therefore H-free WN stars. As some of the early He burning will take place in the supergiant phase, even the H-free WN population is not perfectly homogeneous, but the structural differences due to the progress of core-He burning are expected to be small \citep[e.g.][]{L1989}. Nonetheless, we expect some scatter when comparing particular results of this work to present observations of WR stars, mainly due to abundance uncertainties and our choice of a fixed $T_\ast$. Moreover, for the highest masses, the close proximity to $\Gamma_\text{e} = 1$ may cause convection \citep{Joss+1973} which is not accounted for in our models. 

Overall, our study yields fundamental insights into the nature of He star and WR-type mass loss and reveals principal trends, such as the complex $L/M$- and $Z$-dependent breakdown of WR-type mass loss, or our discovery of $\log \dot{M} \propto \log \varv_\infty$ as an inherent relation in the pure WR-regime. The extension of our findings to other regimes as well as absolute calibrations will require a considerable amount of follow-up work and a further development of stellar atmosphere modelling. 
The intriguing results obtained from first principles in our study demonstrate the essential role of next-generation atmosphere models to make progress in our fundamental understanding of massive stars and their role as cosmic drivers, e.g.\ by bridging the gap between structure and atmosphere calculations, or providing accurate ingredients for population synthesis and galaxy evolution.

\section*{Acknowledgements}

The authors would like to thank the referee, L. Grassitelli, for helpful comments and suggestions.
The authors further acknowledge fruitful discussions with T.\ Shenar, W.-R.\ Hamann, H.\ Todt, E.\ R.\ Higgins, R.\ Hirschi, S.\ E.\ Woosley, and J.\ M.\ Bestenlehner. A number of figures in this work were created with \textsc{WRplot}, developed by W.-R. Hamann. 
A.A.C.S. is supported by STFC funding under grant number ST/R000565/1. 

\section*{Data availability}
 
The data underlying this article will be shared on reasonable request to the corresponding author.



\bibliographystyle{mnras}
\bibliography{literatur}



\appendix

\section{Relating efficiency and flux-weighted optical depth of WR winds}
  \label{asec:etatau}

\begin{figure}
  \includegraphics[width=\columnwidth]{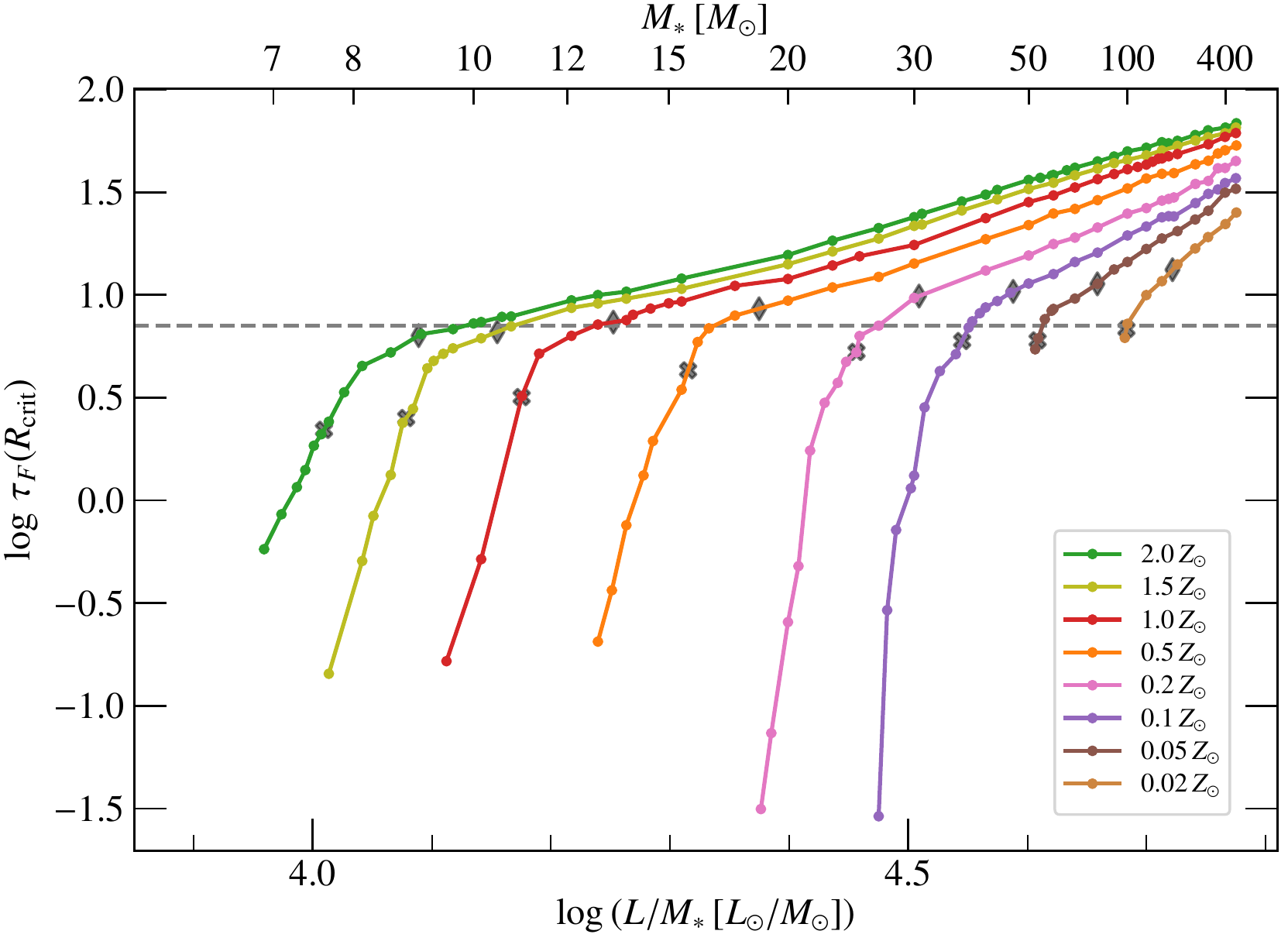}
  \caption{Wind optical depth $\tau_F(R_\text{crit})$ as a function of $L/M$-ratio.}
  \label{fig:taufcldm}
\end{figure}

An important quantity to gain insights into optically thick winds and their driving is the flux-weighted optical depth
\begin{equation}
  \label{eq:tauf}
	\tau_{F}(r) := \int\limits^\infty_{r} \varkappa_{F}(r^\prime)\,\rho(r^\prime)\,\mathrm{d}r^\prime
\end{equation}
which is discussed in more detail in \citetalias{Sander+2020}. Evaluating
$\tau_F$ at the critical point $R_\text{crit}$ provides a measure of the integrated wind density, which we depict in Fig.\,\ref{fig:taufcldm}.
In line with \citetalias{Sander+2020}, the flattening of the curves in Fig.\,\ref{fig:mdotldm} does not coincide with $\tau_F(R_\text{crit}) = 1$, but requires the critical point to be further inwards at $\tau_F(R_\text{crit}) \approx 7$ (dashed line in Fig.\,\ref{fig:mdotldm}).
The bulk of WR-type models in our study is in a regime of high optical depth, placing $R_\text{crit}$ in a location with essentially LTE conditions, which we quantify in more detail in appendix \ref{asec:ltedep}.

\begin{figure}
  \includegraphics[width=\columnwidth]{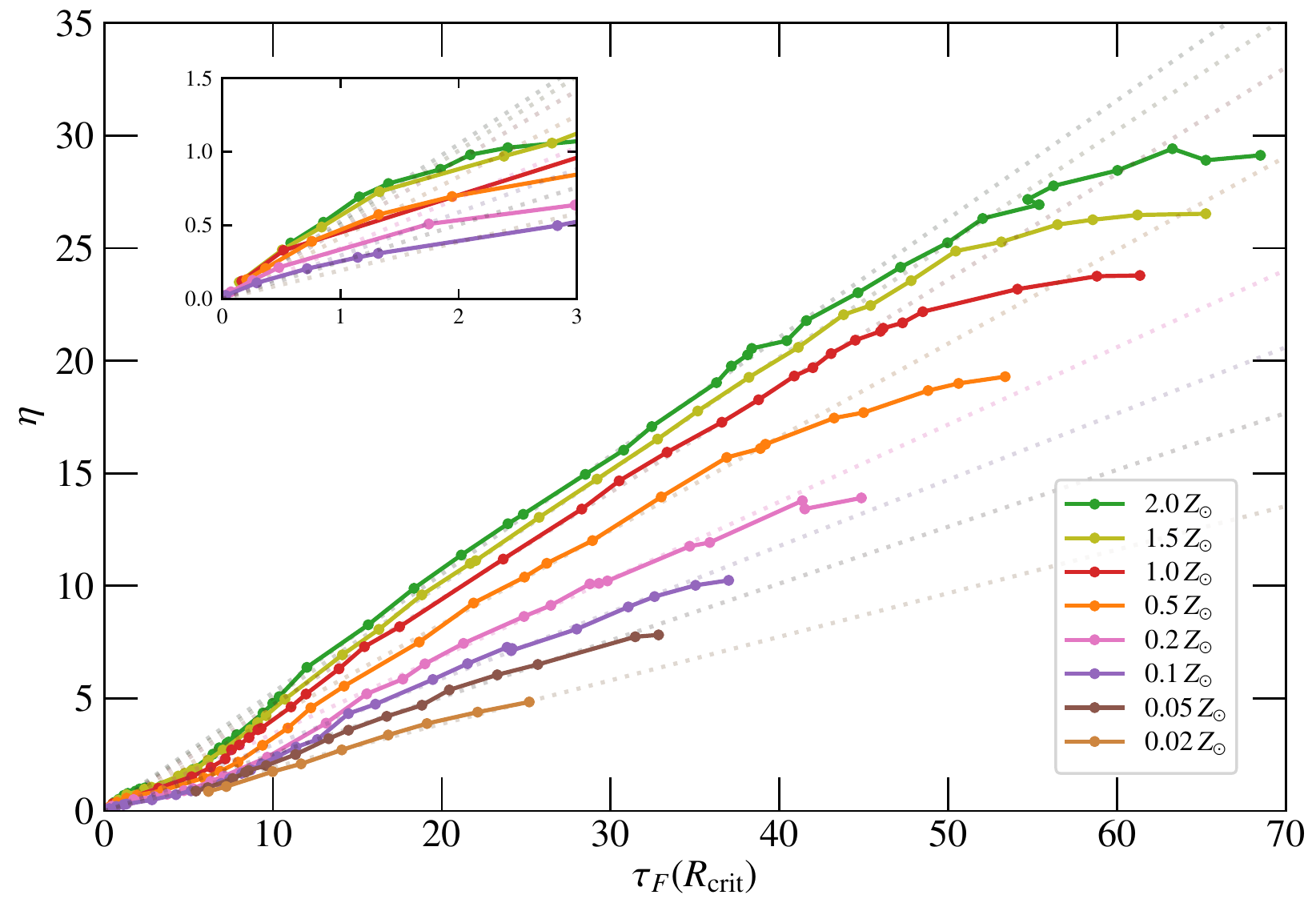}
  \caption{Wind efficiency $\eta$ as a function of wind optical depth $\tau_F(R_\text{crit})$: The grey dashed lines show linear fits for each metallicity. The inlet depicts an enlargement of the lower left region with optically thin winds.}
  \label{fig:etataufc}
\end{figure}

\begin{figure}
  \includegraphics[width=\columnwidth]{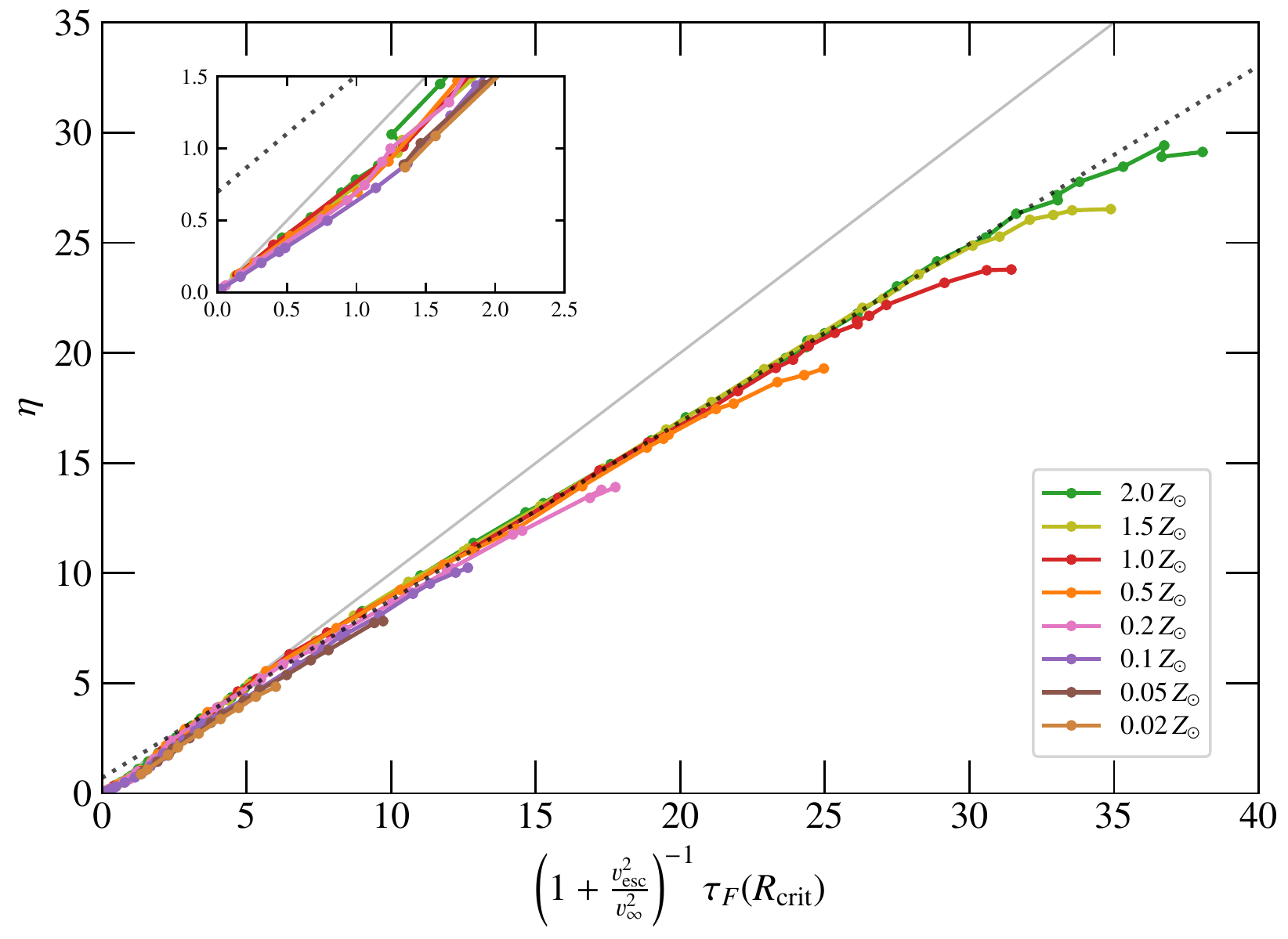}
  \caption{Wind efficiency $\eta$ as a function of the wind optical depth $\tau_F(R_\text{crit})$ multiplied with a factor suggested by \citet{Graefener+2017}. The solid grey line shows the suggested 1:1-relation while the dashed grey line depicts the best fit through the linear part of the curves.}
  \label{fig:etataufcmod}
\end{figure}

\citet{VG2012} connect the wind efficiency number linearly with the (wind) optical depth, i.e.
\begin{equation}
  \label{eq:etataufc}
  \eta = f \cdot \tau_F(R_\text{crit})\text{,}
\end{equation}
and postulate $f$, which is expected to be below unity, to only depend on the ratio of $\varv_\infty$ and the escape velocity $\varv_\text{esc} := \sqrt{2 G M_\ast / R_\text{crit}}$.
The terminal velocity $\varv_\infty$ depends on the metallicity $Z$, so we expect $f$ to be a function of $Z$ as well. In Fig.\,\ref{fig:etataufc}, we plot $\eta$ as a function of the total wind optical depth, i.e.\ $\tau_F(R_\text{crit})$. Beside the immanent $Z$-dependence, we also notice deviations from a linear relation for low and high optical depths. A fit of the linear part yields coefficients $f$ (with uncertainties $< 0.01$) which are monotonically increasing with $Z$.
A further analysis of the coefficients yields that $f$ along the $Z$-dimension can be sufficiently described by a linear fit in $\log Z$. The resulting formula for the factor $f$ in Eq.\,(\ref{eq:etataufc}) thus reads
\begin{equation}
  \label{eq:fzfit}
  f(Z) = 0.168 (\pm 0.004) \cdot \log\frac{Z}{Z_\odot} + 0.470 (\pm  0.003)\text{.}
\end{equation}

As the inlet in Fig.\,\ref{fig:etataufc} highlights, the wind efficiency $\eta$ remains considerably below these linear relations as long as the winds are not sufficiently optically thick, but comes back to it for very low wind optical depths. The reason for this behaviour is rooted in the transition of the wind driving regime. As described for example in \citet{VG2012} and \citet{Graefener+2017}, the linearity of Eq.\,(\ref{eq:etataufc}) roots in the assumption that $\Gamma_\text{rad}$ in the wind domain can be sufficiently described by a constant mean value \citep[denoted $\Gamma_\text{w}$ in][]{Graefener+2012}. An inspection of the models yields that this assumption apparently only holds if the leading driving ion in the outer wind -- which is also the largest contributor to the integrated wind opacity -- does not change. The linear parts for $\tau_F(R_\text{crit}) > 10$ in our model sets can be identified with \ion{Fe}{v} being the lead driving ion in the outer wind. When transitioning to less dense winds, higher Fe ions take over and $\eta$ drops below the linear relation until a thin-wind regime with \ion{Fe}{ix} as a stable leading wind driver is reached.

For models with high wind optical depth, $\eta$ seems to saturate at a value depending on $Z$. In line of the deviation seen for low optical depth, we can once again associate this deviation with a switch in the lead outer wind driver, now to \ion{Fe}{iv} and -- for higher $Z$ -- \ion{Fe}{iii}. Thus, one could speculate whether we approach an actual maximum, i.e.\ a true saturation of $\eta$, or the theoretical possibility of a `super-WR regime' where $\eta$ could return to the linear relation and thus even stronger outflows would be possible. In appendix Sect.\,\ref{asec:etamax}, we also discuss the issue of a maximum $\eta$ implied by our $\dot{M}(L)$-recipe derived in Sect.\,\ref{sec:mdotrecipe}. However, given the overall shape of the $D_\text{mom}(L)$-curves and the fact that both \ion{Fe}{iv} and \ion{Fe}{iii} should in principle be able to provide further stable wind driving regimes, we conclude that $\eta$ indeed approaches a maximum and saturates at a certain value for a given metallicity, no matter how much $R_\text{crit}$ moves further inwards. Still, a saturation of $\eta$ does not immediately imply a saturation of $\dot{M}$, since we also move closer and closer to the Eddington limit ($\Gamma_\text{e} \rightarrow 1$) with higher $L$.

Following up on the conclusion by \citet{VG2012} that $f$ should only depend on ratio of terminal velocity to escape velocity, \citet{Graefener+2017} proposed a completely analytic approximation of the constant $f$, suggesting
\begin{equation}
  \label{eq:goetztauscale}
  \eta \approx \frac{\tau_\text{s}}{1 + \frac{\varv^2_\text{esc}}{\varv^2_\infty}}\text{.}
\end{equation}
Neglecting the tiny difference due to mircro-turbulence, we identify $\tau_\text{s} \equiv \tau_F(R_\text{crit})$ and test their prediction in Fig.\,\ref{fig:etataufcmod}. The scaling factor indeed (almost) unifies the curves by scaling them to match each other apart from the $Z$-dependent saturation for very high $\tau_F(R_\text{crit})$. The re-scaling also considerably reduces the imprint of the `bump' towards lower $\eta$-values in the transition regime at low optical depths. The fact that modifying $\tau_F(R_\text{crit})$ according to Eq.\,(\ref{eq:goetztauscale}) does affect this transition regime, but not the saturation at high optical depths, adds another piece of evidence that the observed maximum of $\eta$ is indeed a true, $Z$-dependent saturation as concluded above. Nonetheless, a depth-dependent discrepancy remains. The total curve is non-linear, especially due to the saturation at the high and the -- albeit much less pronounced -- `bump' at the low end. The linear part can be sufficiently described independent of metallicity by  
\begin{equation}
  \label{eq:goetztaufit}
  \eta = 0.808\,(\pm 0.005) \cdot \frac{\tau_\text{s}}{1 + \frac{\varv^2_\text{esc}}{\varv^2_\infty}} + 0.699\,(\pm 0.077)\text{.}
\end{equation}
Whether this description is preferable to Eqs.\,(\ref{eq:etataufc}) and (\ref{eq:fzfit}) depends on the task at hand as the $Z$-dependence in Eq.\,(\ref{eq:goetztaufit}) is of course implicitly conserved in $\varv_\infty$, which is highly $Z$-dependent.

\section{The maximum wind efficiency and its implications for the mass-loss recipe}
  \label{asec:etamax}

\begin{figure}
  \includegraphics[width=\columnwidth]{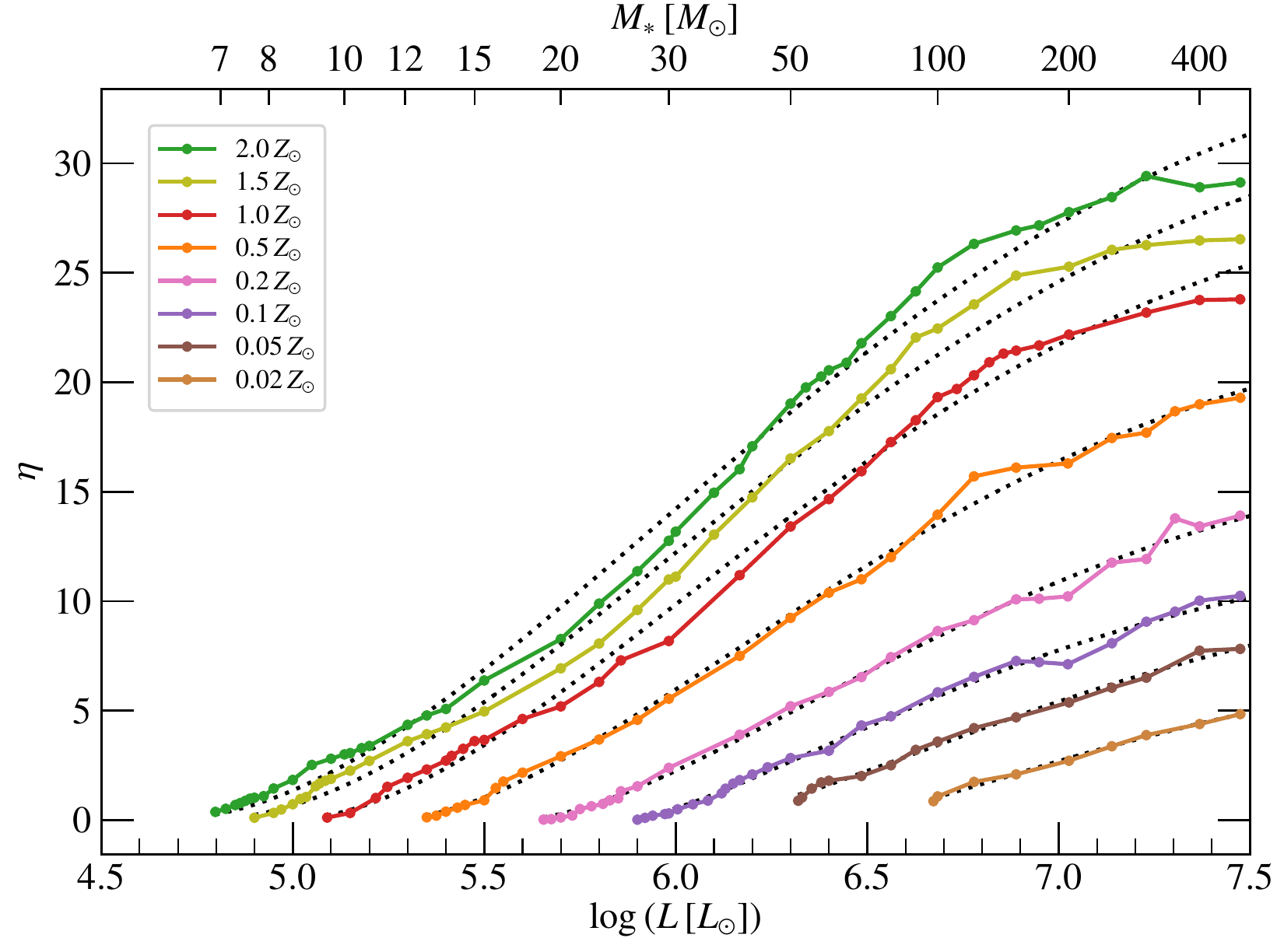}
  \caption{$\eta(L)$ from the models with fits (dashed) according to Eq.\,(\ref{eq:etalformula})}
  \label{fig:etal}
\end{figure}

One can find an analytic description for $\eta(L)$ by combining our mass-loss recipe (Eq.\,\ref{eq:mdotlrecipe})
with the finding of $\varv_\infty \propto \log L$, which is valid in the dense wind regime considered here:
\begin{equation}
  \label{eq:etaofl}
  \eta \propto \dot{M} \varv_\infty L^{-1} = \left(\log\frac{L}{L_0}\right)^{\alpha+1} L^{-1/4}
\end{equation}
The declining $L^{-1/4}$-term competes with the monotonically increasing logarithmic term, leading to a local maximum for $\alpha > 0$.
Determining $L_{\eta,\text{max}}$ is straight forward. For convenience, we can transform the decadic logarithm into a natural logarithm
and substitute all the multiplication factors into a constant $c_{\eta}$. We then obtain:
\begin{align}
  \label{eq:etalformula}
  \eta &= c_{\eta} \left(\ln\frac{L}{L_0}\right)^{\alpha+1} L^{-1/4} \\
	\label{eq:etamaxbed}
	  \frac{\partial\eta}{\partial L} &= \frac{c_{\eta}}{4} \left(\ln\frac{L}{L_0}\right)^{\alpha} L^{-5/4} \left[ 4 \left(\alpha+1\right) - \ln\frac{L}{L_0} \right]
\end{align}
The last Eq.\,(\ref{eq:etamaxbed}) is zero for $L = L_{0}$, which is a minimum for $\alpha > 0$, or if the term in brackets vanishes. The latter yields
  $L_{\eta,\text{max}} = L_0\,e^{4\left(\alpha+1\right)}$\text{.}
With values of $\alpha \approx 1$, this yields a factor of about $e^8 \approx 3000$, i.e. $L_{\eta,\text{max}}$ would be about $3.5\,$dex higher than $L_0$. This is way beyond the covered $L$-regime, so we would not expect to approach a maximum $\eta(L)$ in our study, assuming that Eq.\,(\ref{eq:etaofl}) completely describes the behaviour of $\eta(L)$. As we see from the numerical results depicted in Fig.\,\ref{fig:etal}, this is probably not the case.  
We can further use Eq.\,(\ref{eq:etalformula}) to cross-check the coefficients $\alpha$ and $L_0$, as an alternative to the direct $\dot{M}(L)$-fits in Sect.\,\ref{sec:ztrends}. (The results are shown as olive points connected by dotted lines in Fig.\,\ref{fig:mdlrfit}.) We see a systematic trend towards lower values for both parameters. While the difference for $L_0$ between the two methods is only around $0.1\,$dex, the derived $\alpha$-values from the $\eta$-fit are considerably smaller than those from the direct $\dot{M}$-fit. Assuming that the numerical result of a saturating $\eta$ is real, we would need even lower $\alpha$ values of approximately $\alpha < 0.6$ to get $\log L_{\eta,\text{max}}/L_\odot \approx 7.2...7.5$ for the high-$Z$ curves, which would be almost a factor of $3$ lower than obtained from fitting $\dot{M}(L)$. 
Thus, we conclude that while the double-logarithmic term in $\dot{M}(L)$ is okay to describe the overall behaviour of $\dot{M}$ in the WR-wind regime and its breakdown, this description does not fully represent the underlying physics.

\section{LTE-departure at the critical point}
  \label{asec:ltedep}

\begin{figure}
  \includegraphics[width=\columnwidth]{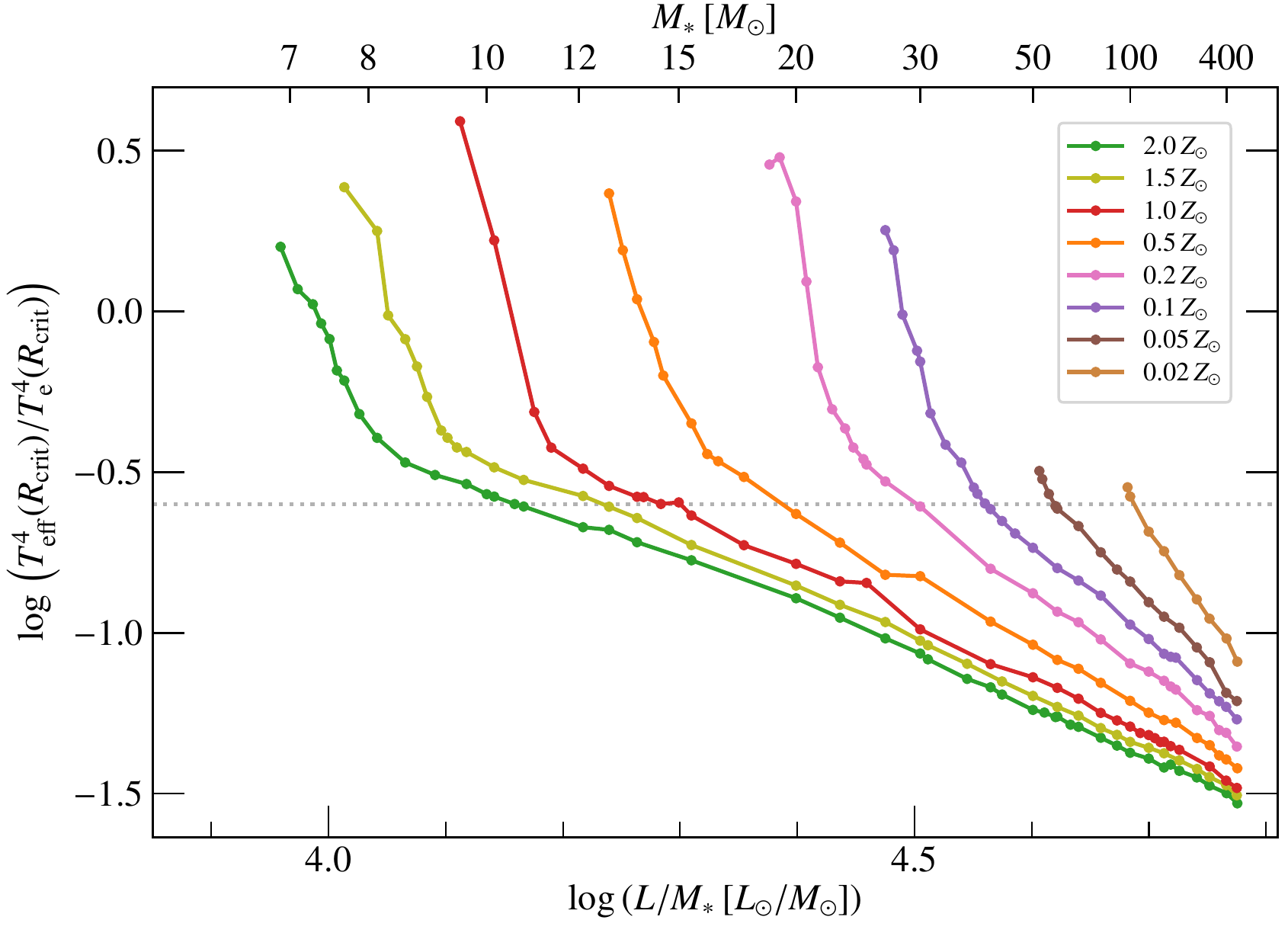}
  \caption{Anisotropy ratio (cf.~Eq.\,\ref{eq:anirat}) at the critical point as a function of $L/M$ for our HD model sequences.}
  \label{fig:anisoldm}
\end{figure}

While the onset of multiple scattering or the breakdown of \ion{He}{ii} ionizing flux already happen at lower $L/M$ (see Sect.\,\ref{sec:ionflux}), the `pure' WR wind regime is marked by the transition in $\dot{M}_\mathrm{t}(L/M)$ towards a shallow power law with a slope that is independent of metallicity. With $R_\text{crit}$ moving further inwards to higher optical depths, the departure from LTE should become lower. In perfect LTE, the radiation field is isotropic, so we can use the ratio of anisotropic to isotropic parts of the radiation field as a benchmark. Following \citet{HuMi2014}, we write
\begin{equation}
  \label{eq:anirat}
  \frac{\text{anisotropic}}{\text{isotropic}} \propto \frac{3H}{B} \propto \left(\frac{T_\text{eff}(r)}{T_\text{e}(r)}\right)^4
\end{equation}
and evaluate this ratio of effective temperature to local (electron) temperature at the critical point $R_\text{crit}$. The result for all model sequences is displayed in Fig.\,\ref{fig:anisoldm} and reveals that the `pure' WR-wind regime indeed occurs once the radiation field is mostly isotropic. All curves change their slope once the ratio in Eq.\,(\ref{eq:anirat}) falls below a value of $0.25$. An additional inspection of the departure coefficient in various models confirms our assumption that for models in the `pure' WR-wind regime, $R_\text{crit}$ is always located in a region with only minor or almost no departure from LTE, while this changes considerably when transitioning to more thin winds. 

In their stellar structure calculations, \citetalias{Grassitelli+2018} made a similar consideration and estimated this term to be on the order of $0.05$ for their models. This non-departure from LTE at the sonic point is an important requirement for their study as structure calculations rely on tabulated Rosseland opacities, while the actual wind driving depends on flux-weighted opacities. In LTE, both opacities are identical, while they can differ by a huge, depth-dependent amount in an expanding, non-LTE atmosphere \citepalias[cf.][]{Sander+2020}.  Our HD atmospheres now confirm the assumptions in \citetalias{Grassitelli+2018} to be justified as long as one is in the `pure' WR-wind regime. Unfortunately, A decent amount of the observed WR stars could actually be in the `transition regime', where LTE-departures still have to be considered at $R_\text{crit}$. Future modelling efforts and tailored analyses of well-constrained objects with HD atmospheres will be required to set proper benchmarks.   

\section{Application of the recipe suggested for VMS by Bestenlehner (2020)}
  \label{asec:bestencmp}

Using a modified definition of the mass to replace the explicit $M$-dependence with an expression of $\Gamma_\text{e}$ in the $\dot{M}$-prescription resulting from the CAK \citep*[after][]{Castor+1975} theory, \citet{Bestenlehner2020} suggested a $\Gamma_\text{e}$-dependent $\dot{M}$-recipe for the massive and very massive stars in R136. While not targeted for hydrogen-free stars, it also addresses the transition to WR-type mass loss, making it an interesting candidate to consider, in particular as earlier publications discussing this transitions only used a broken power-law approach with two different $\Gamma_\text{e}$-exponents \citep[e.g.][]{Vink+2011,VG2012,Bestenlehner+2014}.
The recipe for $\dot{M}(\Gamma_\text{e})$ in \citet{Bestenlehner2020} instead has the form
\begin{equation}
  \label{eq:BestenForm}
  \log \dot{M} = a + b \log \Gamma_\text{e} - c \log \left(1 - \Gamma_\text{e}\right)\text{.}
\end{equation} 
While this kind of formula in principle has three free parameters, the considerations of CAK and \citet{Bestenlehner2020} reduce these to two, as he replaces the parameters $b$ and $c$ with
\begin{equation}
  b = \frac{1}{\alpha} + \frac{1}{2}  \mbox{\hspace{5mm}\text{and}\hspace{5mm}} c = \frac{1-\alpha}{\alpha} + 2 =  \frac{1}{\alpha} + 1 = b + \frac{1}{2}
\end{equation}

\begin{figure}
  \includegraphics[width=\columnwidth]{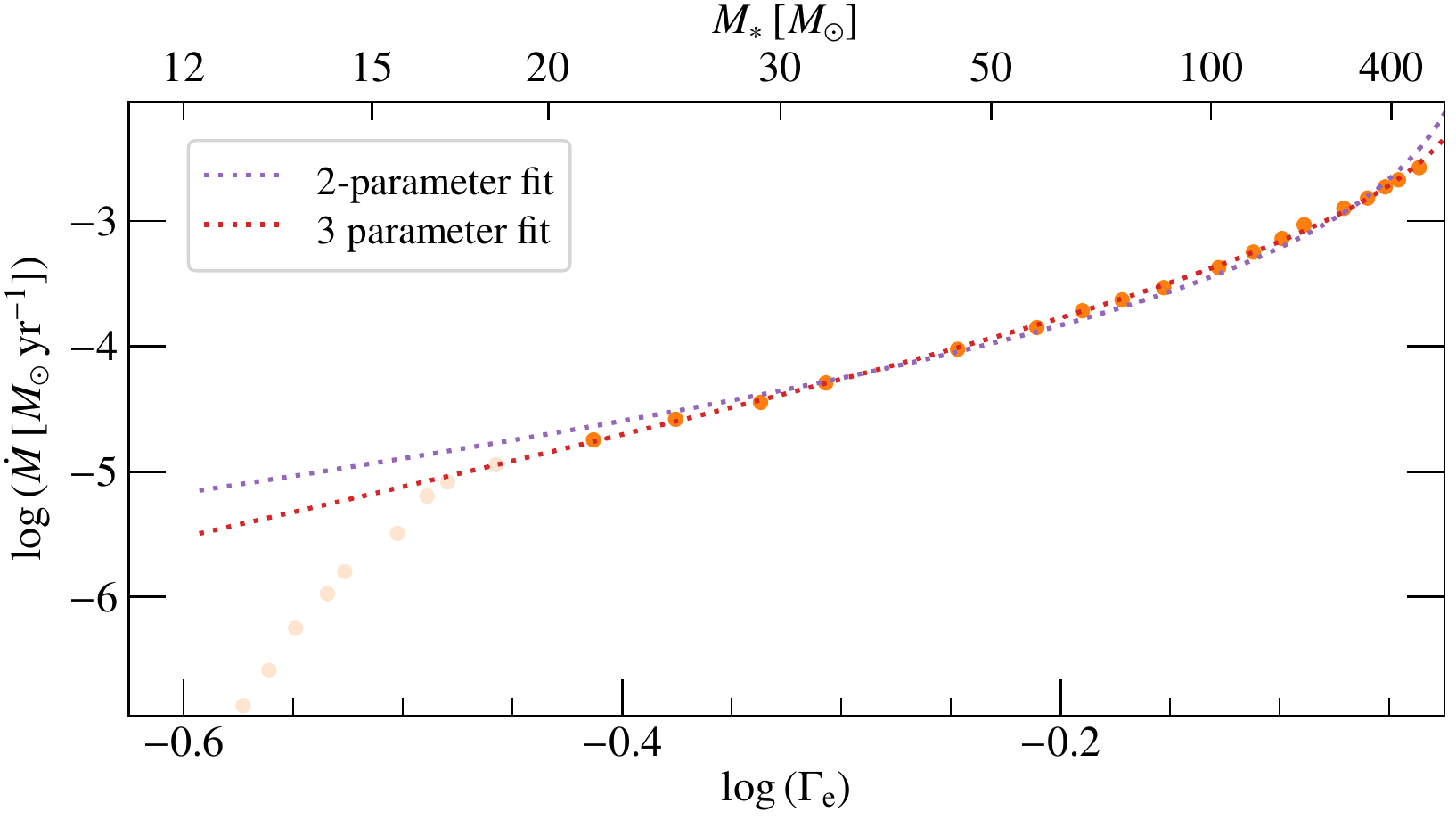}
  \caption{$\dot{M}(\Gamma_\text{e})$ for the model sequence with $Z = 0.5\,Z_\odot$ with fits according to Eq.\,(\ref{eq:BestenForm}): The purple dashed line denotes the fit with two free parameters ($a, \alpha$) as suggested by \citet{Bestenlehner2020}, while the red dashed line represents a fit with three independent parameters ($a, b, c$). The breakdown part of $\dot{M}(\Gamma_\text{e})$ (lighter points) has been omitted in both fits.}
  \label{fig:best05}
\end{figure}

\begin{figure}
  \includegraphics[width=\columnwidth]{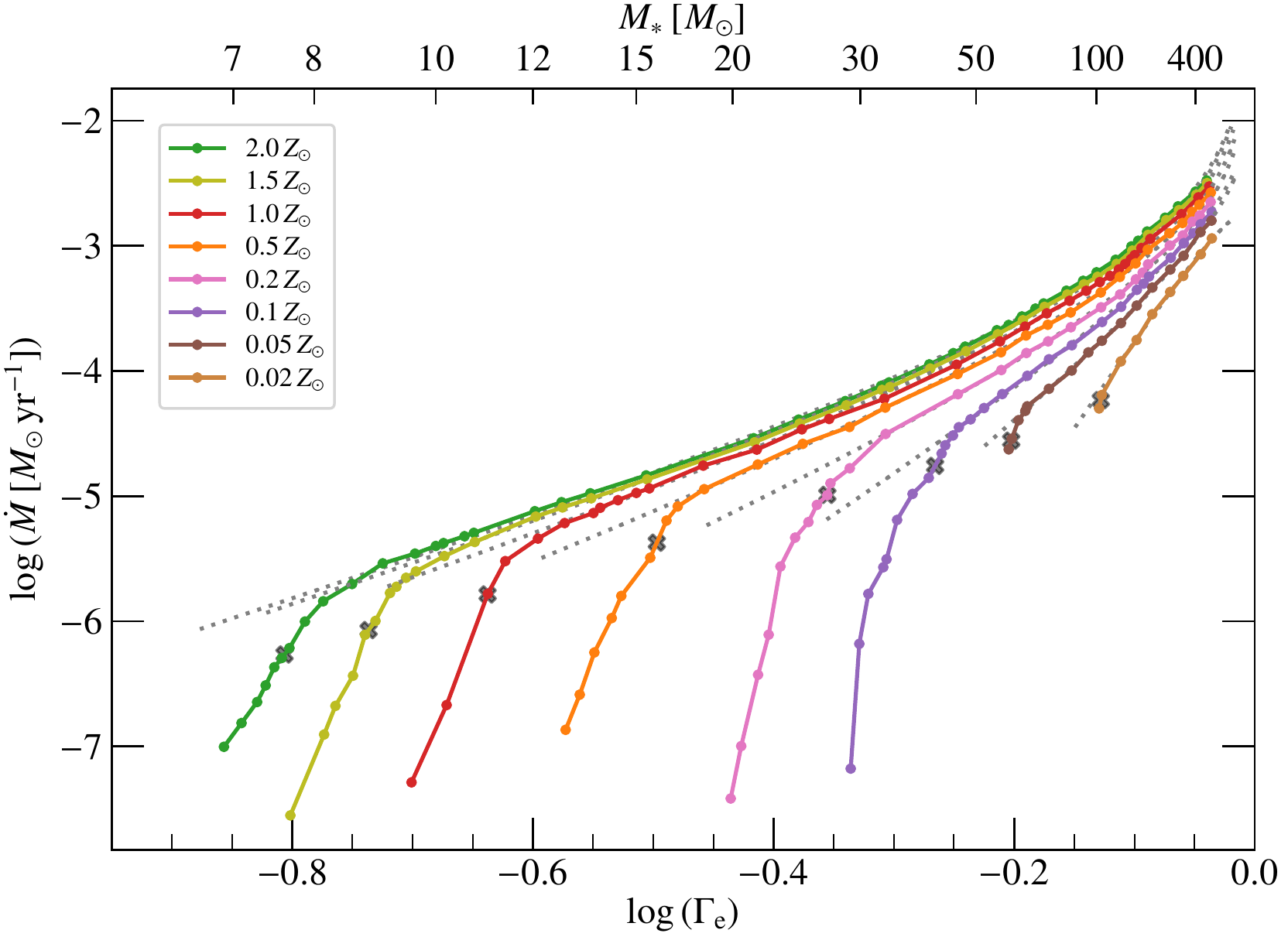}
  \caption{$\dot{M}(\Gamma_\text{e})$ for all model sequences with fits according to Eq.\,(\ref{eq:BestenForm}) using three independent parameters.}
  \label{fig:mdot-gedd}
\end{figure}
\begin{figure}
  \includegraphics[width=\columnwidth]{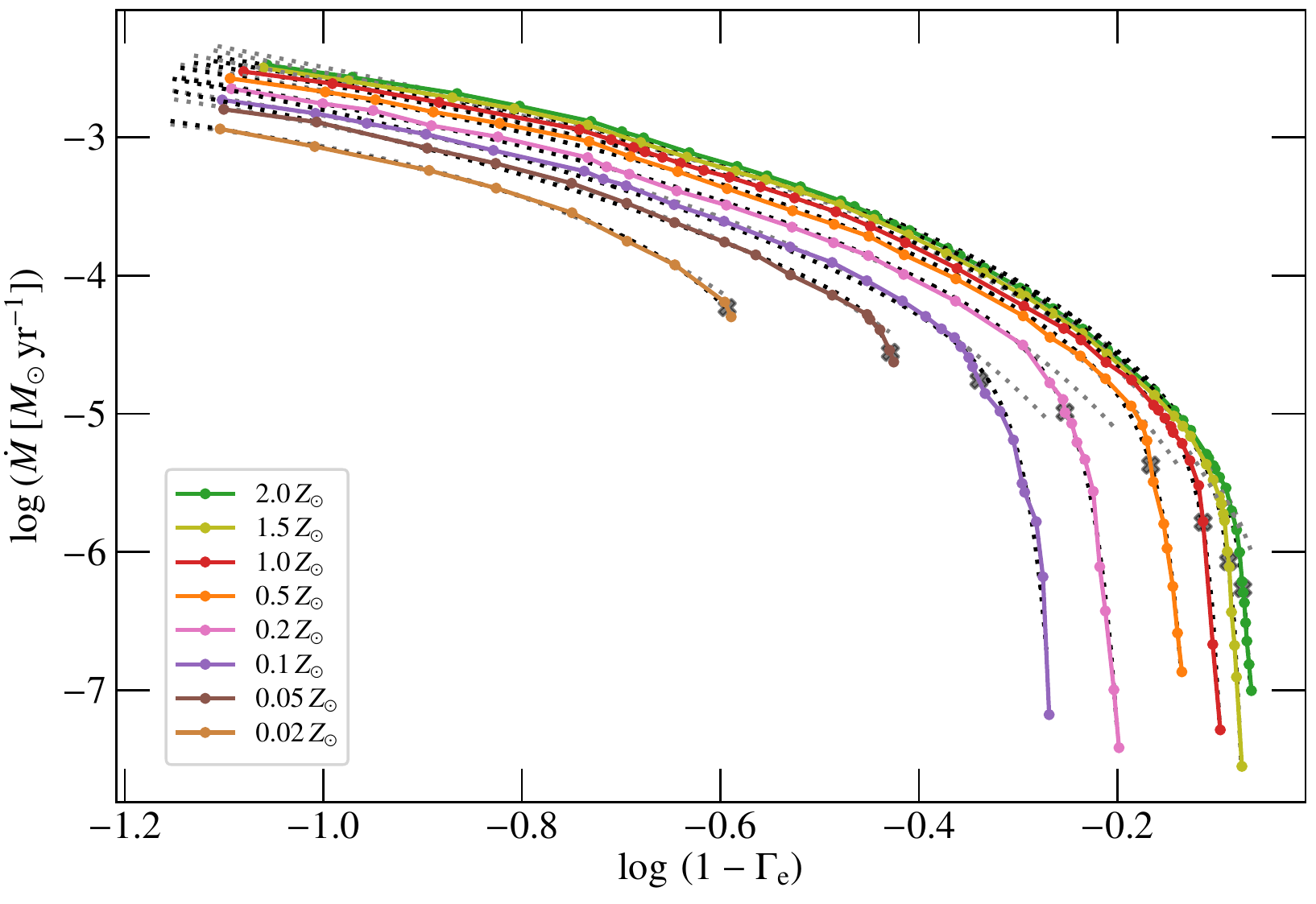}
  \caption{$\dot{M}(1-\Gamma_\text{e})$ for all model sequences with fits according to Eq.\,(\ref{eq:BestenForm}) using three independent parameters (grey dashed lines) and according to our eventual $\dot{M}(\Gamma_\text{e})$-recipe (Eq.\,\ref{eq:mdemrecipe}, black dashed lines).}
  \label{fig:mdot-emgedd}
\end{figure}

Such a recipe is not able to reproduce the steep decline in $\dot{M}$ at lower $\Gamma_\text{e}$, but captures the general behaviour of the remaining trend for optically thick, WR-type winds. The two-parameter version of Eq.\,(\ref{eq:BestenForm}) with a CAK-type $\alpha$-description yields $\alpha \approx 0.8\dots0.9$ with a small increase in $\alpha$ towards higher $Z$. However, as Fig.\,\ref{fig:best05} illustrates, the version with just two free parameters does not provide a good representation of the overall slope and has systematic offsets for the asymptotic behaviour, not just in the pictured example, but for all metallicities. The three-parameter version reproduces the slope much better and also captures the right asymptotic behaviour towards the highest masses. In Fig.\,\ref{fig:mdot-gedd}, we show a 3-parameter fit to each of the detailed datasets. The plot of the multiple datasets as a function of $\Gamma_\mathrm{e}$ illustrates the motivation of requiring two dependencies, namely $\Gamma_\mathrm{e}$ and $1-\Gamma_\mathrm{e}$. When the regime of pure WR-type mass loss is reached, the curves first appear to scale linear with $\log \Gamma_\text{e}$ but then transition into a much steeper dependence when approaching the Eddington Limit of $\Gamma_\mathrm{e} = 1$. The second term with the $\left(1-\Gamma_\mathrm{e}\right)$-dependence tries to address this, assuming that the slope will eventually scale with $\left(1-\Gamma_\mathrm{e}\right)$. While such a $\left(1-\Gamma_\mathrm{e}\right)$-dependency is already in the original CAK recipe, the effect of this term only kicks in for $\Gamma_\text{e}$-values very close to unity, which was the motivation for \citet{Bestenlehner2020} to update this recipe with an adjusted expression for the stellar mass.

Despite its success in the pure WR-wind regime, Eq.\,(\ref{eq:BestenForm}) does not account for the breakdown of WR-type mass loss for lower $\Gamma_\text{e}$, although containing already three free parameters. Nonetheless, it points towards further examining $\dot{M}(1-\Gamma_\mathrm{e})$, which we depict in Fig.\,\ref{fig:mdot-emgedd}. Inspecting only masses up to $\approx 150\,M_\odot$ -- already way beyond the observed ranged for helium stars -- could indeed lead to the conclusion of an asymptotic power-law for $\dot{M}(1-\Gamma_\mathrm{e})$, albeit with different power-law indices. Given that we calculated models up to $500\,M_\odot$ for all the considered metallicities, the smooth bending of all the curves in Fig.\,\ref{fig:mdot-emgedd} instead of an eventual linear turnover is evident. Moreover, the original motivation for Eq.\,(\ref{eq:BestenForm}), namely the broken power-law in $\Gamma_\mathrm{e}$ with a `kink' in a transition regime, is also not favourable from a physical standpoint: There is no further obvious transition inside the pure WR-wind regime, but instead a smooth curvature of $\dot{M}(\Gamma_\mathrm{e})$ for higher $\Gamma_\mathrm{e}$ in Fig.\,\ref{fig:mdot-gedd}. This is eventually reflected in our $\dot{M}(\Gamma_\text{e})$-recipe presented in Sect.\,\ref{sec:mdotgamma}.

\section{Comparison with the metallicity-dependence of WNh stars}
  \label{asec:gh2008}

\begin{figure}
  \includegraphics[width=\columnwidth]{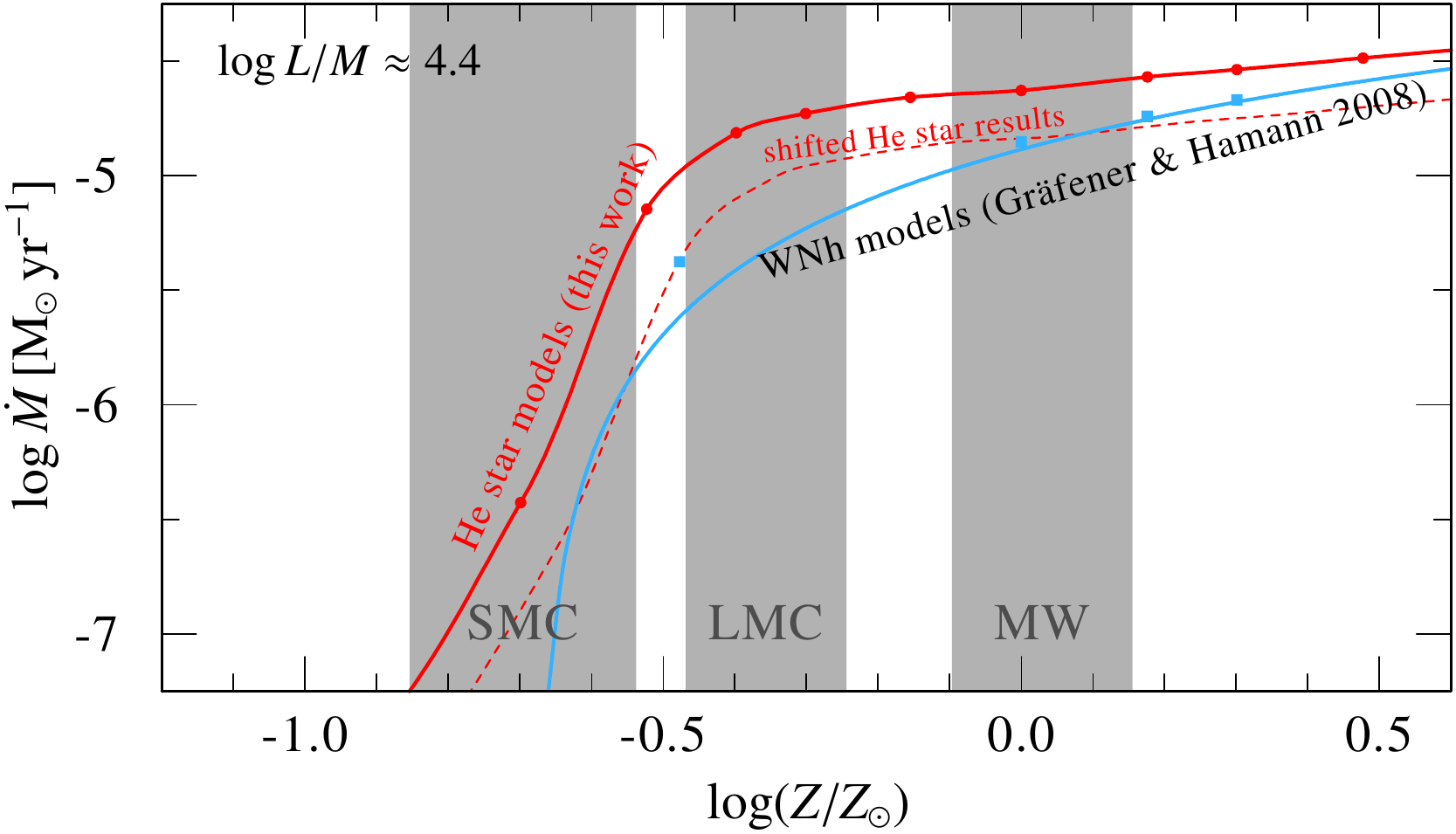}
  \caption{Comparison of the WNh results by \citetalias{GH2008} (blue) with our hot Fe bump'-driven He star results (red) for a similar $\log L/M$.}
  \label{fig:cmpgh2008}
\end{figure}

\citetalias{GH2008} were the first to calculate a series of CMF-based HD WR models. Albeit these models were calculated to describe the winds of the luminous and H-rich WNh stars driven by the so-called `cool iron bump', the results from \citetalias{GH2008} were an early indication for the complexity of the $\dot{M}(Z)$-behaviour in WR-type winds. \citetalias{GH2008} compiled their findings in an $\dot{M}$-recipe of the form
\begin{align}
  \log\dot{M} &= \alpha_\textsc{gh} + \beta_\textsc{gh}(Z) \cdot \log\left[\Gamma_\text{e} - \Gamma_{\textsc{gh},0}(Z)\right] \\
	  \nonumber &\mbox{\hspace{1cm}}- \gamma_\textsc{gh} \log T_\ast + \delta_\textsc{gh} \log L - 0.45 X_\text{H} \\
	\nonumber\text{with} \\
	\nonumber\beta_\textsc{gh}(Z) &= 1.727 + 0.250 \cdot \log(Z/Z_\odot) \\
	\nonumber\Gamma_{\textsc{gh},0}(Z) &= 0.326 - 0.301 \cdot \log(Z/Z_\odot) - 0.045 \left[ \log(Z/Z_\odot) \right]^2\text{.}
\end{align}
This formula accounts for the dramatic change in the slope of $\dot{M}$, as their CMF calculations avoid an incorrect assignment of outer wind opacities to artificially boost $\dot{M}$. However, despite the complexity in $Z$, the formula of \citetalias{GH2008} is already a compromise fit to their data, which is evident in Fig.\,\ref{fig:cmpgh2008} where we plot their dataset for $\Gamma_\text{e} = 0.55$ as well as the relation from the corresponding recipe. For their lowest data point ($Z = Z_\odot/3$), their calculated $\dot{M}$ is approximately $0.25\,$dex higher than derived by their recipe. The formula further predicts a steep drop in $\dot{M}$ with a critical $\Gamma_{\textsc{gh},0}$ where the mass loss would become zero. This is an artefact of the chosen formula and thus \citetalias{GH2008} state that the validity of their description is only limited down to $\log \dot{M} \approx -5.5$. A shift of the relation from our He star models (dashed line in Fig.\,\ref{fig:cmpgh2008}) hints that WNh winds might show a $Z$-trend similar to classical WR stars, albeit with a slightly steeper slope in the pure WR-wind regime.

\bsp	
\label{lastpage}
\end{document}